\PassOptionsToPackage{square,numbers,comma,sort&compress,super}{natbib} %
\documentclass[aps,prl,twocolumn,a4paper,superscriptaddress]{revtex4-2}

\usepackage[paperwidth=215mm,paperheight=300mm,centering,hmargin=1.6cm,vmargin=2cm]{geometry}
\usepackage{graphicx}
\graphicspath{{./}}%
\usepackage{amssymb,amsmath,amsthm,enumitem}

\usepackage{latexsym,stmaryrd}
\usepackage[dvipsnames]{xcolor}
\usepackage[normalem]{ulem}
\usepackage{float}

\usepackage{siunitx} %
\usepackage{physics} %
\usepackage{chemformula} %
\usepackage{titlesec}
\usepackage{gensymb}
\usepackage{xfrac} %
\usepackage[T1,OT1]{fontenc}
\usepackage{dcolumn}
\usepackage{bm}
\usepackage{booktabs}
\usepackage{rotating}

\renewcommand{\topfraction}{0.95}
\renewcommand{\bottomfraction}{0.95}
\renewcommand{\textfraction}{0.05}
\renewcommand{\floatpagefraction}{0.95}

\titleformat{\section}[hang]{\bfseries}{}{1em}{}
\titlespacing*{\section}{0pt}{2.5ex plus 1ex minus .2ex}{0.2ex plus .2ex}
\titleformat{\subsection}[hang]{\bfseries}{}{1em}{}
\titlespacing*{\subsection}{0pt}{2.5ex plus 1ex minus .2ex}{0.2ex plus .2ex}

\renewcommand{\figurename}{\textbf{Fig.}}

\makeatletter
\renewcommand*{\fnum@figure}{{\normalfont\bfseries \figurename~\thefigure}}
\makeatother

\newcommand{\EPFL}{Institute of Physics, Swiss Federal Institute of Technology Lausanne (EPFL), CH-1015 Lausanne, Switzerland}
\newcommand{\CfQS}{Center for Quantum Science and Engineering, EPFL, Lausanne, Switzerland}
\newcommand{\IEM}{Institute of Electrical and Micro Engineering, Swiss Federal Institute of Technology Lausanne (EPFL), Lausanne, Switzerland}
\newcommand{\Purdue}{OxideMEMS Lab, Purdue University, West Lafayette, Indiana 47907, USA}

\newcommand{\Vpzt}{V_\text{pzt}}
\newcommand{\Pdet}{P_\text{det}}
\newcommand{\fm}{f_\text{m}} 
\newcommand{\SiNxy}{\text{Si}_x\text{N}_y} %
\newcommand{\Omeas}{\tilde{\Omega}_\text{m}}                          %
\newcommand{\SOmeas}{S_{\tilde{\Omega}_\text{m}\tilde{\Omega}_\text{m}}} %

\newcommand{\avg}[1]{\langle#1\rangle}

\newcommand{\thmeas}{\tilde{\theta}}                              %
\newcommand{\Sthmeas}{S_{\tilde{\theta}\tilde{\theta}}}           %
\newcommand{\Ogmeas}{\tilde{\Omega}}                              %
\newcommand{\SOgmeas}{S_{\tilde{\Omega}\tilde{\Omega}}}           %

\usepackage[hidelinks]{hyperref}

\begin{document}
\makeatletter
\let\SI@addcontentsline\addcontentsline
\let\addcontentsline\@gobblethree
\let\auto@bib@innerbib\@empty
\makeatother
\pdfbookmark[0]{Main text}{main}
\raggedbottom %

\title{A photonic-integrated optomechanical mid-infrared detector}

\author{Guillermo Arregui}\thanks{These authors contributed equally to this work.}
\affiliation{\EPFL}
\affiliation{\IEM}
\affiliation{\CfQS}

\author{Amirali Arabmoheghi}\thanks{These authors contributed equally to this work.}
\affiliation{\EPFL}
\affiliation{\IEM}
\affiliation{\CfQS}

\author{Alessio Zicoschi}\thanks{These authors contributed equally to this work.}
\affiliation{\EPFL}
\affiliation{\IEM}
\affiliation{\CfQS}

\author{Xinru Ji}
\affiliation{\EPFL}
\affiliation{\IEM}
\affiliation{\CfQS}

\author{Scott E. Kenning}
\affiliation{\Purdue}

\author{Marco Golino}
\affiliation{\EPFL}
\affiliation{\IEM}
\affiliation{\CfQS}

\author{Shivaprasad U. Hulyal}
\affiliation{\EPFL}
\affiliation{\IEM}
\affiliation{\CfQS}

\author{Yi Xia}
\affiliation{\EPFL}
\affiliation{\IEM}
\affiliation{\CfQS}

\author{Sunil A. Bhave}
\affiliation{\Purdue}

\author{Tobias J. Kippenberg}
\thanks{Corresponding authors: guillermo.arregui@epfl.ch, tobias.kippenberg@epfl.ch}
\affiliation{\EPFL}
\affiliation{\IEM}
\affiliation{\CfQS}

\homepage{}

\small
\maketitle

\noindent{\bfseries\boldmath\sisetup{detect-all} Tracking the resonance frequency of a mechanical oscillator is the basis of atomic force microscopy, mass sensing and thermometry. Nanomechanical thermal detectors exploit this principle to sense absorbed radiation, most effectively with tensile-stressed resonators, whose tension converts thermal expansion into a large frequency shift and dilutes dissipation. Approaching their fundamental sensitivity limit, however, has required free-space or fibre-coupled interferometers, hindering compact deployment and scaling to arrays. Here we remove this bottleneck by monolithically integrating a high-$Q$ nanomechanical resonator with a silicon nitride photonic integrated circuit. A tensile-stressed silicon nitride trampoline with an impedance-matched platinum absorber is dispersively coupled to a photonic integrated microresonator beneath it. At room temperature and with microwatt-level optical power, the detector reaches a noise-equivalent power of 3.4\,pW/$\sqrt{\text{Hz}}$ for broadband thermal radiation (3--12\,\si{\micro\meter}), referred to incident power. Referred to absorbed power, this is within a factor of 1.5 of its fundamental thermal-fluctuation limit. Using our detector in a state-of-the-art commercial Fourier-transform infrared spectrometer, we resolve the vibrational fingerprint of a polymer film across 2--20\,\si{\micro\meter} and reach the precision of the built-in pyroelectric detector with about an order of magnitude lower incident power. This architecture opens a route to multiplexed optomechanical focal-plane arrays.\par}
\medskip

\noindent Mid-infrared (mid-IR) detection is central to science and engineering, from environmental monitoring~\cite{patel_spectroscopic_1974} to the study of planet formation~\cite{van_boekel_building_2004}. While scientific applications push for ever higher sensitivity, emerging ones such as autonomous navigation~\cite{bhadoriya_vehicle_2022} and smart cities~\cite{napolitano_smart_2021} require detectors that combine high performance with low size, weight, power and cost. Narrow-bandgap photon detectors typically meet the performance requirement only with cryogenic cooling, as their dark current rises exponentially with temperature~\cite{gordon_room-temperature_2021,rogalski_infrared_2019}. Thermal detectors, which sense the heat generated by absorbed radiation through a temperature-dependent property, range from cryogenic bolometers, which revealed the near-perfect blackbody spectrum of the cosmic microwave background~\cite{mather_preliminary_1990}, to uncooled microbolometer arrays. The latter operate at room temperature and are volume-manufactured on CMOS readout circuits~\cite{rogalski_progress_2012}, but they remain at least several times above the fundamental limit set by thermodynamic temperature fluctuations~\cite{rogalski_infrared_2019,kohin_performance_2004,deng_silicon_2026}.

Since Langley's bolometer of 1880~\cite{langley_bolometer_1880}, most thermal detectors have relied on a suspended, low-heat-capacity absorber linked to its surroundings by thin supports of low thermal conductance~\cite{bock_novel_1995}, with its temperature read by an electrical thermometer. It is this thermometer that limits uncooled microbolometers, in two ways. First, the bias current needed to lift the signal above the Johnson noise of the readout, carried through supports that must double as electrical leads, also heats the element it measures. Second, the few, thermally activated carriers that give semiconducting films their high temperature sensitivity make their resistance flicker, with a $1/f$ noise that no bias can remove~\cite{kohin_performance_2004,yadav_advancements_2022}. Alternative thermometers that avoid resistive readout have therefore been explored, including the dispersion of exciton polaritons~\cite{paschos_exciton-polariton_2018}, magnetization switching rates~\cite{bauer_exploiting_2025}, and the resonance frequency of optical~\cite{xia_integrated_2025,follet_all-optical_2025} and mechanical resonators~\cite{kao_micromachined_2009}.

Among these, tensile-stressed nanomechanical resonators are particularly suited, as their low-frequency modes combine enhanced temperature sensitivity~\cite{pandey_performance_2010} with quality factors far exceeding material limits~\cite{verbridge_high_2006}, which has led to mechanical thermal detectors (MTDs) based on silicon nitride~\cite{chien_single-molecule_2018,piller_nanoelectromechanical_2019,vicarelli_micromechanical_2022} and graphene~\cite{blaikie_fast_2019}. Because the signal is carried by a resonance frequency rather than by a resistance at DC, even electrically transduced MTDs reach noise-equivalent powers (NEPs) comparable to those of uncooled microbolometers~\cite{piller_thermal_2023}. Electrical transduction, however, still requires conductive layers or electrodes at the resonator, and its imprecision is set by the amplification chain~\cite{ekinci_electromechanical_2005,schmid_fundamentals_2023}. Optical readout removes both: it requires no contact with the resonator and can approach the quantum limits of displacement measurement~\cite{schilling_near-field_2016}. Optically read MTDs have accordingly approached their fundamental limit~\cite{martini_uncooled_2025}, with frequency stability limited by thermodynamic temperature fluctuations~\cite{zhang_demonstration_2023,zhang_enhanced_2025}. These devices, however, rely on laboratory-scale free-space or fibre-based interferometers~\cite{he_uncooled_2026}, requiring per-device alignment that is at odds with compact and scalable sensing architectures.

\begin{figure*}
    \centering
    \includegraphics[width=\textwidth]{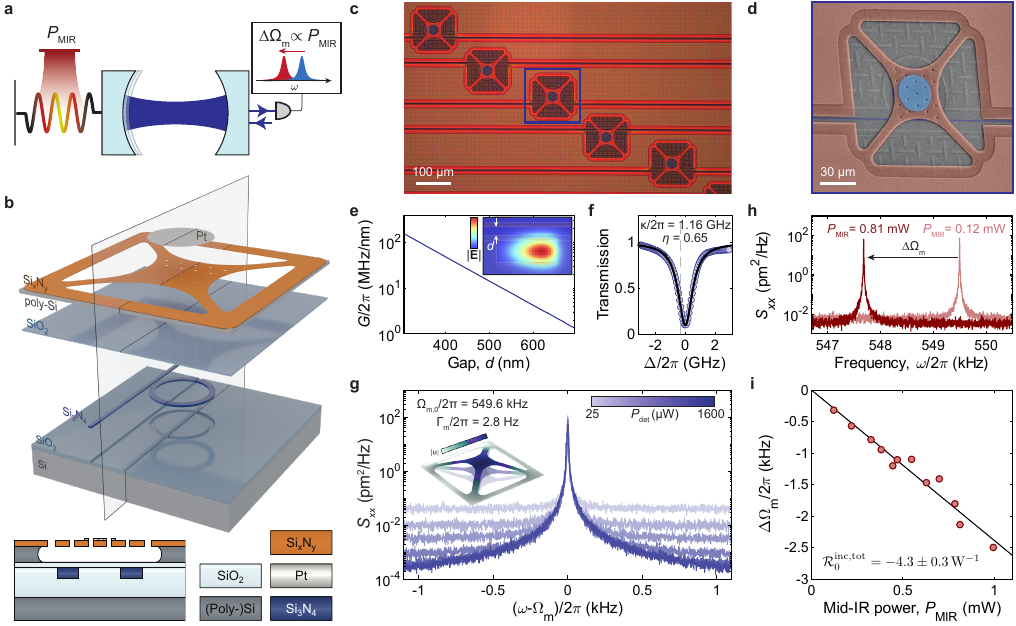}
    \caption{\textbf{Monolithic integration of a mechanical thermal detector (MTD) with a photonic integrated circuit.} \textbf{a}, Schematic of the cavity-optomechanical MTD (OMTD) concept, whereby cavity-enhanced optical detection of mechanical motion is used to determine mechanical frequency changes, $\Delta\Omega_\text{m}$, proportional to the incident mid-infrared radiation, $P_\text{MIR}$. \textbf{b}, Exploded view of the device, which consists of a platinum-coated trampoline nanomechanical resonator suspended hundreds of nanometres above a Si$_3$N$_4$ photonic ring microresonator coupled to a bus waveguide. Bottom: cross-section along the vertical plane shown. \textbf{c}, Optical microscope image of a few OMTDs suspended above respective ring resonators. The bus waveguides are visible through the air channels opened above them. \textbf{d}, False-coloured SEM micrograph of a single OMTD. Uncoloured: SiO$_2$ cladding with Si$_3$N$_4$ filler pattern, navy blue: Si$_3$N$_4$ PICs, red: Si$_x$N$_y$ nanomechanical resonator layer, light blue: Pt disk absorber. \textbf{e}, Finite-element-method (FEM) simulation of the frequency-pull parameter $G$ of the ring resonance as a function of the gap $d$. The inset shows the cross-section of the electric-field magnitude, $|\mathbf{E}|$, of the mode and the electric field lines. \textbf{f}, Measured and fitted normalized transmission of the TM$_{00}$ resonance used in the experiments, with $\eta = \kappa_\text{e}/\kappa$. The dashed line marks the maximum-slope detuning used in \textbf{g} and \textbf{h}. \textbf{g}, Calibrated displacement noise power spectral density (PSD) for the fundamental mechanical mode of the trampoline as a function of the detection power $\Pdet$ sent to the chip. The inset shows the mode profile. \textbf{h}, Same as \textbf{g} for two different incident mid-IR powers, $P_\text{MIR}$. \textbf{i}, Measured dependence of the frequency shift on the incident mid-IR power and linear fit, whose slope gives the fractional responsivity to the total incident power, $\mathcal{R}_0^\text{inc,tot} = (\Delta\Omega_\text{m}/\Omega_\text{m,0})/P_\text{MIR}$.}
    \label{fig:figure1}
\end{figure*}

Photonic integrated circuits (PICs) are a natural route to integrating such readout. They already serve as the optical interface to trapped ions for quantum computing~\cite{niffenegger_integrated_2020,mehta_integrated_2020}, atomic vapours for optical clocks~\cite{newman_architecture_2019} and suspended membranes for ultrasound detection~\cite{westerveld_sensitive_2021}. With propagation losses now below 1\,dB/m~\cite{liu_high-yield_2021}, ultra-low-loss Si$_3$N$_4$ PICs also host chip-scale frequency combs~\cite{gaeta_photonic-chip-based_2019}, erbium-doped amplifiers~\cite{liu_photonic_2022} and low-noise lasers~\cite{lihachev_low-noise_2022,ji_wafer-scale_2026}, the building blocks of a self-contained, multiplexed optical readout. For nanomechanical resonators, chip-scale near-field optical readout is well established, both with resonators positioned next to stand-alone microcavities~\cite{anetsberger_near-field_2009} and with resonators co-fabricated with the optics on dedicated chips~\cite{fong_high_2010,gavartin_hybrid_2012,schilling_near-field_2016,yu_uncooled_2026}. In each case, however, the photonic and mechanical elements are designed and fabricated together for the purpose. A modular architecture, in which the mechanical layer is added to an existing PIC platform, inheriting its ecosystem while remaining free to be optimized on its own, has not yet been demonstrated for high-$Q$, tensile-stressed nanomechanical resonators.

Here we demonstrate such an architecture for mid-IR detection. A high-$Q$, tensile-stressed Si$_x$N$_y$ trampoline resonator, functionalized with an impedance-matched platinum absorber, is surface-micromachined with standard MEMS processes above a foundry-compatible Si$_3$N$_4$ PIC, and a microring resonator of the circuit, directly beneath the trampoline, reads out its motion through near-field coupling. At microwatt-level on-chip power, this cavity-enhanced readout leaves the detector limited, at the frequencies of best sensitivity, by intrinsic fluctuations of the resonance frequency, which also follow a $1/f$ spectrum~\cite{bachtold_mesoscopic_2022}. With a pixel-scale absorber and a millisecond response, the detector reaches a room-temperature minimum NEP of 3.4\,\si{\pico\watt/\sqrt{\hertz}} referred to incident power and, referred to absorbed power, lies within a factor of 1.5 of its fundamental thermal-fluctuation limit. Deployed at an auxiliary port of a commercial Fourier-transform infrared (FTIR) spectrometer, alongside its built-in pyroelectric detector, it records broadband mid-IR transmission spectra in quantitative agreement with those of the reference instrument.

\section{Si$_3$N$_4$ PIC-based optomechanical readout}
\label{sec:platform}
Figure~\ref{fig:figure1}(a) illustrates our cavity-optomechanical approach to MTDs with the canonical representation of cavity optomechanics~\cite{aspelmeyer_cavity_2014}: a laser drives an optical cavity whose end mirror is suspended on a spring, and the dispersive coupling between the intracavity field and the mirror motion allows the displacement of the mirror to be read out. Upon absorption of incident radiation, the resonator heats up, altering its spring constant and thereby shifting its resonance frequency by $\Delta\Omega_\text{m}$~\cite{liu_room-temperature_2022}. Figure~\ref{fig:figure1}(b) shows our implementation in an exploded view and a cross-section. The bottom layers form the SiO$_2$-clad Si$_3$N$_4$ PIC used for readout, a ring microresonator side-coupled to a bus waveguide. Above them, a trampoline resonator made from silicon-rich Si$_x$N$_y$ with moderate tensile stress (about 170\,MPa, versus about 1\,GPa for stoichiometric Si$_3$N$_4$), which enhances the photothermal frequency shift arising from thermal expansion~\cite{pandey_performance_2010}, is anchored through polysilicon (poly-Si) supports and suspended hundreds of nanometres above the thin cladding covering the ring. A platinum (Pt) disk on its central pad serves as an impedance-matched mid-IR absorber~\cite{hadley_reflection_1947} while adding little heat capacity and mechanical dissipation~\cite{piller_thermal_2023,zhang_high_2024,martini_uncooled_2025}.

\begin{figure*}[t]
    \centering
    \includegraphics[width=\textwidth]{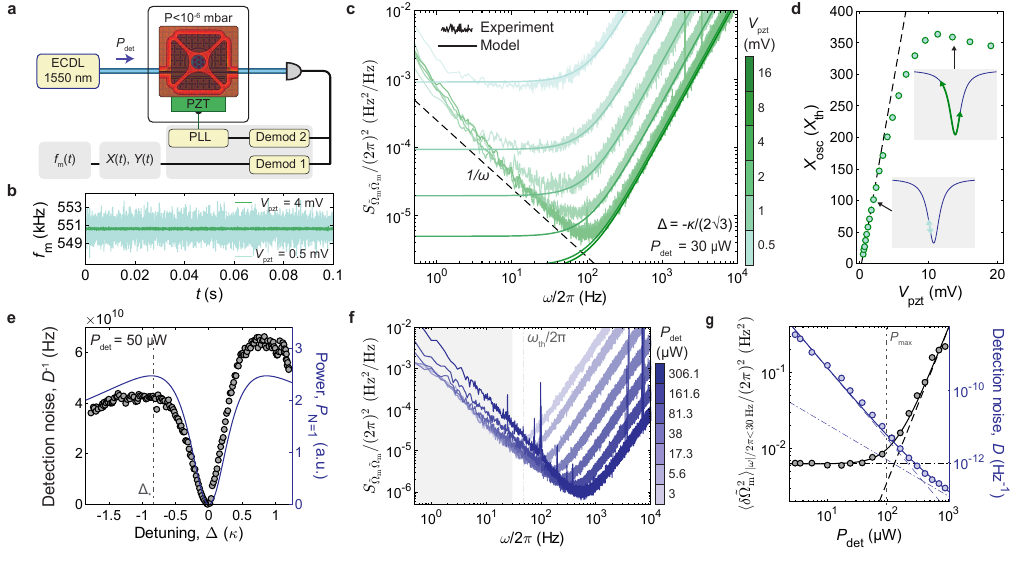}
        \caption{\textbf{Intrinsic, detection and photothermal contributions to the frequency noise.} \textbf{a}, Schematic of the setup used for measuring the frequency $\fm$ of the fundamental mode of the trampoline resonator over time. ECDL: external-cavity diode laser, PZT: piezoelectric transducer, PLL: phase-locked loop. Demod~2 is part of the PLL, while $\fm(t)$ is extracted from the quadratures $X(t)$, $Y(t)$ of the out-of-loop Demod~1. \textbf{b}, Representative time traces of $\fm$ for two driving voltages, $\Vpzt$. \textbf{c}, Power spectral density (PSD) of the frequency fluctuations for varying $\Vpzt$ at a fixed laser--cavity detuning, $\Delta$, and off-chip optical power, $\Pdet$. The solid lines represent Eq.~(\ref{eq:Sff}) and the dashed line is a $1/\omega$ guide to the eye. \textbf{d}, Detected oscillation amplitude, $X_\text{osc}\equiv\sqrt{\langle X(t)^2+Y(t)^2\rangle}$, calibrated assuming linear transduction, as a function of $\Vpzt$. The dashed line represents a linear fit to the data at $\Vpzt< 5$ mV. Schematics: motion-induced modulation of the cavity transmission in the linear ($GX_\text{osc}\ll\kappa/2$) and nonlinear ($GX_\text{osc}\gtrsim\kappa/2$) transduction regimes. \textbf{e}, Inverse of the detection noise coefficient ($D$ from a fit of the PSD at large sideband frequencies $\omega$ with $D\omega^2$) as a function of $\Delta$ (grey dots, left axis) and theoretical maximum of the detected optical power in the first harmonic, $P_\text{N=1}$, assuming a perfect Lorentzian cavity response (blue solid line, right axis). The detuning $\Delta_*$ of subsequent measurements is highlighted with a dash-dotted line. \textbf{f}, PSD of the frequency fluctuations measured at varying $\Pdet$. The shaded area marks the range $\left|\omega\right|/2\pi < 30$\,\si{\hertz} and the dash-dotted line the thermal cutoff $\omega_\text{th}/2\pi$. \textbf{g}, Band-limited (shaded area in \textbf{f}) variance of the measured frequency, $\langle\delta\Omeas^2\rangle_{\left|\omega\right|/2\pi < 30\,\si{\hertz}}$ (grey, left axis), and detection noise coefficient $D$ (blue, right axis) as a function of $\Pdet$. Solid lines are fits to $s_0+s_2\Pdet^2$ and $d_2/\Pdet^2+d_1/\Pdet$, respectively, and broken lines their individual contributions.} 
    \label{fig:figure2}
\end{figure*}

The Si$_3$N$_4$ PICs are defined by the photonic damascene process~\cite{pfeiffer_photonic_2016,liu_high-yield_2021} and buried under a thin SiO$_2$ cladding and a sacrificial poly-Si layer that nominally sets the gap $d$ between the trampoline and the cladding. The Pt-coated Si$_x$N$_y$ resonators are patterned above this stack and released by XeF$_2$ etching of the poly-Si (Methods). Figures~\ref{fig:figure1}(c) and (d) show a microscope image of several optomechanical thermal detectors (OMTDs) on a chip and a false-colour scanning electron micrograph of one of them. Above the bus waveguide, the high-refractive-index poly-Si is removed along an air channel for efficient guiding. The chip is fibre-packaged and operated in vacuum, and, unless stated otherwise, all results reported here are for one of its OMTDs (Methods).

The evanescent field of the photonic microresonator modes, overlapping with the trampoline (inset of Fig.~\ref{fig:figure1}(e)), gives rise to near-field dispersive optomechanical coupling via the gradient optical force~\cite{anetsberger_near-field_2009}, with a frequency-pull parameter $G = |\partial\omega_0/\partial d|$ (with $\omega_0$ the optical resonance frequency) that decays exponentially with $d$ (Fig.~\ref{fig:figure1}(e)). We use the fundamental transverse-magnetic (TM$_{00}$) mode family, which provides about 50\% more coupling than the fundamental transverse-electric family (Supplementary Section~1.3), and drive a single resonance at $\lambda_0$ = 1555.24\,nm with a continuous-wave laser ($\kappa/2\pi$ = 1.16\,GHz, $\eta = \kappa_\text{e}/\kappa = 0.65$, Fig.~\ref{fig:figure1}(f) and Supplementary Section~1.2). Direct detection of the transmitted intensity on the cavity slope (dashed line in Fig.~\ref{fig:figure1}(f)) resolves the thermomechanical motion of the fundamental out-of-plane flexural mode (inset of Fig.~\ref{fig:figure1}(g)) with microwatt-level on-chip power, as shown in Fig.~\ref{fig:figure1}(g) for varying optical power sent to the chip, $\Pdet$. The mode has an unperturbed frequency $\Omega_\text{m,0}/2\pi$ = 549.6\,kHz and a linewidth $\Gamma_\text{m}/2\pi$ = 2.80 $\pm$ 0.09\,Hz, corresponding to $Q\approx 2\times10^{5}$, and its calibrated vacuum optomechanical coupling rate, $g_0/2\pi = 71 \pm 5$\,Hz, is within 20\% of the value from finite-element-method (FEM) simulations (Methods and Supplementary Sections~1.3--1.5).

Illuminating the OMTD with a focused mid-IR thermal source (Methods) shifts its thermomechanical spectrum (Fig.~\ref{fig:figure1}(h)) linearly with the incident power, by up to nearly a thousand mechanical linewidths (Fig.~\ref{fig:figure1}(i)).

\section{Frequency tracking and noise budget}
\label{sec:freqNoise}

The shifts in Figs.~\ref{fig:figure1}(h,i) are inferred from thermomechanical spectra averaged over long times. Operating the detector instead requires tracking the resonance frequency continuously. The most established approach drives the resonator coherently and extracts its frequency from the phase of the demodulated displacement signal~\cite{albrecht_frequency_1991}. We implement it with the scheme of Fig.~\ref{fig:figure2}(a), in which a phase-locked loop (PLL) keeps an off-chip piezoelectric drive (PZT) on resonance and the frequency estimate $\Omeas$ is obtained from a separate, out-of-loop demodulator (Methods). In this scheme, fluctuations of $\Omeas$ combine real fluctuations of $\Omega_\text{m}$, which can have several physical origins, with two measurement artefacts~\cite{fong_frequency_2012,sadeghi_frequency_2020,kanellopulos_comparative_2025,schmid_fundamentals_2023}: the thermomechanical and detection noise imprinted on the measured oscillation phase by the Brownian motion of the resonator and by the finite displacement imprecision. Their double-sided power spectral density (PSD) reads (Supplementary Section~2.1)
\begin{equation}
\label{eq:Sff}
    \SOmeas(\omega) = \underbrace{S_{\Omega_\text{m}\Omega_\text{m}}(\omega)}_{\text{\makebox[0pt]{intrinsic}}}+\underbrace{\frac{X_\text{th}^2}{X_\text{osc}^2}\frac{\Gamma_\text{m}}{2}}_{\text{\makebox[0pt]{thermomechanical}}} + \underbrace{\Big(\omega^2+\frac{\Gamma_\text{m}^2}{4}\Big)\frac{S_\text{xx}^\text{imp}}{X_\text{osc}^2}}_{\text{\makebox[0pt]{detection}}}.
\end{equation}
Here, $X_\text{th}$ is the root-mean-square thermomechanical amplitude, $X_\text{osc}$ the amplitude of the coherent oscillation and $S_\text{xx}^\text{imp}$ the displacement imprecision, visible as the flat background of the spectra in Fig.~\ref{fig:figure1}(g). Increasing $X_\text{osc}$ through the actuation voltage $\Vpzt$ suppresses both artefacts, while $S_\text{xx}^\text{imp}$ is set by the laser--cavity detuning $\Delta = \omega_\text{L}-\omega_0$ and the optical power $\Pdet$.

At fixed $\Pdet = 30$~\si{\micro\watt} and $\Delta = -\kappa/(2\sqrt{3})$, the maximum slope of the cavity transmission, increasing $\Vpzt$ reduces the variance of the frequency record (Fig.~\ref{fig:figure2}(b)) and both the thermomechanical and detection contributions to its PSD (Fig.~\ref{fig:figure2}(c)), in quantitative agreement with Eq.~(\ref{eq:Sff}). At large drive amplitudes, however, two floors remain. At small Fourier (sideband) frequencies $\omega$, the intrinsic frequency noise $S_{\Omega_\text{m}\Omega_\text{m}}$, whose flicker-type spectrum ($\propto 1/\omega$, dashed line in Fig.~\ref{fig:figure2}(c)) is commonly attributed to thermally activated material defects~\cite{bachtold_mesoscopic_2022}, bounds the achievable frequency stability. At large $\omega$, the detection contribution stops decreasing with drive, because the transduction gain drops once the motion-induced modulation of the ring resonance becomes comparable to its half-linewidth ($GX_\text{osc}\gtrsim\kappa/2$)~\cite{javid_cavity_2021}, as evidenced by the saturation of the detected amplitude in Fig.~\ref{fig:figure2}(d).

In this nonlinear regime, the conversion of motion into the demodulated signal is no longer maximal at the maximum-slope detuning but at $\left|\Delta_*\right|\simeq 0.85\kappa$, where the power in the first harmonic of the transmitted signal peaks (Supplementary Section~1.7). Fitting the PSD at large $\omega$ to $D\omega^2$ while varying $\Delta$, with the drive amplitude that maximizes the transduction at each detuning, supports this prediction, as the inverse detection-noise coefficient $1/D$ qualitatively follows the detuning dependence of the first-harmonic power (Fig.~\ref{fig:figure2}(e)). The blue-detuned branch yields lower detection noise, but reaching it requires larger drive amplitudes, at which we observe drive-induced excess frequency noise, consistent with thermomechanical amplitude fluctuations converted by the Duffing nonlinearity of the resonator. We therefore operate at $\Delta\approx-\left|\Delta_*\right|$ (Supplementary Sections~1.7, 1.8 and 2.3).

The optical power offers a second handle on the detection noise. Increasing $\Pdet$ reduces it (Fig.~\ref{fig:figure2}(f)), with the coefficient $D$ falling as $\Pdet^{-2}$, as expected for detector-noise-limited imprecision, and crossing over to $\Pdet^{-1}$ at the highest powers as the measurement approaches the shot-noise limit (Fig.~\ref{fig:figure2}(g)). The low-frequency noise remains unchanged up to $P_\text{max} \simeq 100$\,\si{\micro\watt}, consistent with an intrinsic origin, but its band-limited variance (shaded area in Fig.~\ref{fig:figure2}(f)) grows as $\Pdet^{2}$ above $P_\text{max}$ (Fig.~\ref{fig:figure2}(g)). We attribute this excess to the residual absorption of the readout light ($\kappa_\text{abs}/2\pi \approx 2.4$\,\si{\mega\hertz}, Supplementary Section~1.6), which produces a steady-state frequency shift proportional to the intracavity power and converts power fluctuations into true frequency fluctuations, as observed in other temperature-sensitive resonators probed optically~\cite{sadeghi_frequency_2020,martini_uncooled_2025}. These fluctuations stem mostly from intensity noise of the delivered light, with an additional contribution from frequency noise converted into intensity noise by the slope of the microring response (Supplementary Section~2.2). As this photothermal noise dominates below the thermal cutoff $\omega_\text{th}$ (Fig.~\ref{fig:figure2}(f)), we keep $\Pdet < P_\text{max}$. The frequency-noise measurements that set the NEP below use direct detection at $\Delta\approx-\left|\Delta_*\right|$, $\Pdet = 56\,\si{\micro\watt}$ and $\Vpzt = 30$\,mV, the drive that maximizes the first-harmonic signal.

 \begin{figure*}[t]
    \centering
    \includegraphics[width=\textwidth]{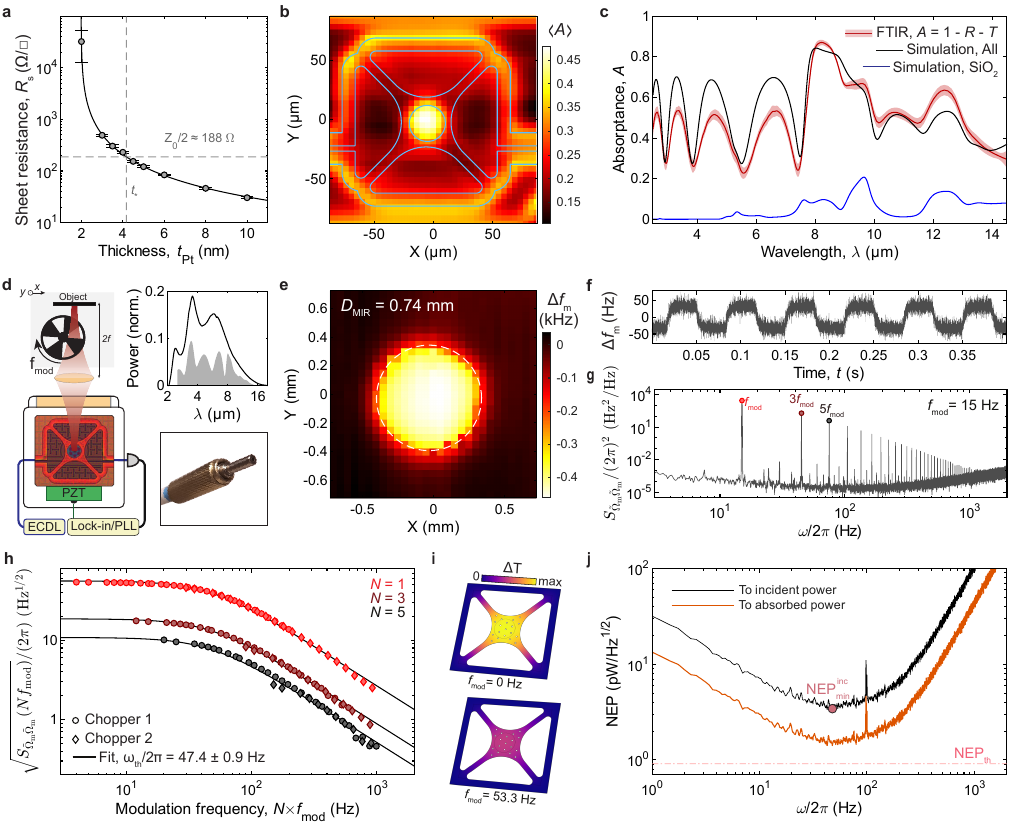}
        \caption{\textbf{Mid-infrared response characterization.} \textbf{a}, Sheet resistance ($R_\text{s}$) of Pt films evaporated on Si$_x$N$_y$ as a function of thickness, $t_\text{Pt}$. \textbf{b}, Example spatially resolved absorptance, spectrally averaged over the wavelength range $\lambda_\text{MIR} \in [2.5,14.5]$ \si{\micro\meter}. \textbf{c}, Measured absorptance ($A$) spectra at the centre of the Pt disk, averaged over eight devices on a chip released together with that of the measured detector (solid red). The red shade represents $\pm2$ standard deviations over the eight devices. Solid black and blue lines represent respectively the absorptance in all layers and in the SiO$_2$ cladding. \textbf{d}, Schematic of the setup used to characterize the mid-IR response, in which the output of a polycrystalline fibre connected to a thermal source (object plane) is imaged onto the device. Insets: photograph of the fibre output and spectrum incident on the device, with the absorbed fraction shaded. \textbf{e}, Spatially resolved steady-state frequency shift, $\Delta f_\text{m}(x,y)$, measured upon raster scanning ($xy$) the fibre output over the device. The radius of the dashed circle results from fitting a smooth top-hat function to the data. \textbf{f}, Time-resolved $\Delta f_\text{m}$ upon modulation of the fibre output at $f_\text{mod}$ = 15 \si{\hertz} with a mechanical chopper, at reduced source power. \textbf{g}, PSD of the time trace in \textbf{f}. The first three odd harmonics ($N$ = 1, 3, 5) are highlighted in the spectrum. \textbf{h}, Reconstruction of the transfer function from the peak amplitudes highlighted in \textbf{g} for varying $f_\text{mod}$. Data from two choppers with different blade densities are used. \textbf{i}, Temperature field obtained from finite-element heat transport simulations including a harmonic boundary heat source at frequency $f_\text{mod}$ = 0 \si{\hertz} and $f_\text{mod}$ = 53.3 \si{\hertz}, above the thermal cutoff. The colour bar range is common to both colour maps. \textbf{j}, NEP as a function of modulation frequency $\omega/2\pi$, referred to the incident power for the incident spectrum in \textbf{d} (black) and to the absorbed power (orange). The marker indicates the minimum incident-referred NEP, NEP$^\text{inc}_\text{min}$, and the dash-dotted line the fundamental thermal-fluctuation limit of the detector, NEP$_\text{th}$.}
    \label{fig:figure3}
\end{figure*}

\section{Broadband impedance-matched absorption}
\label{sec:Platinum}

The absorption of mid-IR radiation relies on the Pt disk, whose thickness is chosen for impedance matching. From the sheet resistance $R_\text{s}$ of Pt films evaporated on Si$_x$N$_y$ as a function of thickness $t_\text{Pt}$ (Fig.~\ref{fig:figure3}(a)), we select $t_\text{Pt} = t_* = 4.2$\,nm, for which $R_\text{s} \approx Z_0/2$, with $Z_0\approx 377\,\Omega$ the impedance of free space. At this deep-subwavelength thickness ($t_*\ll\lambda_\text{MIR}$), impedance matching yields a theoretical 50\% absorptance for a free-standing resistive film or one on a dielectric layer much thinner than the wavelength~\cite{hadley_reflection_1947}, which our films approach for $\lambda_\text{MIR} \in [5,25]$\,\si{\micro\meter} on reference membranes (Supplementary Section~3.4). Hyperspectral FTIR microscopy of OMTDs on a second chip, released together with that of the measured detector, shows only weak contrast in the Pt-free tethers and pad, whereas the Pt disk produces a clear enhancement, with an average absorptance $\langle A\rangle \approx 0.49$ for $\lambda_\text{MIR} \in [2.5,14.5]$\,\si{\micro\meter} (Fig.~\ref{fig:figure3}(b) and Supplementary Section~3.5). The absorptance spectrum at the centre of the disk, averaged over eight devices (Fig.~\ref{fig:figure3}(c)), is reproduced by a normal-incidence transfer-matrix model (Methods), including the Fabry--P\'erot fringes formed by the multilayer stack below $\lambda_\text{MIR} \sim 8$\,\si{\micro\meter}. Subtracting the simulated absorption of the SiO$_2$ cladding (blue line in Fig.~\ref{fig:figure3}(c)), which is at most 20\%, yields an average resonator absorptance of 0.441 $\pm$ 0.010 (standard deviation over the eight devices), to which the Si$_x$N$_y$ layer contributes only weakly, near its absorption band at 12\,\si{\micro\meter}.

\section{Thermal responsivity and bandwidth}
\label{sec:MIRresponse}

We study the mid-IR response with the setup of Fig.~\ref{fig:figure3}(d), in which a fibre-coupled thermal source ($T$ = 1550\,K) is imaged onto the chip and the frequency is tracked as in Fig.~\ref{fig:figure2} (Methods). Raster-scanning the source over the device while recording the steady-state frequency shift $\Delta f_\text{m}$ maps the fibre output (inset of Fig.~\ref{fig:figure3}(d)) with negligible blurring, as the output is much larger than the absorber (Fig.~\ref{fig:figure3}(e)). The map is well represented by a smoothed top-hat profile of diameter $D_\text{MIR} = 738.1 \pm 1.1$\,\si{\micro\meter} (fit uncertainty), which, together with the independently measured incident power (Fig.~\ref{fig:figure1}(i)) and the absorber area, yields a fractional frequency responsivity to incident power $\mathcal{R}_0^\text{inc} = -(2.65 \pm 0.18)\times 10^{3}$~\si{\watt}$^{-1}$ (uncertainty from the fits in Figs.~\ref{fig:figure1}(i) and \ref{fig:figure3}(e)). With the resonator absorptance averaged over the incident spectrum (inset of Fig.~\ref{fig:figure3}(d)), $\langle A\rangle_\text{S} = 0.420\pm 0.010$ (Supplementary Section~3.6), the responsivity to absorbed power is $\mathcal{R}_0^\text{abs} = -(6.3 \pm 0.5)\times 10^{3}$~\si{\watt}$^{-1}$, which lies within the range obtained from FEM simulations using literature values of the material parameters, $-(4.9\pm2.0)\times 10^{3}$~\si{\watt}$^{-1}$ (Supplementary Section~3.1). This range is dominated by the thermal expansion coefficient and the thermal conductivity of thin Si$_x$N$_y$ films, which vary considerably between reports~\cite{martini_uncooled_2025,ftouni_thermal_2015,snell_heat_2022}. The same scheme also images a resistively heated element to scale (Supplementary Section~3.7).

For time-varying signals, the responsivity becomes $\mathcal{R}^\text{abs}(\omega) = \mathcal{R}_0^\text{abs} H_\text{th}(\omega)$, with $H_\text{th}(\omega)$ the thermal transfer function, which we measure by chopping the incident radiation (Methods). The mechanical frequency responds synchronously (Fig.~\ref{fig:figure3}(f) for $f_\text{mod}$ = 15\,Hz), with peaks at the odd harmonics of $f_\text{mod}$ as expected for a square-wave modulation (Fig.~\ref{fig:figure3}(g)). The extracted $\left|H_\text{th}(\omega)\right|$ follows a first-order low-pass response, $\left|H_\text{th}(\omega)\right| = 1/\sqrt{1+(\omega/\omega_\text{th})^2}$, with a cutoff frequency $\omega_\text{th}/2\pi = 47.4 \pm 0.9$\,\si{\hertz} (Fig.~\ref{fig:figure3}(h)). An independent measurement confirms this cutoff to within about 11\% and shows that it does not depend on the wavelength of the absorbed radiation (Supplementary Section~4.4). Frequency-domain FEM simulations of heat transport confirm its thermal origin (Fig.~\ref{fig:figure3}(i)), yielding $\omega^\text{FEM}_\text{th}/2\pi = 48 \pm 8$\,Hz, where the uncertainty reflects the spread of the literature values of the material parameters (Supplementary Section~3.2).

\section{Noise-equivalent power}

The NEP combines the frequency-noise PSD with the responsivity into a single figure of merit. Referred to the power incident on the absorber, it reads
\begin{equation}
\label{eq:NEP}
    \text{NEP}^\text{inc}(\omega) = \frac{\sqrt{2\SOmeas(\omega)}}{\Omega_\text{m,0}\left|\mathcal{R}^\text{inc}(\omega)\right|},
\end{equation}
with $\mathcal{R}^\text{inc}(\omega) = \mathcal{R}_0^\text{inc} H_\text{th}(\omega)$. Figure~\ref{fig:figure3}(j) shows $\text{NEP}^\text{inc}$ for the incident spectrum of Fig.~\ref{fig:figure3}(d), together with the NEP to absorbed power, $\text{NEP}^\text{abs} = \langle A\rangle_\text{S}\,\text{NEP}^\text{inc}$. The best detection limit is reached for signals modulated at $\omega/2\pi \sim$ 40--50\,\si{\hertz}, although the frequency noise is lowest at higher sideband frequencies (Fig.~\ref{fig:figure2}(f)), because $\left|H_\text{th}(\omega)\right|$ rolls off as $\omega^{-1}$ above $\omega_\text{th}$ while $\sqrt{2\SOmeas(\omega)}$ falls only as $\sim\omega^{-1/2}$. The minimum is $\text{NEP}^\text{inc}_\text{min} = 3.4\pm0.2$\,\si{\pico\watt/\sqrt{\hertz}} referred to incident power ($\text{NEP}^\text{abs}_\text{min} = 1.4\pm0.1$\,\si{\pico\watt/\sqrt{\hertz}} referred to absorbed power), with uncertainties propagated from $\mathcal{R}_0^\text{inc}$ and $\mathcal{R}_0^\text{abs}$. The NEP to incident power for monochromatic radiation, $\text{NEP}^\text{inc}_\text{min}(\lambda_\text{MIR}) = \text{NEP}^\text{abs}_\text{min}/A(\lambda_\text{MIR})$, reaches $1.8\pm0.1$\,\si{\pico\watt/\sqrt{\hertz}} at $\lambda_\text{MIR}$ = 8.17\,\si{\micro\meter} and remains below 8\,\si{\pico\watt/\sqrt{\hertz}} across $\lambda_\text{MIR} \in [2.5,14.5]$\,\si{\micro\meter} (Supplementary Section~3.8). Referring the NEP to the absorber presumes that the radiation can be focused onto it, for example with bulk mid-IR optics or integrated microlenses~\cite{gordon_application_1991}, and that the responsivity does not depend on the focusing conditions, which FEM simulations confirm to within 0.3\% (Supplementary Section~3.3).

The fundamental limit of a thermal detector is set by thermal-fluctuation noise: random phonon and photon exchanges between the resonator and its environment drive temperature fluctuations that appear as irreducible frequency noise through the thermal responsivity~\cite{zhang_enhanced_2025}. The corresponding limit is $\text{NEP}_\text{th} = \sqrt{4k_BT_0^2G_\text{th}}$~\cite{richards_bolometers_1994}, with $k_B$ the Boltzmann constant, $T_0$ the ambient temperature and $G_\text{th}$ the thermal conductance of the resonator to its surroundings. Evaluated at $T_0 = 298$\,K with $G_\text{th} = (1.69\pm0.10)\times 10^{-7}$~\si{\watt/\kelvin}, obtained from the measured cutoff and the simulated heat capacity (Supplementary Section~3.9), it is shown as the dash-dotted line in Fig.~\ref{fig:figure3}(j). The minimum NEP to absorbed power lies within a factor of $1.5\pm0.1$ of this limit, which is itself a lower bound~\cite{zhang_enhanced_2025}. Since radiative exchange contributes only $\approx 2$\% of $G_\text{th}$, the limit is set by the tether geometry, and for a purely radiation-limited resonator it would drop to $\text{NEP}_\text{rad}\approx 0.14$~\si{\pico\watt/\sqrt{\hertz}} (Supplementary Section~3.9).

 \begin{figure*}[t]
    \centering
    \includegraphics[width=\textwidth]{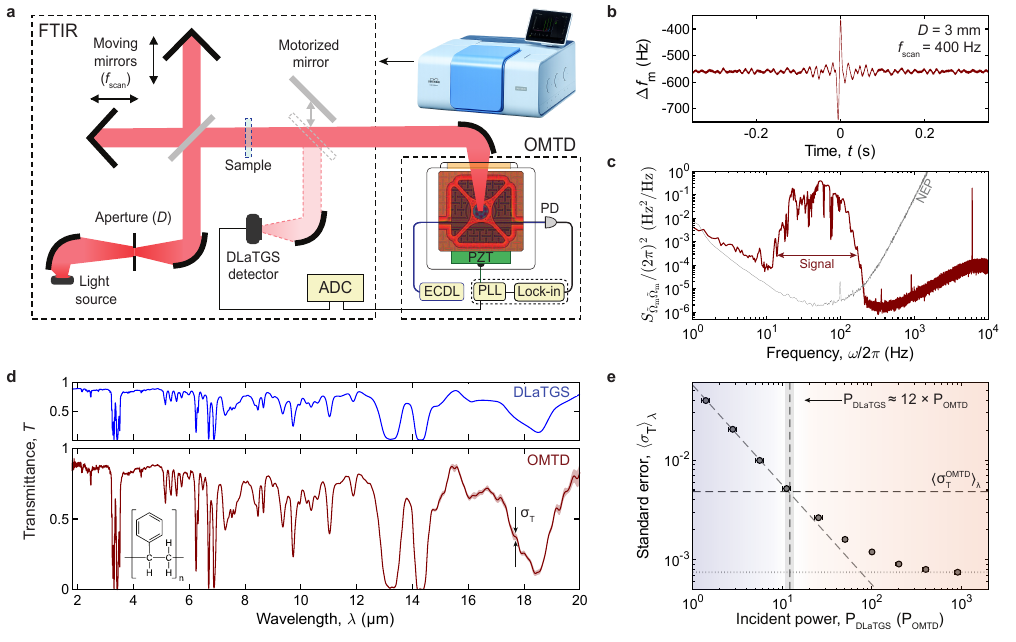}
        \caption{\textbf{Fourier-transform infrared (FTIR) spectroscopy using an optomechanical thermal detector (OMTD).} \textbf{a}, Schematic of the setup employed for FTIR spectroscopy, where the dashed box corresponds to a commercial Bruker Invenio spectrometer. A motorized mirror directs the infrared beam either to the internal detector (DLaTGS) or through an external port towards the OMTD. PD: photodetector, ADC: analog-to-digital converter. \textbf{b}, Example interferogram trace (aperture $D$ = 3 mm, scan rate $f_\text{scan}$ = 400 \si{\hertz}) measured with the PLL frequency-tracking scheme. \textbf{c}, Frequency noise PSD obtained from a continuous trace of five forward and five backward scans, of which \textbf{b} shows one. For reference, NEP($\omega$) is also shown (arbitrary units). \textbf{d}, FTIR transmittance spectrum ($T$) of a 38-\si{\micro\meter}-thick polystyrene foil, the internal reference sample of the spectrometer, obtained using the built-in detector (DLaTGS) and our detector (OMTD). Solid lines are means obtained from ten forward scans and the shades around them have width $2\sigma_T$, with $\sigma_T$ the standard error of $T$. \textbf{e}, Wavelength-averaged $\langle\sigma_T\rangle_\lambda$ (over the spectral range shown in \textbf{d}) measured with the DLaTGS detector as a function of incident optical power, $P_\text{DLaTGS}$, referred to the power incident on our Pt absorber for the measurement in \textbf{d}, $P_\text{OMTD}$. The value of $\langle\sigma_T\rangle_\lambda$ for our detector is highlighted as a horizontal dashed line, and the vertical dashed line marks the power at which the DLaTGS detector reaches the same value, with its uncertainty shown as a grey band. Grey dashed and dotted lines indicate the $1/P_\text{DLaTGS}$ dependence at low power and the floor reached at high power. Horizontal error bars show the uncertainty of the power calibrations (Supplementary Section~4.3). Blue (orange) shading indicates regions where our detector outperforms (underperforms) the DLaTGS detector.}
    \label{fig:figure4}
\end{figure*}

\section{System-level demonstration in a commercial FTIR}

We integrate the OMTD into a commercial FTIR spectrometer (Bruker Invenio), whose built-in motorized mirror routes the thermal-source beam either to the reference pyroelectric detector (deuterated L-alanine-doped triglycine sulfate, DLaTGS) or out through an auxiliary port onto the OMTD (Fig.~\ref{fig:figure4}(a) and Methods). Scanning the optical path difference of the interferometer produces an interferogram in the tracked frequency (Fig.~\ref{fig:figure4}(b)), which, for the chosen scan speed, falls around the minimum of the NEP (Fig.~\ref{fig:figure4}(c)). Transmittance spectra of the spectrometer's internal reference sample, a 38-\si{\micro\meter}-thick polystyrene foil, obtained from ten scans with the foil and ten reference scans without it, agree well with those of the DLaTGS detector (Fig.~\ref{fig:figure4}(d)), although the standard error $\sigma_T$ of the OMTD spectra is far larger. This is expected: from the steady-state frequency shift (Fig.~\ref{fig:figure4}(b)) and $\mathcal{R}_0^\text{inc}$, we estimate the power incident on the Pt disk to be nearly three orders of magnitude lower than that incident on the $2\,\si{\milli\meter}\times 2\,\si{\milli\meter}$ DLaTGS detector. Measurements with the DLaTGS detector at varying power show that our detector reaches the same wavelength-averaged standard error, $\langle\sigma_T\rangle_\lambda$, with roughly an order of magnitude less power incident on its absorber (Fig.~\ref{fig:figure4}(e) and Supplementary Section~4.3). The single OMTD is thus throughput-limited, as its micron-scale absorber intercepts only a small fraction of the millimetre-scale beam, but per unit of incident power it already compares favourably with the commercial pyroelectric detector.

\section{Discussion and conclusions}

In summary, we have demonstrated a photonic-integrated mid-infrared detector in which a high-$Q$, tensile-stressed Si$_x$N$_y$ trampoline is micromachined directly above a foundry-compatible Si$_3$N$_4$ PIC, such that the mechanical and photonic layers are designed and fabricated largely independently, sharing only the cladding and sacrificial-layer thicknesses that set their separation and the footprint of the ring beneath the pad. This decoupling allows the resonator to carry an impedance-matched platinum absorber, whose residual absorption of the readout light, $\kappa_\text{abs}/2\pi\approx 2.4$\,MHz, is negligible compared with the intrinsic loss rate of the ring, $\kappa_\text{i}/2\pi\approx 0.4$\,GHz (Supplementary Section~1.6). The microring beneath the trampoline reads out its motion without any free-space alignment. With microwatt-level optical power at room temperature, the detector reaches a noise-equivalent power of 3.4~\si{\pico\watt/\sqrt{\hertz}} for a broadband thermal source and, referred to absorbed power, lies within a factor of 1.5 of its fundamental thermal-fluctuation limit, a regime so far attained only with free-space or fibre-based interferometric readout of larger and about 27 times slower resonators~\cite{zhang_demonstration_2023,zhang_enhanced_2025} (Supplementary Section~5). The remaining factor is set by the intrinsic frequency noise of the resonator, which, at the chosen operating point, is independent of drive amplitude, optical power and detection scheme (Supplementary Section~2.4).

Further improvements follow from the architecture. Because the motion is read out optically, the tethers need not carry electrical leads, unlike the legs of uncooled microbolometers built above CMOS readout circuits~\cite{rogalski_progress_2012,fusetto_readout_2023}, and can be designed for thermal isolation and mechanical performance alone. Narrower tethers, for instance, would raise the responsivity below the present cutoff without degrading it above, while lowering the cutoff. The minimum NEP then improves until photothermal noise from the absorbed readout light takes over. In our setup, this noise stems mostly from intensity fluctuations added by the fibre delivery, typically more than 20\,dB above those of the laser itself (Supplementary Section~2.2), which limits the reduction of the minimum NEP to a factor of about 2, against about 4 with the noise of the laser alone (Supplementary Section~3.10). With the readout already on a Si$_3$N$_4$ PIC, this delivery can be removed altogether by co-integrating a low-noise laser with the circuit, either as a laser diode self-injection locked to an on-chip resonator~\cite{reichler_endlessly_2026} or as a monolithic erbium-doped laser~\cite{ji_wafer-scale_2026}. With the laser on chip, the local oscillator of a balanced homodyne receiver could also be routed on chip. In our fibre-based implementation, homodyne detection already lowers the detection noise about 500-fold at equal optical power, but picks up spurious tones in the non-common fibre arms (Supplementary Section~2.4). As each layer is fabricated with the standard processes of its own ecosystem, the remaining components can follow, towards a fully integrated, self-contained detector: III--V photodetectors~\cite{yu_heterogeneous_2020} can join the lasers on the photonic side, while monolithic piezoelectric actuation~\cite{liu_monolithic_2020} and wafer-level or chip-scale vacuum packaging with integrated getters~\cite{nicolas_high_2014,jo_wafer-level_2022} suit the mechanical layer, whose quality factor halves only at about $10^{-3}$\,mbar (Supplementary Section~1.9).

Integration with ultra-low-loss PICs also changes how optically read nanomechanical sensors scale. Free-space readout requires an aligned beam for each resonator, and this beam competes for optical access with stimuli such as the mid-infrared illumination used here~\cite{thijssen_parallel_2014,zheludev_reconfigurable_2016,yeh_infrared_2023}. On-chip, many resonators can instead share a single bus, each read out with microwatt-level power and addressed by its optical~\cite{pirmoradi_integrated_2025} and mechanical~\cite{gregorat_highly_2024} frequencies, so that their number is set by routing and tone-tracking electronics rather than by alignment, up to limits imposed by mode crowding, crosstalk and nonlinear intermodulation~\cite{fedorov_thermal_2020}. For example, co-adding such resonators into a single large-area detector, as in silicon photomultipliers~\cite{gundacker_silicon_2020} or tiled superconducting-nanowire detectors~\cite{cheng_100-pixel_2023}, would lift the \'etendue limitation of our FTIR demonstration. Replacing the absorber with magnetostrictive~\cite{gotardo_silicon-photonic_2026} or sorbent~\cite{li_nanoelectromechanical_2010} coatings would likewise extend the same readout to magnetic-field and mass sensing. Our results thus establish the monolithic integration of tensile-stressed nanomechanics with ultra-low-loss PICs as a route towards scalable, high-performance optomechanical sensing.

\section*{Methods}

\subsection*{Device fabrication}

\begingroup
\renewcommand{\figurename}{\textbf{Extended Data Fig.}}
\setcounter{figure}{0}
\renewcommand{\theHfigure}{ED\arabic{figure}}
\begin{figure*}[t]
    \centering
    \includegraphics[width=0.8\textwidth]{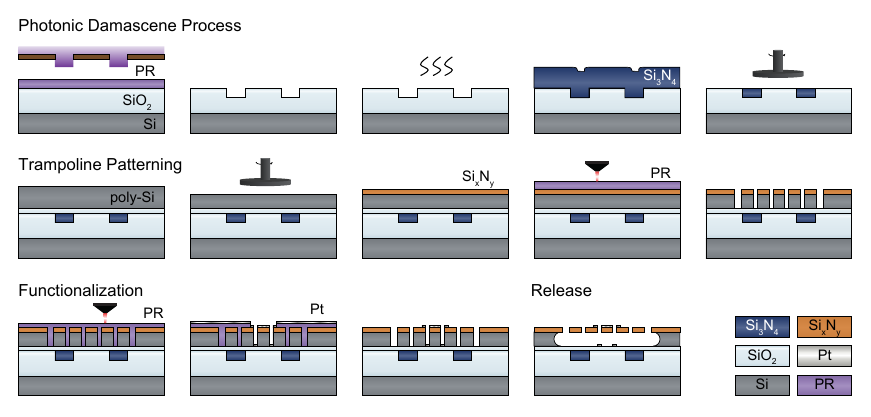}
    \caption{\textbf{$\mid$ Fabrication process flow.} Step-by-step fabrication of the photonic-integrated optomechanical thermal detector. The photonic damascene process defines the Si$_3$N$_4$ waveguide and ring resonator layer, buried in SiO$_2$. An LPCVD polysilicon sacrificial layer sets the air gap, above which the Si$_x$N$_y$ trampoline resonator is patterned. A 4.2\,nm Pt disk is deposited by electron-beam evaporation and lift-off to form the mid-IR absorber. After dicing, the resonators are released by isotropic XeF$_2$ vapour etching of the polysilicon, suspending the Si$_x$N$_y$ trampoline above the photonic ring. PR: photoresist.}
    \label{fig:figED1}
\end{figure*}
\endgroup

The devices are fabricated monolithically on a 4-inch silicon wafer in two layers: a Si$_3$N$_4$ PIC made with a foundry-compatible photonic process and, above it, a Si$_x$N$_y$ resonator made by MEMS surface micromachining (Extended Data Fig.~\ref{fig:figED1}). The photonic layer is formed first by the photonic damascene process~\cite{pfeiffer_photonic_2016, liu_high-yield_2021}: a thermal SiO$_2$ cladding (4~\si{\micro\meter}) is grown on a single-side-polished Si wafer, waveguide trenches are defined by DUV-stepper lithography and ICP-RIE, and the trenches are filled by LPCVD stoichiometric Si$_3$N$_4$. Chemical-mechanical planarization (CMP) leaves the Si$_3$N$_4$ PIC inlaid flush with the SiO$_2$ surface, followed by burial under a 190-nm-thick PECVD SiO$_2$ top cladding, which also serves as an etch stop that protects the ring during the later release.
A 480-nm-thick LPCVD polysilicon layer is then deposited on the planarized photonic chip to define the sacrificial gap. A 150-nm-thick (silicon-rich) Si$_x$N$_y$ film is deposited by LPCVD and patterned by optical lithography and dry etching into the trampoline geometry. The built-in stress of the as-deposited Si$_x$N$_y$ layer is $\sigma_\text{dep} = 235\,$MPa from wafer-bow measurements. The stress in the released resonators, inferred from the flexural frequencies of test string resonators of varying length, is $\approx 170\,$MPa and reproduces the measured mode frequencies of the device.
The mid-IR absorber is deposited by electron-beam evaporation of a 4.2\,nm Pt disk on the central pad, patterned by lift-off using a double-layer photoresist. The disk edges on the lithographic mask are laterally recessed from the microring inner edge by 5\,\si{\micro\meter} to suppress photothermal backaction from residual 1550\,nm absorption. The wafer is then diced into individual $5\times5\,$mm$^2$ chips by deep reactive ion etching and backside grinding.
Release of the Pt-coated resonators is performed via a low-pressure XeF$_2$ vapour etch (Xactix X4, 5\,s pulses at $P \approx 0.5$\,Torr) after a brief buffered-HF dip and a 115\,\si{\celsius} bake to remove the native oxide on the polysilicon surface. XeF$_2$ removes the polysilicon isotropically through the release holes and the trampoline-window perimeter, suspending the Si$_x$N$_y$ trampoline above the buried ring. Note that byproducts of the XeF$_2$ silicon etch can also attack the Si$_x$N$_y$ membrane being released~\cite{ronde_manipulating_2021}, leading to a finite Si:Si$_x$N$_y$ selectivity of $\approx 350:1$.
The exposure of the Si$_x$N$_y$ surfaces to the etching byproduct depends on the region (platinum-covered or uncovered) and location within the resonator. Consequently, the Si$_x$N$_y$ layer exhibits thickness steps and moderate thickness gradients, evidenced by visible interference effects in the optical micrograph of Fig.~\ref{fig:figure1}(c). For modelling purposes, we assume fixed thicknesses of $t_1 = 140$\,nm in the Pt-covered areas and $t_2 = 100$\,nm in the uncovered areas, values consistent with optical reflectometry measurements of the suspended devices. The same reflectometry measurements indicate an air gap of $d = 465$\,nm, which differs from the polysilicon sacrificial layer thickness due to the steady-state deformation associated with the built-in stress release. The device characterized in the main text has a ring microresonator of radius $R$ = 20 \si{\micro\meter} and measured waveguide cross-section 0.76$\times$2.16 \si{\micro\meter^2}, and a trampoline resonator of window size $L$ = 120 \si{\micro\meter}, pad size $L_\text{p}$ = 42 \si{\micro\meter}, tether width $w_\text{t}$ = 8 \si{\micro\meter}, and Pt disk radius $R_\text{Pt}$ = 15.2 \si{\micro\meter}. Complementary free-space interferometric characterization of the trampoline resonators and swept-wavelength spectroscopy of the coupled ring resonators are provided in Supplementary Sections~1.1 and~1.2.

\subsection*{Photonic packaging and optical readout}

The devices are packaged by gluing ultra-high numerical aperture fibres (UHNA7), butt-coupled at the chip facets, with a low-shrinkage epoxy (EPO-TEK OG154-1), which mitigates drift and low-frequency vibrations that affect optical alignment~\cite{bereyhi_perimeter_2022}. From the fibre-to-fibre transmission of the packaged chip away from resonance (9.5\%), calibrated against a through-fibre reference, we infer a per-facet coupling efficiency $\eta_\text{f} \approx 0.31$, assuming identical facets. The packaged chips are then operated in a vacuum chamber ($P < 10^{-6}$\,mbar) to reduce mechanical dissipation through air damping. The microresonator is driven by a continuous-wave laser (Toptica CTL 1550) via the bus waveguide and the trampoline motion is read out via direct transduction of cavity-frequency fluctuations onto the amplitude of the transmitted optical field~\cite{aspelmeyer_cavity_2014}, detected with a low-noise photoreceiver (New Focus 1811). A bias tee is used to separate DC and AC photodetected signals, respectively used to measure the optical and mechanical responses. The AC signal is fed to a digital lock-in amplifier (Zurich Instruments UHFLI) and demodulated at $f_\text{D2} \approx \Omega_\text{m,0}/2\pi$ with a demodulation bandwidth of $\Gamma_\text{D2}/2\pi = 50$\,kHz. Displacement spectra are calibrated using the variance of the thermal motion, which, together with the cavity transduction efficiency, also yields $g_0/2\pi$, as detailed in Supplementary Section~1.5.

\subsection*{Mechanical frequency tracking}

A digital oscillator drives the fundamental mode of the resonator through an off-chip piezoelectric (PZT) actuator, and a digital phase-locked loop (PLL) continuously adjusts the actuation frequency, $f_\text{PZT}$, to maintain a fixed phase between the drive and the measured response, so that $f_\text{D2}(t) = f_\text{PZT}(t) \approx \Omega_\text{m}(t)/2\pi$. The PLL PID controller parameters are selected following Ref.~\cite{demir_fundamental_2020} targeting a PLL bandwidth of $\Gamma_\text{PLL}/2\pi = 10$\,kHz. The combined low-pass transfer function of the demodulation and PLL, $H_\text{LP}(\omega)$, multiplies all terms of Eq.~(\ref{eq:Sff}) as $|H_\text{LP}(\omega)|^2$ (Supplementary Section~2.1). Since $|H_\text{LP}(\omega)|\approx 1$ over the frequency range of interest, it is omitted in the main text. The reported frequency estimate $\Omeas$ is instead obtained from a separate, out-of-loop demodulator by numerically differentiating the phase of the displacement quadratures~\cite{besic_schemes_2023}. Its reference is offset from the in-loop demodulator ($f_\text{D1} = \Omega_\text{m,0}/2\pi - 3$\,kHz, $\Gamma_\text{D1}/2\pi = 50$\,kHz) to prevent interference between the two and to place the demodulated signal at a finite intermediate frequency. The centre frequency of the PLL and the demodulation frequency of the out-of-loop demodulator are set before each measurement from a thermomechanical spectrum, as the mechanical frequency differs by a few kHz between pump-downs and increases slowly over time, which we attribute to desorption from the resonator surfaces. The demodulated quadratures are recorded at 54.9\,kSa/s over records of up to a few minutes, and frequency-noise PSDs are obtained by averaging periodograms of non-overlapping 2--10\,s segments of the frequency record.

\subsection*{Absorptance measurements and modelling}

Prior to device characterization, the optical constants of the employed Pt and Si$_x$N$_y$ films are retrieved from FTIR absorption spectroscopy (Bruker Vertex 70) on centimetre-scale Pt-coated Si$_x$N$_y$ membranes with the silicon substrate removed, which also validates the impedance-matched absorption of the ultra-thin Pt film (Supplementary Section~3.4). The spectrally resolved absorptance of the device is then obtained as $A(\lambda_\text{MIR}) = 1-R-T$ from separately acquired reflectance and transmittance maps measured by FTIR microscopy (PerkinElmer Spotlight 400, Fig.~\ref{fig:figure3}(b,c) and Supplementary Section~3.5), and the SiO$_2$ cladding contribution is subtracted using a normal-incidence transfer-matrix model, leaving the resonator absorptance. The model is built from these optical constants, literature data for SiO$_2$~\cite{franta_optical_2016} and Si~\cite{franta_temperature-dependent_2017}, the nominal Pt thickness and the dielectric layer and air-gap thicknesses obtained by UV--Vis reflectometry (Filmetrics F54-XY-200-UV).

\subsection*{Mid-infrared sources and power calibration}

Mid-infrared radiation from a SiC globar source (Arcoptix ARCLIGHT-MIR, $T = 1550$\,K) is delivered to the chip through a polycrystalline infrared fibre (Arcoptix FIB-PIR-900-100, 900\,\si{\micro\meter} core) and focused onto the device plane by a ZnSe aspheric lens (Thorlabs AL72550-E3, $f = 50$\,mm, clear aperture 23\,mm) in a $2f$--$2f$ ($1\times$) relay, entering the vacuum chamber through a ZnSe viewport (Kurt J.~Lesker VPZL-450UZ, 2-mm-thick). Mid-infrared powers are measured with a thermal power sensor (Thorlabs S401C). As a second, extended source, a 56\,\si{\ohm} metal-film resistor powered by a 9\,V battery (7.5\,V measured across it, $\approx 1$\,W dissipated, surface temperature $T \approx 443$\,K) is imaged onto the chip in the same relay. The absolute power spectrum incident on the device is obtained from the measured globar emission spectrum, weighted by the transmittances of the fibre, lens, and viewport and by the spatial profile of the fibre output at the focal plane (Supplementary Section~3.6).

\subsection*{Responsivity mapping and bandwidth measurements}

The frequency shifts induced by the mid-infrared illumination are tracked with the PLL scheme described above. Because of the slow drift of the mechanical frequency, all responsivities are fractional, referred to the frequency at the time of each measurement. Spatially resolved responsivity maps (Fig.~\ref{fig:figure3}(e)) are acquired by raster-scanning the source across the chip plane in 50-\si{\micro\meter} steps over $1.4\times1.4$\,mm$^2$ while recording the steady-state frequency shift $\Delta f_\text{m}$. Since the polycrystalline fibre core is much larger than the Pt disk, these maps provide, to a good approximation, an object-plane image of the fibre output. The fitted top-hat diameter is smaller than the 900\,\si{\micro\meter} core, consistent with multimode propagation in which the excited modes do not fill the full core. Because the absorptance is not spectrally flat, the responsivity to incident power depends on the incident spectrum, and it is converted into the responsivity to absorbed power using the absorptance integrated over that spectrum (Supplementary Section~3.6). The thermal transfer function is measured by intensity-modulating the fibre-coupled source with a mechanical chopper at frequency $f_\text{mod}$, producing a quasi-square-wave modulation, and reconstructing $|H_\text{th}(\omega)|$ from the peak heights at the fundamental and odd harmonics ($N = 1, 3, 5$) of the $\Delta f_\text{m}(t)$ spectrum across a range of $f_\text{mod}$. Two choppers (Thorlabs MC1F2 and MC1F15) of different blade density cover the full range.

\subsection*{FTIR spectroscopy measurements}

The OMTD is operated as the detector of a commercial Fourier-transform infrared spectrometer (Bruker Invenio). A built-in motorized mirror routes the spectrometer's collimated thermal emission either to the internal DLaTGS pyroelectric reference detector (DigiTect D301, active area $2\,\si{\milli\meter}\times 2\,\si{\milli\meter}$, specified detectivity $D^* > 2\times 10^8 $\,cm\,Hz$^{1/2}$\,W$^{-1}$) or out through an auxiliary output port, where an off-axis parabolic mirror (Thorlabs MPD139-M03, $f = 76.2$ mm) focuses it onto the OMTD to a millimetre-scale spot, comparable to the source aperture diameter ($D = 3$ mm). Measurements with the DLaTGS use the spectrometer's standard internal optics. The scan velocity is set through the fringe frequency of the built-in 632.8-nm HeNe reference laser, $f_\text{scan}$. For the data of Fig.~\ref{fig:figure4}, $f_\text{scan} = 400$\,\si{\hertz}, so that the interferogram signal falls around the minimum of the NEP of the OMTD, and Supplementary Section~4.4 shows the effect of varying it. The same scan velocity is used for the reference detector, whose responsivity---unlike the first-order roll-off of the OMTD---rises to a peak near 50\,\si{\hertz}, so that this band is near-optimal for both. Spectra are acquired at a resolution of 4\,\si{\per\centi\meter}. Unlike the out-of-loop records used for all other measurements, the OMTD interferograms are taken from the in-loop PLL frequency $f_\text{PZT}(t)$: an auxiliary signal $V(t) = K(f_\text{PZT}(t)-\langle f_\text{PZT}(t)\rangle)$ is synthesized in the lock-in and routed through an auxiliary output to the built-in analog-to-digital converter (ADC) of the spectrometer, with the gain $K$ set to span the full ADC dynamic range. The DLaTGS interferograms are digitized by the same ADC, and all spectra are generated with the vendor software (OPUS). To quantify spectral precision, $M = 10$ independent forward-scan transmittance spectra are acquired and the standard error of the mean transmittance at each wavenumber, $\sigma_T = s/\sqrt{M}$ (with $s$ the sample standard deviation), is averaged over wavelength to give the figure of merit $\langle\sigma_T\rangle_\lambda$. The power sweep in Fig.~\ref{fig:figure4}(e) is obtained with neutral-density filters (nominal transmission 50\%, 25\%, 12.5\% and 6.25\%) and two source apertures ($D = 3$ and $0.5$\,\si{\milli\meter}, the former matching the OMTD measurement). The aperture serves here only to set the incident power, independently of its usual effect on spectral resolution. The power of each setting is computed from the measured filter transmission, $0.88\pm0.02$ times the nominal value (Supplementary Section~4.1), and from the scaling $P_\text{DLaTGS}\propto D^2$, whose 8\% deviation at $D = 0.5$\,mm is included in the horizontal error bars (Supplementary Sections~4.2 and~4.3).

\subsection*{Numerical simulations}

All finite-element simulations are performed in COMSOL Multiphysics: the frequency-pull parameter $G$ via moving-boundary perturbation theory, and the trampoline mechanical eigenmodes, photothermal responsivity, and thermal conductance $G_\text{th}$ via a coupled heat-diffusion, thermoelastic, and pre-stressed-eigenvalue analysis with conduction-only heat transport. The mid-infrared absorptance is modelled with a normal-incidence transfer-matrix method formalism. Full simulation details and the material parameters used are given in Supplementary Sections~1.3, 3.1--3.2 and~3.4--3.5.

\section*{Acknowledgements}

The authors thank Curdin Maissen (Bruker) and Laetitia Monbaron (Universit\'e de Lausanne, UNIL) for assistance with the system-level demonstration in an FTIR and for FTIR microscopy, respectively. This work was supported by the Defense Advanced Research Projects Agency (DARPA) through the OPTIM program (HR00112320026). The views, opinions and/or findings expressed are those of the authors and should not be interpreted as representing the official views or policies of the Department of Defense or the U.S. Government. All samples were fabricated in the Center of MicroNanoTechnology (CMi) at EPFL. A.Z. acknowledges access to the XeF$_2$ release facilities at Chalmers University of Technology through the INFRACHIP project (access project IC045), which has received funding from the European Union's Horizon Europe research and innovation programme under grant agreement No.~101131822.

\section*{Author contributions}
G.A., A.A. and A.Z. contributed equally to this work.
A.A., G.A., A.Z. and Y.X. conceived the cavity-optomechanical platform.
X.J. fabricated the photonic circuits, whose GDS lithography mask was made by S.U.H.
A.Z. and G.A. fabricated the platinum-coated silicon nitride mechanical resonators.
G.A., A.A. and Y.X. built the experimental setup.
G.A. performed the measurements with assistance from A.A. and M.G.
G.A., A.A., A.Z. and Y.X. developed the finite element models.
S.E.K. and G.A. analysed the FTIR data with transfer-matrix models. 
S.A.B. contributed to the conceptualization of the project.
G.A. wrote the paper with input from all authors. 
G.A. and T.J.K. supervised the project.

\section*{Data availability}
The data that support the findings of this study, including raw experimental data and calibration records, will be made available in a Zenodo repository at the time of publication.

\section*{Code availability}
The scripts producing all figures and the finite-element simulation models (COMSOL Multiphysics with LiveLink for MATLAB) will be made available in a Zenodo repository at the time of publication.

\section*{Competing interests}
The authors declare no competing interests.

\makeatletter
\def\urlprefix{}%
\def\url{\begingroup
  \catcode`\#=12 \catcode`\%=12 \catcode`\&=12 \catcode`\~=12
  \catcode`\_=12 \catcode`\$=12 \catcode`\^=12 \gobbleURL}%
\def\gobbleURL#1{\endgroup\@ifnextchar.{\@gobble}{}}%
\makeatother

\onecolumngrid
\clearpage
\newgeometry{centering,textwidth=468pt,vmargin=2cm}
\makeatletter
\def\normalsize{%
  \@setfontsize\normalsize\@xiipt{14pt}%
  \abovedisplayskip 12\p@ \@plus3\p@ \@minus7\p@
  \belowdisplayskip \abovedisplayskip
  \abovedisplayshortskip \z@ plus3\p@
  \belowdisplayshortskip 6.5\p@ \@plus3.5\p@ \@minus3\p@
  \let\@listi\@listI
}%
\def\small{%
  \@setfontsize\small\@xipt{14.5pt}%
  \abovedisplayskip 8\p@ \@plus3\p@ \@minus6\p@
  \belowdisplayskip \abovedisplayskip
  \abovedisplayshortskip \z@ \@plus3\p@
  \belowdisplayshortskip 6.5\p@ \@plus3.5\p@ \@minus3\p@
  \def\@listi{\leftmargin\leftmargini \topsep 9\p@ \@plus3\p@ \@minus5\p@ \parsep 4.5\p@ \@plus2\p@ \@minus\p@ \itemsep \parsep}%
}%
\def\footnotesize{%
  \@setfontsize\footnotesize\@xpt{14.5pt}%
  \abovedisplayskip 10\p@ \@plus2\p@ \@minus5\p@
  \belowdisplayskip \abovedisplayskip
  \abovedisplayshortskip \z@ \@plus3\p@
  \belowdisplayshortskip 6\p@ \@plus3\p@ \@minus3\p@
  \def\@listi{\leftmargin\leftmargini \topsep 6\p@ \@plus2\p@ \@minus2\p@ \parsep 3\p@ \@plus2\p@ \@minus\p@ \itemsep \parsep}%
}%
\def\scriptsize{\@setfontsize\scriptsize\@viiipt{9.5pt}}%
\def\tiny{\@setfontsize\tiny\@vipt{7pt}}%
\def\large{\@setfontsize\large\@xivpt{18pt}}%
\def\Large{\@setfontsize\Large\@xviipt{22pt}}%
\def\LARGE{\@setfontsize\LARGE\@xxpt{25pt}}%
\def\huge{\@setfontsize\huge\@xxvpt{30pt}}%
\let\Huge\huge
\def\@listI{\leftmargin\leftmargini \parsep 5\p@ plus2.5\p@ minus\p@ \topsep 10\p@ plus4\p@ minus6\p@ \itemsep 5\p@ plus2.5\p@ minus\p@}%
\def\@listii{\leftmargin\leftmarginii \labelwidth\leftmarginii\advance\labelwidth-\labelsep \topsep 5\p@ plus2.5\p@ minus\p@ \parsep 2.5\p@ plus\p@ minus\p@ \itemsep \parsep}%
\leftmargini 30pt \leftmarginii 26pt \leftmarginiii 22pt \leftmarginiv 20pt
\labelsep 6pt
\parskip 0pt plus 1pt
\parindent 15pt
\emergencystretch 8\p@
\partopsep 3pt plus 2pt minus 2pt
\floatsep 14pt plus 2pt minus 4pt
\textfloatsep 20pt plus 2pt minus 4pt
\intextsep 14pt plus 4pt minus 4pt
\@fptop 0pt plus 1fil \@fpsep 10pt plus 2fil \@fpbot 0pt plus 1fil
\def\baselinestretch{1.5}%
\fontencoding{T1}\def\encodingdefault{T1}%
\normalfont\normalsize
\flushbottom
\pagestyle{preprint}
\let\section\section@preprintsty
\let\subsection\subsection@preprintsty
\let\subsubsection\subsubsection@preprintsty
\let\sin\sine
\let\cos\cosine
\def\figurename{FIG.}%
\def\fnum@figure{\figurename~\thefigure}%
\setlength{\abovecaptionskip}{2pt plus 1pt minus 1pt}
\setlength{\belowcaptionskip}{2pt plus 1pt minus 1pt}
\setcounter{topnumber}{2}
\setcounter{bottomnumber}{1}
\setcounter{totalnumber}{3}
\renewcommand{\topfraction}{0.98}
\renewcommand{\bottomfraction}{0.98}
\renewcommand{\textfraction}{0.02}
\renewcommand{\floatpagefraction}{0.98}
\setcounter{secnumdepth}{4}
\setcounter{section}{0}
\setcounter{figure}{0}
\setcounter{table}{0}
\setcounter{equation}{0}
\renewcommand{\thesection}{\arabic{section}}
\renewcommand{\thesubsection}{\thesection.\arabic{subsection}}
\renewcommand{\thesubsubsection}{\thesection.\arabic{subsection}.\arabic{subsubsection}}
\def\p@subsection{}\def\p@subsubsection{}%
\renewcommand{\thefigure}{S\arabic{figure}}
\renewcommand{\thetable}{S\arabic{table}}
\renewcommand{\theequation}{S\arabic{equation}}
\renewcommand{\theHsection}{S\arabic{section}}
\renewcommand{\theHfigure}{S\arabic{figure}}
\renewcommand{\theHtable}{S\arabic{table}}
\renewcommand{\theHequation}{S\arabic{equation}}
\setcitestyle{numbers,square}
\def\NATx@bibnumfmt#1{[#1]}%
\def\@extra@b@citeb{@SI}%
\renewcommand\NAT@wrout[5]{%
 \if@filesw{\let\protect\noexpand\let~\relax
  \immediate\write\@auxout{\string\bibcite{#5\@extra@b@citeb}{{#1}{#2}{{#3}}{{#4}}}}}\fi
 \ignorespaces}%
\global\let\@FMN@list\@empty
\def\bibsection{%
 \par
 \onecolumngrid@push
 \begingroup
  \baselineskip26\p@
  \bib@device{\textwidth}{245.5\p@}%
 \endgroup
 \nobreak\@nobreaktrue
 \addvspace{19\p@}%
 \par
 \onecolumngrid@pop
}%
\def\NAT@bibsetnum#1{\setlength{\topsep}{\z@}\NATx@bibsetnum{\ref{LastBibItemSI}}}%
\def\endthebibliography{%
 \edef\@currentlabel{\arabic{NAT@ctr}}%
 \label{LastBibItemSI}%
 \endNAT@thebibliography
 \aftergroup\auto@bib@empty
}%
\let\addcontentsline\SI@addcontentsline
\pdfbookmark[0]{Supplementary Information}{SI}
\frontmatter@title@above
\begingroup\large\bfseries\centering\parskip\z@skip
Supplementary information for:\\A photonic-integrated optomechanical mid-infrared detector\par
\endgroup
\makeatother
\bigskip
\tableofcontents

\section{Optical, mechanical and optomechanical properties}\label{sec:OM_properties}
In this section, details on the optical, mechanical and optomechanical characterization and numerical modelling are provided.

\subsection{Free-space interferometric characterization of trampoline mechanical resonators}
\label{subsec:freespace_char}

\begin{figure}[!htbp]
    \centering
    \includegraphics[width=\textwidth]{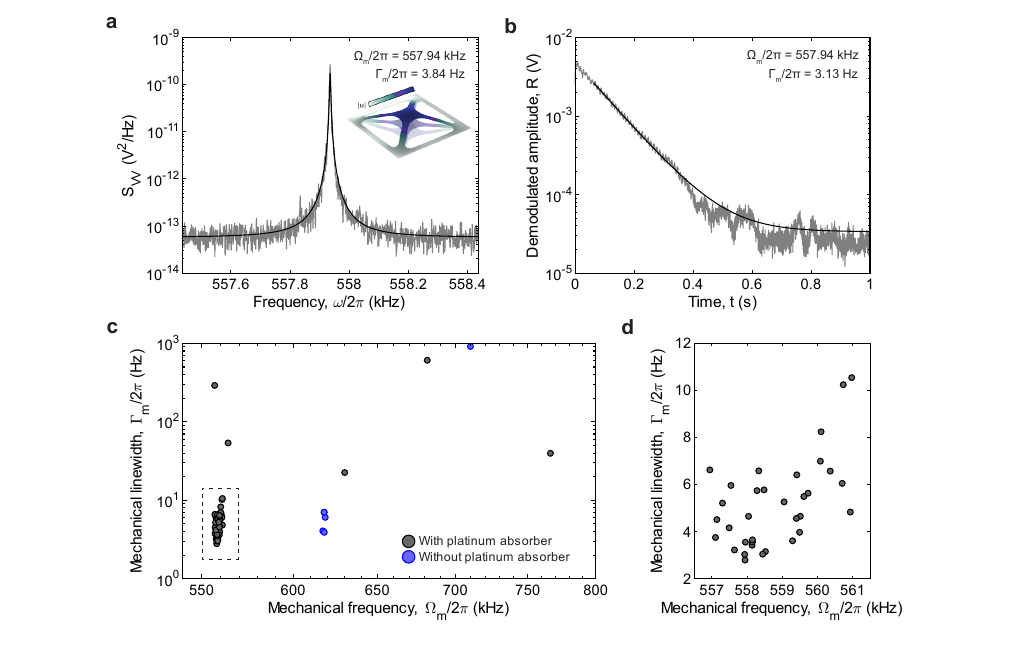}
        \caption{\textbf{Free-space characterization of trampoline resonators in a single chip.} \textbf{a}, Uncalibrated displacement noise PSD (solid grey) and fit to a lorentzian lineshape (black) for the fundamental out-of-plane flexural mode. The inset shows the fitted parameters and simulated mode shape. \textbf{b}, Ringdown measurement of the mode in \textbf{a} and its fit to an exponential decay. The demodulation frequency and fitted loss rate are reported. \textbf{c}, Mechanical frequencies and linewidths measured for all resonators on a chip, some with platinum absorber (grey dots) and some without it (blue dots). \textbf{d}, Zoom-in of the region highlighted in \textbf{c}.}
    \label{fig:figPreMec}
\end{figure}

Before performing photonic packaging, we pre-characterize all 47 mechanical resonators on a 5x5 \si{\milli\meter^2} chip via a free-space optical interferometer ($\lambda$ = 765 nm) with balanced homodyne detection (see Ref.~\cite{bereyhi_perimeter_2022} for setup details). Figures~\ref{fig:figPreMec}(a) and (b), respectively, show the thermomechanical spectrum and a ringdown measurement after excitation with an off-chip piezo-actuator of the fundamental flexural mode of a trampoline resonator. We observe a slight difference in the extracted linewidth, $\Gamma_\text{m}$, between the two methods. We attribute this to inhomogeneous broadening in the spectrum due to photothermal frequency fluctuations induced by the joint effect of absorption of the platinum (Pt) absorber and laser amplitude noise. Such effect is observed despite the fact that the detection laser spot is on the trampoline tether, probably due to sufficient uncontrolled scattering ending up at the absorber. The effect is much more pronounced when the laser spot is at the centre of the platinum absorber, where a large photothermal frequency shift is also observed. 

We perform the same characterization for the fundamental mode of all trampoline resonators\textemdash each associated with one waveguide and ring resonator\textemdash and report in Fig.~\ref{fig:figPreMec}(c) the measured resonant frequencies and linewidths. All trampolines have nominally-identical geometry except for the presence or absence of the Pt absorber. First, we observe that the presence of the ultrathin Pt absorber does not affect the linewidth of the mode of interest and only reduces the mechanical frequency due to mass loading. Second, we observe the data to be clustered except for a few resonators, for which we observe simultaneously an increase in frequency and linewidth. This positive correlation indicates that the reason for the dysfunctional devices is not simple mass loading but rather likely a pinning of the resonator motion which modifies the effective resonator length. Overall, the dispersion within the major cluster is limited ($\sigma_{\Omega_\text{m}}/2\pi$ = 1.19 kHz, $\sigma_{\Gamma_\text{m}}/2\pi$ = 1.9 Hz), but the few outliers make this characterization advisable before packaging.

\subsection{Optical spectroscopy of the bus-waveguide-coupled photonic ring resonators}
\label{subsec:ring_spectroscopy}

 \begin{figure}[!htbp]
    \centering
    \includegraphics[width=0.90\textwidth]{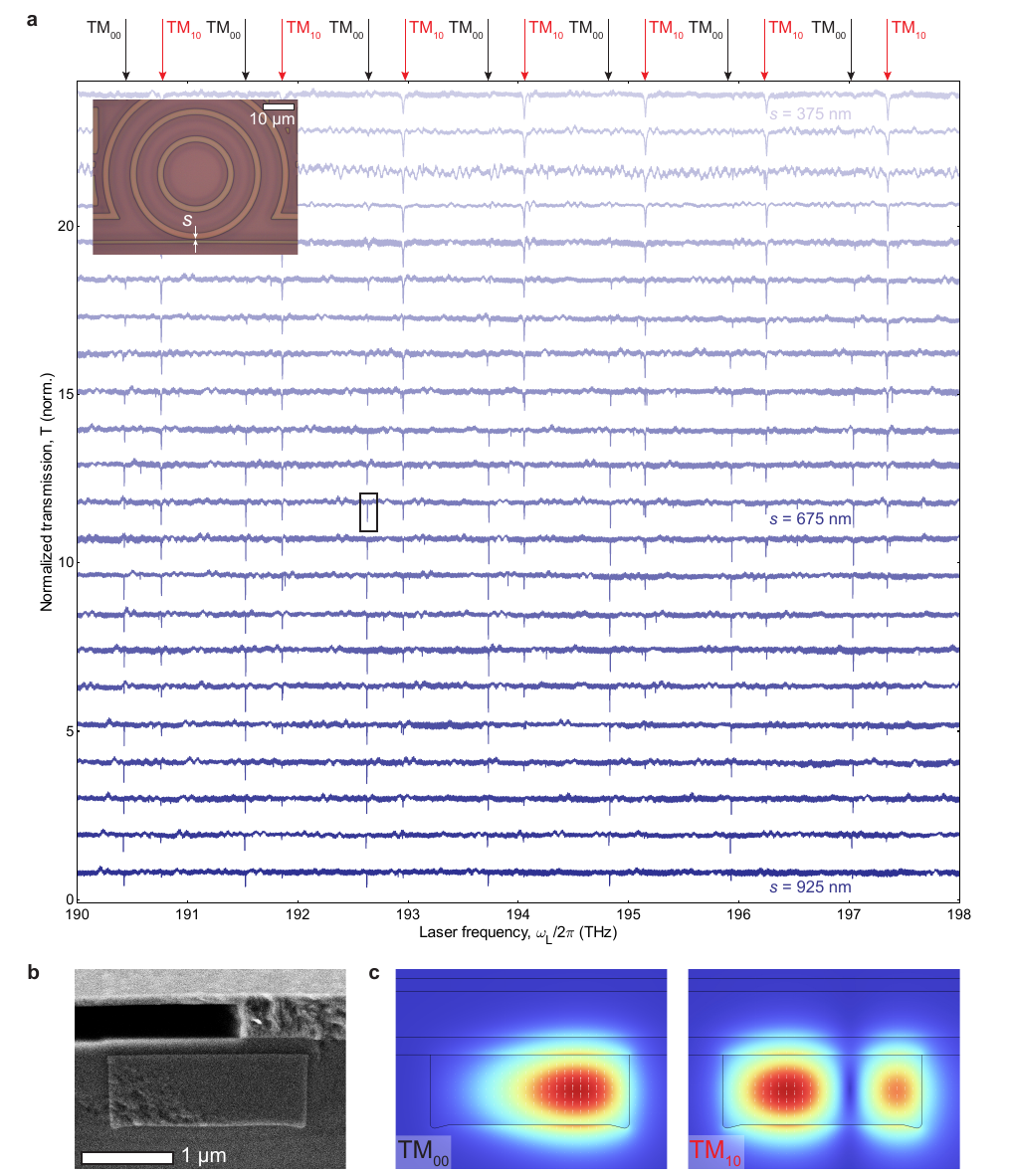}
        \caption{\textbf{Optical spectra of ring resonators coupled to a bus waveguide, measured before packaging with lensed fibres.} \textbf{a}, Normalized transmission spectra as a function of the waveguide-to-ring distance, $s$, for vertical input polarization. Spectra are shifted vertically by an equal amount and horizontally by an amount ranging from $-494$ to 559\,GHz. Black (red) arrows mark the TM$_{00}$ (TM$_{10}$) resonances, and the rectangle marks the resonance used in the main text. Inset: microscope image of an unclad ring. \textbf{b}, Tilted (5$\degree$) SEM of the ring cross-section. \textbf{c}, Simulated electric field (arrows) and its intensity (map) for TM$_{00}$ and TM$_{10}$ modes.}
    \label{fig:figPreOpt}
\end{figure}

Before photonic packaging, we also characterize the ring resonators of each chip by transmission spectroscopy, coupling light into and out of the bus waveguides with lensed fibres. The spectra in this section are therefore normalized, and their absolute level does not reflect the insertion loss of the packaged device. Although the different mechanical resonators on a chip are identical, each photonic circuit below them differs in the exact distance between the bus waveguide and the ring resonator, $s$ (see inset in Fig.~\ref{fig:figPreOpt}(a)). This gap is swept over a broad range ($s \in$ [300,1475] \si{\nano\meter}) to ensure that the range from highly over-coupled to highly under-coupled ring resonances is covered. Figure~\ref{fig:figPreOpt}(a) shows the normalized transmission spectra for vertical input polarization\textemdash perpendicular to the chip surface\textemdash and $s$ varying from 375 to 925 nm. All spectra are shifted for clarity: vertically by the same amount and horizontally by selected frequency amounts to align the observed resonant frequencies of the different mode orders. The spectrum of the waveguide selected for photonic packaging ($s$ = 675 nm) and the specific optical mode used for the full characterization in the main text are highlighted. By inspecting the whole data set, we identify two main families, with a free-spectral-range (FSR) of $\langle\Delta\nu/2\pi\rangle$ = 1.1013 and 1.096 THz respectively. Through FEM simulations of the effective refractive index, $n_\text{eff}$, and group index, $n_\text{g}$, of curved waveguides ($R$ = 20 \si{\micro\meter}) with cross-section as imaged with scanning electron microscopy (SEM) (Fig.~\ref{fig:figPreOpt}(b)), we identify the two sets of modes as the fundamental and first higher-order transverse-magnetic-like (TM-like) modes, that we denote as TM$_{00}$ and TM$_{10}$. Their electric-field magnitude, $\left| \mathbf{E}(r,z) \right|$, and field lines are shown in Fig.~\ref{fig:figPreOpt}(c).\\

\begin{figure}[!htbp]
    \centering
    \includegraphics[width=\textwidth]{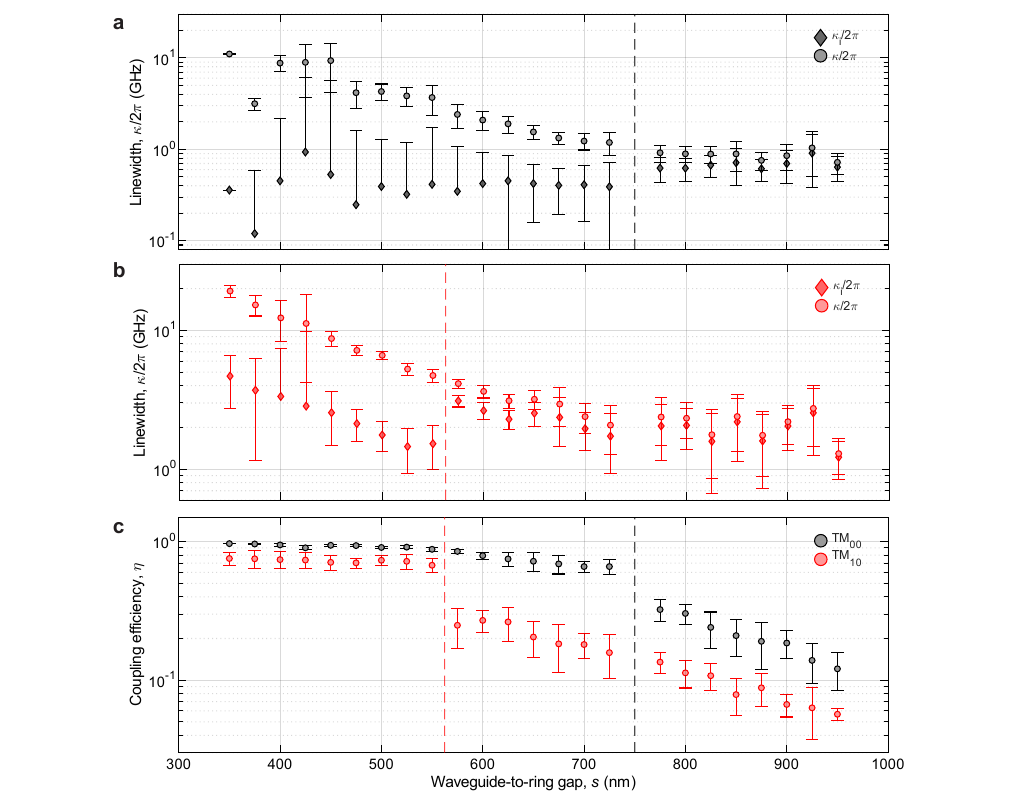}
        \caption{\textbf{Characterization of ring resonance parameters as a function of the waveguide-to-ring distance, $s$.} \textbf{a}, Total (circles) and intrinsic (diamonds) losses for the TM$_{00}$ resonances. The dots represent averages for resonances $\omega_c/2\pi\in[190, 200]$\,\si{\tera\hertz} and the error bar amounts to $\pm \sigma$, with $\sigma$ the standard deviation over the subset. \textbf{b}, Same as \textbf{a} for TM$_{10}$. \textbf{c}, External coupling fraction, $\eta$, for TM$_{00}$ (grey) and TM$_{10}$ (red). In all three panels, dashed lines represent the global gap $s_*$ selected for the over- to under-coupling transition.}
    \label{fig:figPreOpt2}
\end{figure}

To understand the coupling between the bus waveguide and the ring resonators, we fit every resonant dip in the spectra in Fig.~\ref{fig:figPreOpt}(a) with a Lorentzian lineshape,
\begin{equation}
\label{eq:lorentzian}
    T(\Delta) = \frac{\Delta^2 + \left[\frac{\kappa}{2}(1-2\eta)\right]^2}{\Delta^2+\left(\frac{\kappa}{2}\right)^2}
\end{equation}
where $\Delta = \omega-\omega_\text{c}$ is the laser-cavity detuning, $\kappa = \kappa_\text{i}+\kappa_\text{e}$ the total linewidth, $\kappa_\text{i}$ the intrinsic loss rate, $\kappa_\text{e}$ the waveguide-to-ring coupling rate, and $\eta = \kappa_e/\kappa \in [0,1]$ the external coupling fraction. Since Eq.~(\ref{eq:lorentzian}) satisfies $T(\Delta,\eta)=T(\Delta,1-\eta)$, we address whether the resonances are undercoupled ($\eta$ < 0.5) or overcoupled ($\eta$ > 0.5) by observing the behaviour of the whole set of resonances as a function of $s$ and set a fixed $s_*$ above (below) which all resonances are considered undercoupled (overcoupled). The behaviour of the loss rates averaged over the TM$_{00}$ (TM$_{10}$) resonances in the range $\omega_c/2\pi\in[190, 200]$\,\si{\tera\hertz} is shown in Fig.~\ref{fig:figPreOpt2}(a) (Fig.~\ref{fig:figPreOpt2}(b)). The dashed lines represent the gap $s_*$ chosen, which depends on the mode family. For TM$_{10}$ we observe that, for $s$< 500 nm, the presence of the bus waveguide not only affects $\kappa_\text{e}$ but also $\kappa_\text{i}$. For larger gaps, we observe the intrinsic linewidth to settle to a value $\kappa_\text{i}/2\pi\sim $ 2 \si{\giga\hertz}. In contrast, the presence of the bus waveguide does not significantly affect $\kappa_\text{i}$ for the TM$_{00}$ resonances, which we measure to be around 4 times smaller than for TM$_{10}$. We attribute this to the larger confinement, which reduces intrinsic radiative bending losses and extrinsic scattering-induced losses. Figure~\ref{fig:figPreOpt2}(c) depicts the coupling fraction $\eta$ for both modes, evidencing that TM$_{00}$ resonances remain overcoupled over a broader range of gaps. This is due to their smaller intrinsic losses, as our measurements show comparable $\kappa_\text{e}$ for both modes. We note that our approach for determining the coupling fraction is not completely accurate as not all resonances are over- or under-coupled for a fixed gap, which biases the inferred average rates for intermediate gaps. This is clear from the steps observed in all three panels in Fig.~\ref{fig:figPreOpt2} across $s=s_*$. To ensure overcoupling for most TM$_{00}$ resonances without incurring excessive degradation of $\kappa$, we select the waveguide with $s$ = 675 nm for photonic packaging. This dataset also resolves the ambiguity $\eta \leftrightarrow 1-\eta$ of Eq.~(\ref{eq:lorentzian}) for the packaged device: since TM$_{00}$ resonances are overcoupled at this gap, we assign $\eta = 0.65$ rather than 0.35 to the resonance used in the main text, whose parameters are measured after packaging.\\

\subsection{Dispersive optomechanical coupling between a trampoline and ring resonators}
\label{subsec:OM_trampoline_ring}

We consider the dispersive optomechanical coupling between an optical mode of a ring resonator and the out-of-plane mechanical modes of a trampoline resonator placed above it.  Here we analyse which mechanical modes couple to the optical resonance and how the nature of the optical field (travelling-wave versus standing-wave) determines the coupling strength. We then derive a minimal approximate expression for the single-photon vacuum coupling rate, $g_0$, in terms of the sensitivity of the waveguide mode to changes in the resonator position, $\partial n_\mathrm{eff}/\partial d$, and an azimuthal mechanical participation factor.

We define the mechanical displacement of the eigenmode as $\mathbf{u}(\mathbf{r})=q\,\mathbf{s}(\mathbf{r})$, where $\mathbf{s}(\mathbf{r})$ is its dimensionless mode shape. We normalize the modes such that $\max_{\mathbf{r}}|\mathbf{s}(\mathbf{r})|=1$, so that the generalized coordinate $q$ equals the maximum displacement amplitude of the mode. The corresponding zero-point fluctuation amplitude is $q_\mathrm{zpf}=\sqrt{\hbar/(2m_\mathrm{eff}^{(q)}\Omega_m)}$, where $m_\mathrm{eff}^{(q)}$ is the effective mass associated with this choice of coordinate and $\Omega_m$ is the mechanical resonance frequency. For our near-field cavity optomechanical transducer, the dispersive optomechanical coupling arises from the moving-boundary (MB) effect, whereby a mechanical displacement of a dielectric interface shifts the optical resonance frequency. For a single optical mode with angular frequency $\omega_c$, the frequency pull per unit of generalized displacement coordinate $q$ can be computed from first-order perturbation theory to~\cite{johnson_perturbation_2002}
\begin{equation}
\frac{\partial \omega_c}{\partial q}
=
-\frac{\omega_c}{2}
\frac{
\displaystyle
\oint_{\partial V}
(\mathbf{s}\cdot\mathbf{n})
\left[
\Delta \varepsilon\, |\mathbf{E}_{\parallel}|^2
-
\Delta(\varepsilon^{-1})\, |\mathbf{D}_{\perp}|^2
\right]
\, dA
}{
\displaystyle
\int_V \varepsilon |\mathbf{E}|^2 \, dV
},
\label{eq:MB_s}
\end{equation}
where $\partial V$ denotes the perturbed optical boundary, $\mathbf{n}$ is the outward normal, $\mathbf{E}_{\parallel}$ and $\mathbf{D}_{\perp}$ are the tangential electric field and normal electric displacement at the boundary, and $\Delta\varepsilon$ is the permittivity contrast across the interface. For compactness, we define the standard MB optical weight function as
$\mathcal{W}(\mathbf{r}) \equiv \Delta \varepsilon\, |\mathbf{E}_{\parallel}|^2 - \Delta(\varepsilon^{-1})\, |\mathbf{D}_{\perp}|^2$.

Because the trampoline membrane is thin, both its top and bottom interfaces contribute to the MB overlap. We therefore decompose the perturbed boundary as $\partial V=\partial V_\mathrm{top}\cup \partial V_\mathrm{bot}$ and write the numerator in Eq.~(\ref{eq:MB_s}) as
\begin{equation}
\oint_{\partial V}(\mathbf{s}\cdot\mathbf{n})\,\mathcal{W}\,dA
=
\oint_{\partial V_\mathrm{top}}(\mathbf{s}\cdot\mathbf{n})\,\mathcal{W}\,dA
+
\oint_{\partial V_\mathrm{bot}}(\mathbf{s}\cdot\mathbf{n})\,\mathcal{W}\,dA .
\label{eq:MB_topbot_split}
\end{equation}
We then rewrite each surface integral in cylindrical coordinates $(r,z,\phi)$ centred at the ring centre. The top and bottom membrane interfaces are (locally) at fixed heights $z=z_\mathrm{top}$ and $z=z_\mathrm{bot}$, so we parameterize each swept surface by $(r,\phi)$ with area element $dA=r\,dr\,d\phi$. The two contributions become
\begin{align}
\oint_{\partial V_\mathrm{top}}(\mathbf{s}\cdot\mathbf{n})\,\mathcal{W}\,dA
&=
\int_{0}^{2\pi}\! d\phi
\int_{r\in\mathcal{R}(\phi)}
(\mathbf{s}\cdot\mathbf{n})\big(r,z_\mathrm{top},\phi\big)\,
\mathcal{W}_\mathrm{top}\big(r,z_\mathrm{top},\phi\big)\,
r\,dr,
\label{eq:MB_top_rphi}
\\
\oint_{\partial V_\mathrm{bot}}(\mathbf{s}\cdot\mathbf{n})\,\mathcal{W}\,dA
&=
\int_{0}^{2\pi}\! d\phi
\int_{r\in\mathcal{R}(\phi)}
(\mathbf{s}\cdot\mathbf{n})\big(r,z_\mathrm{bot},\phi\big)\,
\mathcal{W}_\mathrm{bot}\big(r,z_\mathrm{bot},\phi\big)\,
r\,dr,
\label{eq:MB_bot_rphi}
\end{align}
where $\mathcal{W}_\mathrm{top}$ and $\mathcal{W}_\mathrm{bot}$ denote the weight function evaluated on the top and bottom membrane interfaces, respectively. The radial integration domain $\mathcal{R}(\phi)$ indicates the portion of the membrane interface that overlaps with the optical mode at each azimuth; since the ring mode is confined to a narrow annulus, $\mathcal{R}(\phi)$ is effectively restricted to radii near the ring radius, $R$.

In addition, the displacement is well approximated by $\mathbf{s}(\mathbf{r})\simeq s_z(r,\phi)\,\hat{\mathbf{z}}$ for out-of-plane modes of thin membranes. The outward normals on the membrane surfaces satisfy $\mathbf{n}_\mathrm{top}\approx +\hat{\mathbf{z}}$ and $\mathbf{n}_\mathrm{bot}\approx -\hat{\mathbf{z}}$, so that $(\mathbf{s}\cdot\mathbf{n})\approx \pm s_z$. Over the radial extent sampled by the optical field, we further approximate $s_z(r,\phi)$ by its value at $r$ = $R$ and represent it by a single scalar function of azimuth, denoted $s_\perp(\phi)$:
\begin{equation}
(\mathbf{s}\cdot\mathbf{n})\big(r,z_\mathrm{top},\phi\big)\simeq +\,s_\perp(\phi),
\qquad
(\mathbf{s}\cdot\mathbf{n})\big(r,z_\mathrm{bot},\phi\big)\simeq -\,s_\perp(\phi),
\qquad r\in\mathcal{R}(\phi).
\label{eq:sperp_signs_r}
\end{equation}
Substituting Eq.~(\ref{eq:sperp_signs_r}) into Eqs.~(\ref{eq:MB_top_rphi})--(\ref{eq:MB_bot_rphi}) yields
\begin{equation}
\oint_{\partial V}(\mathbf{s}\cdot\mathbf{n})\,\mathcal{W}\,dA
\simeq
\int_{0}^{2\pi} s_\perp(\phi)\,
\Big[\mathcal{F}_\mathrm{top}(\phi)-\mathcal{F}_\mathrm{bot}(\phi)\Big]\,d\phi,
\label{eq:MB_num_topbot_F}
\end{equation}
where we define the top and bottom radial functionals
\begin{equation}
\mathcal{F}_\mathrm{top}(\phi)\equiv \int_{r\in\mathcal{R}(\phi)} \mathcal{W}_\mathrm{top}\big(r,z_\mathrm{top},\phi\big)\,r\,dr,
\qquad
\mathcal{F}_\mathrm{bot}(\phi)\equiv \int_{r\in\mathcal{R}(\phi)} \mathcal{W}_\mathrm{bot}\big(r,z_\mathrm{bot},\phi\big)\,r\,dr.
\label{eq:Ftop_Fbot}
\end{equation}
Inserting Eq.~(\ref{eq:MB_num_topbot_F}) into Eq.~(\ref{eq:MB_s}) gives
\begin{equation}
\frac{\partial \omega_c}{\partial q}
=
-\frac{\omega_c}{2}
\frac{
\displaystyle
\int_{0}^{2\pi} s_\perp(\phi)\,\Big[\mathcal{F}_\mathrm{top}(\phi)-\mathcal{F}_\mathrm{bot}(\phi)\Big]\,d\phi
}{
\displaystyle
\int_V \varepsilon |\mathbf{E}|^2 \, dV
}.
\label{eq:dwdq_topbot_F}
\end{equation}
We now define the normalized azimuthal optical weighting
\begin{equation}
W(\phi)\equiv
\frac{\mathcal{F}_\mathrm{top}(\phi)-\mathcal{F}_\mathrm{bot}(\phi)}
{\displaystyle\int_V \varepsilon |\mathbf{E}|^2 \, dV},
\label{eq:Wphi_topbot}
\end{equation}
so that the coupling reduces to the compact overlap form
\begin{equation}
\frac{\partial \omega_c}{\partial q}
=
-\frac{\omega_c}{2}
\int_{0}^{2\pi}
s_\perp(\phi)\,W(\phi)\,d\phi .
\label{eq:overlap_general_topbot}
\end{equation}

Although the trampoline modes are two-dimensional and defined on a geometry without axial symmetry, the coupling depends only on the restriction of the mode to the azimuthal coordinate sampled by the ring. The function $s_\perp(\phi)$ is therefore $2\pi$-periodic and can be expanded as
\begin{equation}
s_\perp(\phi)
=
a_0
+
\sum_{p\ge 1}
\left[
a_p \cos(p\phi) + b_p \sin(p\phi)
\right].
\label{eq:Fourier_sperp}
\end{equation}
In contrast, the dependence of $W(\phi)$ on $\phi$ can be governed by two possible scenarios. For an ideal travelling-wave optical mode (clockwise or counterclockwise), the time-averaged intracavity intensity is uniform along the azimuth and $W(\phi)=W_0$ is constant. Equation~(\ref{eq:overlap_general_topbot}) then reduces to
\begin{equation}
\frac{\partial \omega_c}{\partial q}
=
-\frac{\omega_c}{2}\,W_0
\int_0^{2\pi} s_\perp(\phi)\,d\phi
=
-\frac{\omega_c}{2}\,(2\pi W_0)\,a_0 ,
\label{eq:travellingWave}
\end{equation}
where $a_0 = \langle s_\perp(\phi)\rangle_\phi$ and all Fourier components with $p\ge 1$ integrate to zero by orthogonality. In the presence of backscattering, the clockwise and counterclockwise optical modes hybridize into standing waves with azimuthally-modulated intensity. A convenient parametrization is
\begin{equation}
W(\phi) = W_0 \left[ 1 + \xi \cos(2m\phi+\phi_0) \right],
\qquad 0\le \xi \le 1,
\label{eq:standingW_weight}
\end{equation}
where $m$ is the azimuthal mode number, $\xi$ quantifies the intensity contrast, and $\phi_0$ is the standing-wave phase. Substituting Eq.~(\ref{eq:standingW_weight}) into Eq.~(\ref{eq:overlap_general_topbot}) yields an additional overlap term that selects the $p=2m$ Fourier component of $s_\perp(\phi)$ and depends on $\phi_0$. In the transmission spectra of our fabricated devices (Fig.~\ref{fig:figPreOpt}(a)), we do not observe mode splitting of the resonances, so the optical field experienced by the mechanical resonator is effectively travelling-wave. We therefore use Eq.~(\ref{eq:travellingWave}) and express the coupling in terms of the azimuthal averages $a_0$.

To connect Eq.~(\ref{eq:travellingWave}) to a minimal waveguide calculation, we express the travelling-wave constant $W_0$ in terms of the same top--bottom optical weighting that appears in the cavity MB overlap, and then identify it with the propagation-constant sensitivity $\partial\beta/\partial d$ of the corresponding (locally-bent) waveguide mode. From Eq.~(\ref{eq:Wphi_topbot}), the travelling-wave constant is
\begin{equation}
W_0=
\frac{\mathcal{F}_\mathrm{top}-\mathcal{F}_\mathrm{bot}}
{\displaystyle\int_V \varepsilon |\mathbf{E}|^2 \, dV},
\label{eq:W0_def}
\end{equation}
where in the travelling-wave regime the radial functionals are azimuth-independent. Using Eq.~(\ref{eq:Ftop_Fbot}) and the fact that the optical mode is confined to a narrow annulus around the ring radius $R$, we approximate $r\simeq R$ over $r\in\mathcal{R}(\phi)$ and write
\begin{equation}
\mathcal{F}_\mathrm{top}-\mathcal{F}_\mathrm{bot}
\simeq
R\int_{r\in\mathcal{R}}
\Big[
\mathcal{W}_\mathrm{top}\big(r,z_\mathrm{top}\big)
-
\mathcal{W}_\mathrm{bot}\big(r,z_\mathrm{bot}\big)
\Big]\,dr,
\label{eq:Fdiff_Rapprox}
\end{equation}
where $\mathcal{R}$ denotes the (narrow) radial interaction region sampled by the optical mode.

The integral in Eq.~(\ref{eq:Fdiff_Rapprox}) is precisely the moving-boundary perturbation functional that determines the propagation-constant sensitivity of the corresponding waveguide mode to a uniform gap change $d\to d+\delta d$. For a uniform gap change, the boundary-normal displacement factors are $(\delta\mathbf{r}\cdot\mathbf{n})_\mathrm{top}=+\delta d$ and $(\delta\mathbf{r}\cdot\mathbf{n})_\mathrm{bot}=-\delta d$, and first-order perturbation theory for the guided mode gives
\begin{equation}
\frac{\partial\beta}{\partial d}
=
\frac{\omega_c}{4P}
\int_{r\in\mathcal{R}}
\Big[
\mathcal{W}_\mathrm{top}
-
\mathcal{W}_\mathrm{bot}
\Big]\,dr,
\label{eq:dbeta_dg_MB}
\end{equation}
where $P$ is the guided power of the waveguide mode (evaluated at $\omega=\omega_c$). Combining Eqs.~(\ref{eq:Fdiff_Rapprox}) and (\ref{eq:dbeta_dg_MB}) yields
\begin{equation}
\mathcal{F}_\mathrm{top}-\mathcal{F}_\mathrm{bot}
\simeq
R\,\frac{4P}{\omega_c}\,\frac{\partial\beta}{\partial d}.
\label{eq:Fdiff_to_dbeta}
\end{equation}

\begin{figure}[!htbp]
    \centering
    \includegraphics[width=\textwidth]{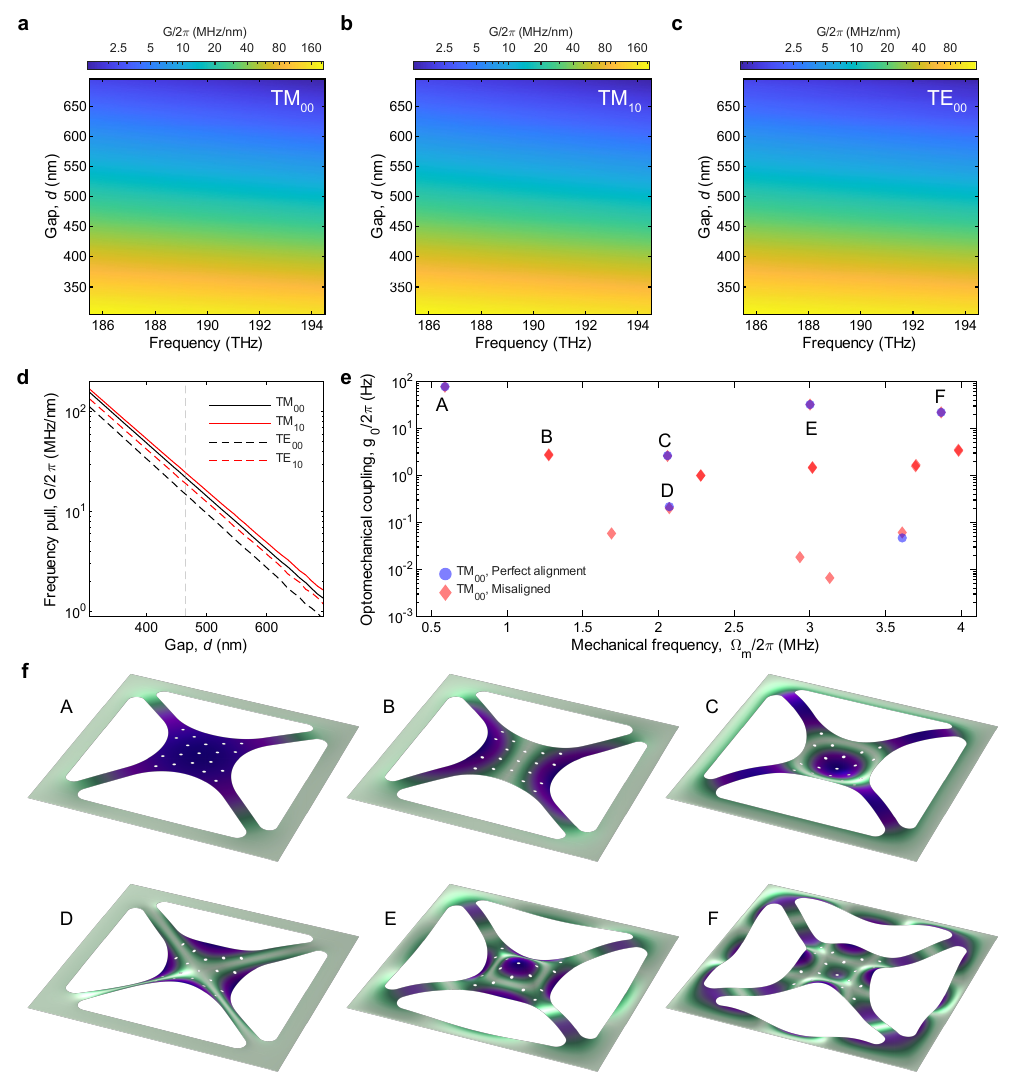}
        \caption{\textbf{Estimation of the vacuum optomechanical coupling rate, $g_0/2\pi$.} \textbf{a}--\textbf{c}, Simulated frequency pull $G/2\pi$ versus optical frequency and gap $d$ for the TM$_{00}$ (\textbf{a}), TM$_{10}$ (\textbf{b}) and TE$_{00}$ (\textbf{c}) ring modes. \textbf{d}, $G/2\pi$ versus gap at $\lambda_c = 1550$\,nm for the TM$_{00}$, TM$_{10}$, TE$_{00}$ and TE$_{10}$ modes. The dashed line marks $d = 465$\,nm, the gap used in \textbf{e}. \textbf{e}, Simulated $g_0/2\pi$ of the trampoline modes coupled to the TM$_{00}$ ring mode versus mechanical frequency, for aligned ring and trampoline centres (blue circles) and for the measured misalignment (red diamonds). \textbf{f}, Displacement profiles of the modes labelled A--F in \textbf{e}.}
    \label{fig:figFEMg0}
\end{figure}

Substituting Eq.~(\ref{eq:W0_def}) and Eq.~(\ref{eq:Fdiff_to_dbeta}) into Eq.~(\ref{eq:travellingWave}) gives
\begin{equation}
\frac{\partial\omega_c}{\partial q}
=
-\frac{\omega_c}{2}\,(2\pi a_0)\,
\frac{R\,\frac{4P}{\omega_c}\,\frac{\partial\beta}{\partial d}}
{\displaystyle\int_V \varepsilon |\mathbf{E}|^2 \, dV}
=
-(2\pi a_0)\,
\frac{RP}{\frac{1}{2}\displaystyle\int_V \varepsilon |\mathbf{E}|^2 \, dV}\,
\frac{\partial\beta}{\partial d}.
\label{eq:dwdq_in_terms_dbeta_1}
\end{equation}
For a travelling-wave cavity, the stored energy $U\equiv \frac{1}{2}\int_V \varepsilon|\mathbf{E}|^2\,dV$ is related to the circulating power $P$ by $U=(P/v_g)\,L$, where $L=2\pi R$ is the round-trip length and $v_g=c/n_g$ is the group velocity. Thus $U=(Pn_g/c)\,2\pi R$, and Eq.~(\ref{eq:dwdq_in_terms_dbeta_1}) reduces to
\begin{equation}
\frac{\partial\omega_c}{\partial q}
=
-\frac{c}{n_g}\,a_0\,\frac{\partial\beta}{\partial d}.
\label{eq:dwdq_in_terms_dbeta_2}
\end{equation}
Using $\beta=(\omega/c)\,n_\mathrm{eff}$ then yields
\begin{equation}
\frac{\partial\omega_c}{\partial q}
=
-\frac{\omega_c}{n_g}\,\frac{\partial n_\mathrm{eff}}{\partial d}\,a_0,
\label{eq:dwdq_final}
\end{equation}
and therefore the single-photon coupling rate
\begin{equation}
g_0
=
\left|\frac{\partial\omega_c}{\partial q}\right|\,q_\mathrm{zpf}
=
\frac{\omega_c}{n_g}\left|\frac{\partial n_\mathrm{eff}}{\partial d}\right|\,|a_0|\,q_\mathrm{zpf} \equiv G\,|a_0|\,q_\mathrm{zpf}
\label{eq:g0_final}
\end{equation}
where we denote $G = \frac{\omega_c}{n_g}\left|\frac{\partial n_\mathrm{eff}}{\partial d}\right|$ the frequency pull parameter. Although approximate, the expression in Eq.~(\ref{eq:g0_final}) allows the calculation of $g_0$ via independent optical and mechanical simulations without the need of calculating specific mode overlaps. In practice, $\partial n_\mathrm{eff}/\partial d$ can be obtained either from perturbation theory using Eq.~(\ref{eq:dbeta_dg_MB}) or more directly by displacing the membrane boundaries by a small $\Delta d$, recomputing $n_\mathrm{eff}(d)$ for the (locally-bent) waveguide mode, and extracting $\partial n_\mathrm{eff}/\partial d$ by finite differences. Figures~\ref{fig:figFEMg0}(a,b,c) show $G$ evaluated for our specific waveguide cross-section (see Fig.~\ref{fig:figPreOpt}(b)) and the radius of the ring ($R$ = 20 \si{\micro\meter}) as a function of optical frequency $\omega_c$ and gap $d$, respectively for modes TM$_{00}$, TM$_{10}$, and TE$_{00}$. For clarity, Fig.~\ref{fig:figFEMg0}(d) also depicts the same quantity at a fixed frequency corresponding to $\lambda_c$ = 1550 nm, including TE$_{10}$. In all panels, we observe the expected exponential behaviour as a function of $d$, while the colour maps show a slight decrease of $G$ with increasing frequency, which stems from the shorter decay length of the evanescent field for shorter wavelengths. Interestingly, TM-like modes exhibit a frequency pull about 50\% larger than TE-like modes, and the first higher-order modes a larger pull than the corresponding fundamental modes, by about 10\% (TM) and 30\% (TE) at $d = 465$\,nm. This motivates our choice of a TM mode. Within the TM family, we use TM$_{00}$ rather than TM$_{10}$, whose intrinsic loss rate is about four times larger (Section~\ref{subsec:ring_spectroscopy}), since the shot-noise-limited displacement imprecision at fixed optical power grows with $\kappa/G$. 
Independently, we perform finite element simulations of the mechanical modes of the trampoline resonator and extract the azimuthally-averaged displacement amplitude $|a_0|$. Using Eq.~(\ref{eq:g0_final}), we calculate $g_0$ between all the flexural out-of-plane modes of the trampoline and a TM$_{00}$ ring resonance at $\lambda_c$ = 1550 nm for a gap of $d$ = 465 nm. The latter is the gap measured using UV--Vis reflectometry on the suspended device prior to packaging and differs from the thickness of the polysilicon sacrificial layer, $t_{pSi}$, by approximately 15 nm. We attribute the difference to the steady-state deformation associated with the stress release (see Section~\ref{subsec:Rf}), which amounts to 14 nm in our simulations. Figure~\ref{fig:figFEMg0}(e) reports the inferred $g_0$ for two scenarios: the nominal case (solid blue circles), in which the ring resonator and trampoline centres are aligned in the $xy$-plane, and for a misalignment of $(\Delta x,\Delta y) = (0.87,-0.81)$ \si{\micro\meter} of the centre of the trampoline relative to the centre of the ring (solid red diamonds). Such misalignment corresponds to that measured on our devices. In the first case, only modes that have a displacement field symmetric relative to both $x$ and $y$ axis of the trampoline ($\sigma_x$ = +1, $\sigma_y$ = +1) have non-vanishing $a_{0}$, e.g., modes labelled A/C/D/E/F in Fig.~\ref{fig:figFEMg0}(f). Among these, one could also expect mode D to have vanishing $a_0$ due to the parity relative to an axis going through the centre of the tether, but the holes used for assisting the release break the symmetry, leading to a finite $g_0$. As shown in Fig.~\ref{fig:figFEMg0}(e), the misalignment between the two lithographies does not considerably alter the coupling strength for the modes with full symmetry, but it does lead to finite $g_0$ for otherwise dark mechanical modes. An example of such mode is that labelled B in Fig.~\ref{fig:figFEMg0}(f), but many others are visible. We also observe that for higher-order mechanical modes, e.g. E/F, the motion of the central trampoline extends to the overhang region at and around the clamping points of the resonator. Many other modes with only motion on the overhang are also found by the finite-element solver, but not shown. As shown in the next section, we observe for particular geometries how this effect leads to multiple mechanical peaks over the same frequency region, likely resulting from hybridized modes between the otherwise perfect trampoline mechanical modes and those of the frame. 

For generality, we note how the properties of the optomechanical transducer scale with size. Under a uniform in-plane scaling of the device geometry\textemdash both the trampoline and the ring\textemdash by a factor $W$ (i.e. $x,y \to W x, W y$), while keeping the material properties, film thickness, optical cross section, and the gap between resonators fixed, $g_0$ scales primarily through the zero-point amplitude $q_{\mathrm{zpf}}$. For sufficiently large rings or racetrack resonators, $n_g$ and $\partial n_{\mathrm{eff}}/\partial d$ are governed by the local straight-waveguide cross section and are therefore invariant under in-plane similarity transformations. Likewise, the mechanical participation factor $a_0$ remains unchanged for the same mode order. In the case of a tension-dominated thin resonator, the effective mass scales as $m_{\mathrm{eff}} \propto W^2$, while the mechanical eigenfrequency scales as $\Omega_m \propto 1/W$. Thus, for geometrically scaled devices operated at fixed optical wavelength, the single-photon coupling rate decreases with the square root of the in-plane scaling factor, $g_0 \propto W^{-1/2}$.

 \begin{figure}[!htbp]
    \centering
    \includegraphics[width=\textwidth]{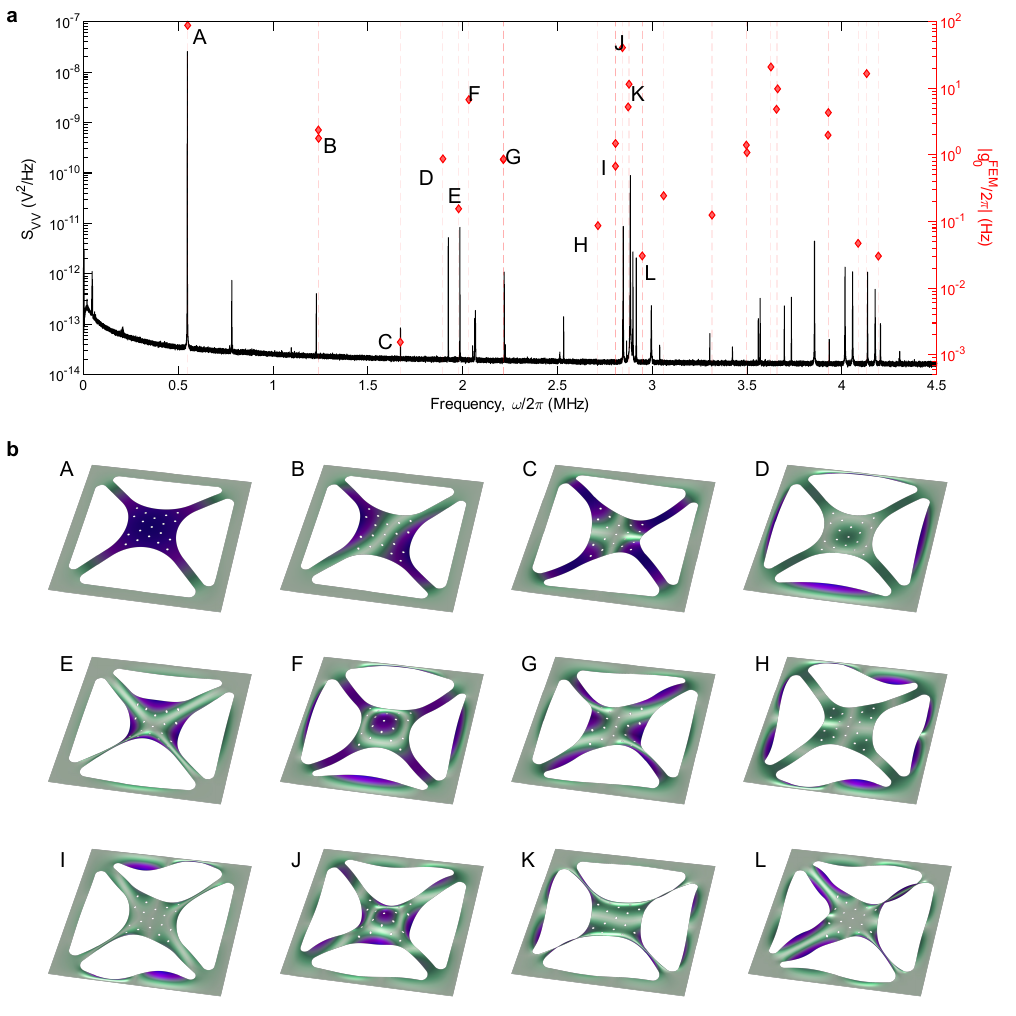}
        \caption{\textbf{Mechanical spectroscopy of a trampoline resonator.} \textbf{a}, (Left axis) Uncalibrated displacement PSD over a broad frequency range measured via direct detection using the same optical resonance as in the main text. (Right axis) Simulated mechanical frequencies and their respective $g_0$ with a TM$_{00}$ ring resonance at $\lambda_c = 1555 \, \mathrm{nm}$, close to the resonance used in the main text ($\lambda_0 = 1555.24$\,nm). Vertical dashed lines mark the simulated mode frequencies as a guide to the eye. \textbf{b}, Simulated displacement profiles for the mechanical modes highlighted in \textbf{a}.}
    \label{fig:figBBMec}
\end{figure}

\subsection{Broadband mechanical characterization with integrated optical readout}
\label{subsec:broadband_mech}

In the main text, we use the fundamental out-of-plane flexural mode of the trampoline resonator for radiation sensing. Nevertheless, our photonic-integrated readout via the ring resonator allows the detection of thermomechanical motion associated with other higher-order flexural modes. Figure~\ref{fig:figBBMec}(a) shows the uncalibrated displacement PSD over a broad frequency range measured via direct detection using the same optical resonance as in the main text. In addition to the fundamental mode at $\Omega_m/2\pi \sim$ 550 \si{\kilo\hertz}, we observe several higher-order modes with smaller peak transduction at higher frequencies. Figure~\ref{fig:figBBMec}(a) also shows the calculated $g_0$ for the different simulated mechanical modes and assuming the ring-to-trampoline in-plane offset described before. We note that compared to the simulation in Fig.~\ref{fig:figFEMg0}(e,f), the trampoline here approximates the fabricated geometry more closely by: i) using a different thickness for the silicon nitride in the region below the platinum disk (140 nm instead of 100 nm) to account for the differences in etching due to XeF$_2$ (see Methods of the main text) and ii) including the misalignment of the platinum disk with respect to the trampoline. For this geometry, the fundamental mode has a simulated frequency of 550.1\,kHz and an effective mass $m_\mathrm{eff} = 729$\,pg, normalized to its maximum displacement at the centre of the trampoline, which we use for the displacement calibrations of the thermomechanical spectra in main text Figs.~1(g,h) and of the Duffing response (Section~\ref{subsec:duffing}). While the correspondence between the evolution of $g_0$ and the peak transduction is not obvious, the mechanical frequencies and mode density agree qualitatively. By inspection of the mode profiles (Fig.~\ref{fig:figBBMec}(b)), we see that the three lowest frequency modes, for which motion is primarily on the trampoline, have measured frequencies that agree to within 5\% with the simulation. In contrast, for frequency regions around $\omega/2\pi$ = 2 MHz and $\omega/2\pi$ = 3 MHz, the one-to-one correspondence of simulated and measured modes is less accurate. The modes in these regions correspond to modes with significant motion on the silicon nitride frame, with some modes mainly vibrating on the frame, e.g., D in Fig.~\ref{fig:figBBMec}(b). This reveals strong interaction and coupling between trampoline modes and frame modes, the specifics of which depend finely on the geometry of the overhang. Since our simulations do not include the overhang sections bounding the waveguide trenches, which extend along the whole chip, the exact frequencies and number of modes in these frequency regions are hard to predict, and the agreement degrades as frequency increases. Nevertheless, the overall eigenspectrum landscape is well reproduced.

\subsection{Calibration of the vacuum optomechanical coupling rate}

In this section, the procedure for calibrating the vacuum optomechanical coupling rate, $g_0$, is explained. Within the linear cavity optomechanical transduction regime and using direct detection, the photocurrent generated by a photodetector with responsivity $R_D$ (A/W) is given by
$$ I(t) = R_D(D^{(0)} +  D^{(1)}\xi(t)),$$
where
\begin{align}
    \xi(t) &= \frac{2G}{\kappa}x(t),\\
    D^{(0)} &= \eta_c P_\text{det}\left(1 - \frac{4\eta(1-\eta)}{1 + \bar\delta^2}\right),\\
    D^{(1)} &= \eta_c P_\text{det}4\eta(1-\eta)\frac{2\bar\delta}{(1 + \bar\delta^2)^2},
\end{align}
and $\bar\delta = \frac{2\bar\Delta}{\kappa}$ is the mean normalized detuning. Here, $P_\text{det}$ is the optical power sent to the chip and $\eta_c = \eta_\text{f}^2\approx 0.095$ the fibre-to-fibre transmission of the packaged chip away from resonance (Methods of the main text). The coordinate $x$ is the displacement of the trampoline averaged over the ring, $x = a_0 q$ (Section~\ref{subsec:OM_trampoline_ring}), so that a displacement $x$ shifts the cavity resonance by $Gx$. We assume that the employed detector has a flat response $R_D$ at the relevant frequencies. The DC and AC components of the detected voltage are therefore given by
\begin{align}
    V_\mathrm{DC} &= Z_\mathrm{DC}R_DD^{(0)},\\
    V_\mathrm{AC}(t) &= Z_\mathrm{AC}R_DD^{(1)}\xi(t),
\end{align}
where $Z_\mathrm{DC}$ and $Z_\mathrm{AC}$ are values of the transimpedance gain used for reading out the DC and AC signals. The double-sided power spectral density (PSD) of the AC signal, defined as $S_{VV}(\omega) = \int e^{i\omega\tau}\langle V_\mathrm{AC}(t)V_\mathrm{AC}(t+\tau)\rangle d\tau$, can therefore be written as
\begin{equation}\label{eq:S_VV_calibration}
    S_{VV}(\omega) = \left[ \frac{Z_\mathrm{AC}}{Z_\mathrm{DC}} \right]^2\left[\frac{V_\mathrm{DC}D^{(1)}(\bar\delta)}{D^{(0)}(\bar\delta)}\right]^2 \left(\frac{2G}{\kappa}\right)^2S_{xx}(\omega).
\end{equation}

\begin{figure}[!htbp]
    \centering
    \includegraphics[width=\textwidth]{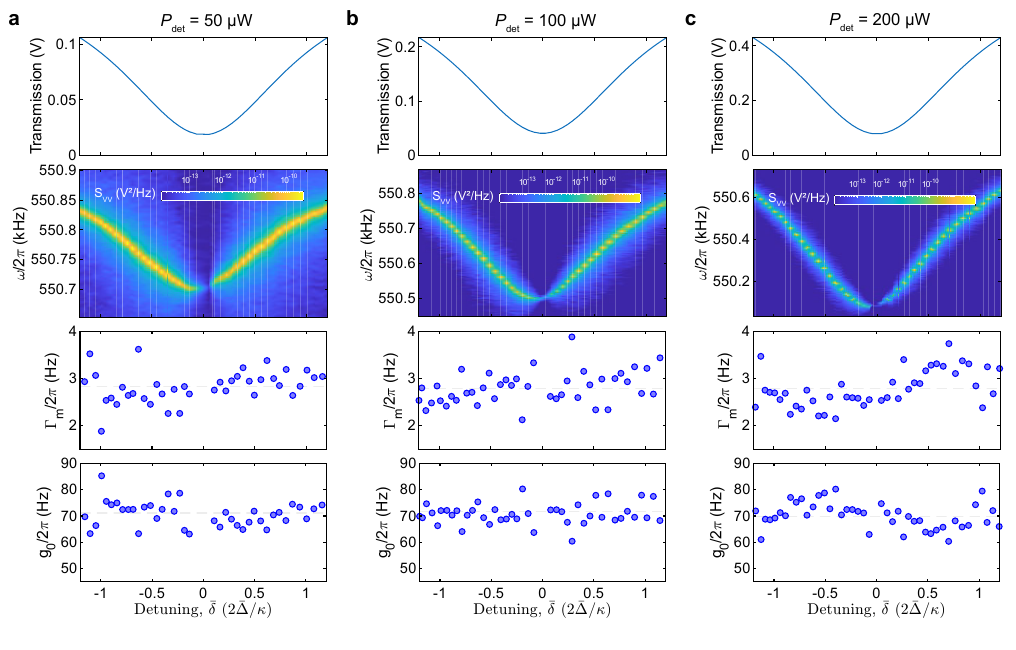}
    \caption{\textbf{Calibration of the vacuum optomechanical coupling rate, $g_0$.} From top to bottom: transmission ($V_\mathrm{DC}$), PSD of the uncalibrated displacement noise (colour), fitted linewidth, $\Gamma_\mathrm{m}$, and extracted $g_0$. All quantities are plotted as a function of the normalized detuning, $\bar\delta$, which is obtained from a scan of the optical resonance. The horizontal dashed lines represent the detuning-averaged values. Data for three different off-chip optical powers are shown: \textbf{a}, $P_\mathrm{det} = 50$\,\si{\micro\watt}, \textbf{b}, $P_\mathrm{det} = 100$\,\si{\micro\watt}, \textbf{c}, $P_\mathrm{det} = 200$\,\si{\micro\watt}.}
    \label{fig:g0_calibration}
\end{figure}

To calibrate the vacuum optomechanical coupling rate $g_0$ one has to integrate both sides of Eq.~(\ref{eq:S_VV_calibration}) around a mechanical sideband. Through integration of $S_{xx}$ around a mechanical mode with frequency $\Omega_m$ one obtains $\langle x^2\rangle = k_{B}T/m_\text{eff}\Omega_m^2$ and using the definition of the zero-point-motion, of the average thermal occupation number $n_\mathrm{th} = k_BT/\hbar\Omega_m$ and $g_0 = Gx_\text{zpf}$ with $x_\text{zpf} = |a_0|q_\text{zpf}$, one finds
\begin{equation}
    g_0 = \frac{1}{\sqrt{2n_\mathrm{th}}}\,\frac{V_\mathrm{rms}}{V_\mathrm{DC}}\,\frac{\kappa}{2}\,\frac{D^{(0)}(\bar\delta)}{D^{(1)}(\bar\delta)}\,\frac{Z_\mathrm{DC}}{Z_\mathrm{AC}}
\end{equation}
where $V_\mathrm{rms}$ is the RMS value of the mechanical sideband in the voltage spectrum.
In practice, the calibration is performed by measuring the optical resonance and fitting the voltage trace, acquired through the same DC path, to a Lorentzian, which allows the determination of $\kappa$, $\eta$ and the off-resonance voltage, $V_\mathrm{BG} = \eta_c P_\mathrm{det} Z_\mathrm{DC}R_D$. Then, at any given detuning, $V_\mathrm{DC}$ and the PSD of the voltage are acquired, and the PSD is fitted to the sum of a Lorentzian function plus a measurement imprecision background to determine $\Omega_m$ and $\Gamma_m$. $V_\text{rms}$ is then calculated from the area under the mechanical peak and $\bar{\delta}$ can be found from $V_\mathrm{DC}$ and $V_\mathrm{BG}$. 

We show in Figure~\ref{fig:g0_calibration} the calibrated $g_0$ between the fundamental out-of-plane mode of the trampoline and the TM$_{00}$ optical resonance used in the main text, as a function of detuning and power. The extracted $g_0$ is consistent across the different powers and detunings, despite two effects of the residual absorption of the readout light (Section~\ref{subsec:kappa_abs}). First, the mechanical frequency redshifts in proportion to the intracavity power. Second, the linewidth responds dispersively to the detuning, with $\Gamma_\mathrm{m}$ decreasing for red and increasing for blue detuning. This becomes visible at $P_\mathrm{det} = 200$\,\si{\micro\watt} (Fig.~\ref{fig:g0_calibration}(c)) and is more pronounced in measurements at higher power (not shown). It is opposite to dynamical radiation-pressure backaction, which would moreover change $\Gamma_\mathrm{m}/2\pi$ by only a few mHz at this power given $g_0$, $\kappa$ and the intracavity photon number, and therefore identifies photothermal backaction.Using the data for $P_\mathrm{det} = 50\,\mu\mathrm{W}$, we obtain $g_0/2\pi = 71 \pm 5 \,\mathrm{Hz}$ (mean and standard deviation over detunings), within 20\% of the simulated value $g^\mathrm{FEM}_0/2\pi = 87.9 \,\mathrm{Hz}$. We attribute the difference to the uncertainty in the determination of the trampoline-to-cladding gap, $d$, to which $g_0$ is exponentially sensitive (Fig.~\ref{fig:figFEMg0}). The mechanical frequency in this and the following datasets differs from the value quoted in the main text by up to a few kHz, because it drifts between pump-downs (Methods of the main text).

\subsection{Photothermal frequency shifts due to residual C-band absorption}
\label{subsec:kappa_abs}

\begin{figure}[!htbp]
    \centering
    \includegraphics[width=\textwidth]{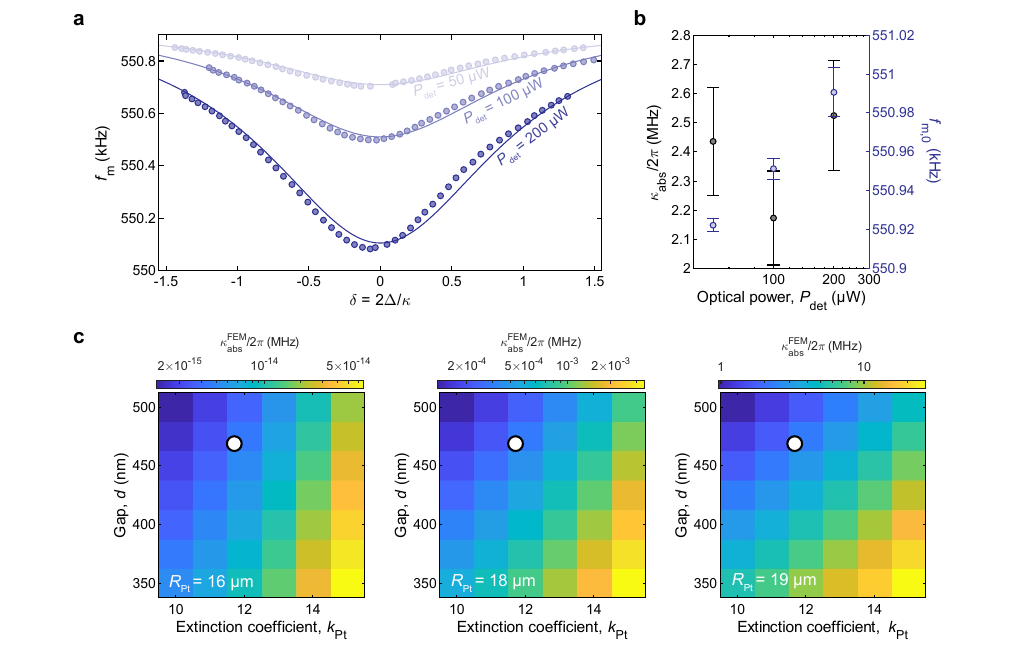}
    \caption{\textbf{Photothermal absorption at the detection wavelength.} \textbf{a}, Mechanical frequency $f_m$ as a function of normalized detuning $\delta = 2\Delta/\kappa$ for three off-chip optical powers $P_\text{det}$. Solid lines are fits to the cavity-lineshape model of Eq.~(\ref{eq:f_m}). \textbf{b}, (Left axis) Effective absorption rate $\kappa_\text{abs}/2\pi$ extracted from the fits in \textbf{a} as a function of $P_\text{det}$. (Right axis) Unperturbed mechanical frequency $f_{m,0}$. \textbf{c}, FEM-simulated absorption rate $\kappa_\text{abs}^\text{FEM}/2\pi$ as a function of the trampoline-to-ring gap $d$ and the Pt extinction coefficient $k_\text{Pt}$, for three Pt disk radii $R_\text{Pt}$. The white circle marks the nominal device parameters in each panel.}
    \label{fig:kappa_abs}
\end{figure}

In addition to its operation as a bolometric detector for externally incident mid-infrared radiation, the device exhibits a photothermal response at the optical wavelength used for readout, $\lambda_\text{det}$. This effect originates from unwanted absorption of the intracavity field, which heats the membrane--absorber system and shifts the mechanical resonance frequency. This effect is parasitic from a detection standpoint and has only been observed in devices including the platinum disk.

We relate the measured steady-state frequency shift to the absorbed power through the same responsivity as for mid-IR radiation, $\mathcal{R}^{\rm abs}_0$. In both cases, absorption occurs predominantly in the platinum layer, and the resulting temperature rise is set by the same thermal conductance of the trampoline tethers. Assuming linear absorption from intracavity photons, the absorbed power is $P_{\rm abs} = \hbar \omega_c \kappa_{\rm abs} n_c$, where $\kappa_{\rm abs}$ is the effective absorption rate and the intracavity photon number is given by
\begin{equation}
n_c =
\frac{4\eta}{1+\delta^2}
\frac{P_{\rm in}}{\hbar \omega_c \kappa},
\end{equation}
with $\eta = \kappa_e/\kappa$ the external coupling fraction, $P_{\rm in} = \eta_\text{f}P_\text{det}$ the power in the bus waveguide and $\delta = 2\Delta/\kappa$ the normalized detuning. The photothermally shifted mechanical frequency as a function of $\delta$ is then given by
\begin{equation}
    \label{eq:f_m}
    f_\mathrm{m}(\delta) = f_{m,0} + \mathcal{R}^{\rm abs}_0 f_{m,0} \frac{4\eta \kappa_{\rm abs}}{1+\delta^2}
\frac{P_{\rm in}}{\kappa},
\end{equation}
where $f_{m,0}$ denotes the mechanical resonance frequency in the absence of photothermal effects and $\mathcal{R}^{\rm abs}_0 = -(6.3\pm0.5)\times10^{3}$\,W$^{-1}$ is the measured responsivity to absorbed power (main text). Figure~\ref{fig:kappa_abs}(a) shows, for the three optical powers, the mechanical frequencies $f_\text{m}(\delta_i)$ extracted from Lorentzian fits to the thermomechanical displacement spectra of Fig.~\ref{fig:g0_calibration}, plotted as a function of the detuning at which each spectrum is acquired. Fits of this dependence to Eq.~(\ref{eq:f_m}), with $\kappa$, $\eta$ and $P_\mathrm{in}$ known from optical spectroscopy and insertion-loss calibration (Methods of the main text), yield $\kappa_{\rm abs}$ and $f_{m,0}$ as the only free parameters. Figure~\ref{fig:kappa_abs}(b) shows the extracted $\kappa_{\rm abs}/2\pi$ at the three powers alongside the unperturbed mechanical frequency $f_{m,0}$. The extracted absorption rate varies by less than 15\% over a four-fold change in power, confirming linearity, and gives $\kappa_{\rm abs}/2\pi = 2.4\pm0.2$\,MHz (mean and standard deviation over the three powers, left axis). This value is negligible compared with the cavity linewidth ($\kappa/2\pi = 1.16$\,GHz), yet as discussed in the main text and in Section~\ref{subsec:power_fluctuations} here, large enough to produce significant photothermal frequency noise through fluctuations of the delivered optical power. In addition, we observe a small increase in $f_\text{m,0}$ with $P_\text{det}$ (right axis), which we simply attribute to long-term drift during the minutes-long step scans from which the data are obtained.

To assess the physical origin of this absorption, we compute $\kappa_{\rm abs}^{\rm FEM}$ from FEM simulations of the TM$_{00}$ mode, using optical constants of platinum at $\lambda_\text{det}$ = 1550\,nm extracted from UV-Vis spectroscopy of reference Pt-coated Si$_x$N$_y$ membranes (complex refractive index $n_\text{Pt} + i k_\text{Pt} = 1.24 + i\,11.75$). The absorption rate is computed from the simulated mode as $\kappa_{\rm abs}^{\rm FEM} = P_{\rm abs,Pt}/U$, where $P_{\rm abs,Pt}$ is the power dissipated in the platinum volume and $U$ is the total electromagnetic energy stored in the mode, which captures the direct evanescent overlap between the guided field and the absorbing metal. Figure~\ref{fig:kappa_abs}(c) shows $\kappa_{\rm abs}^{\rm FEM}/2\pi$ as a function of gap $d$ and Pt extinction coefficient $k_{\rm Pt}$ for three disk radii $R_\text{Pt}$, with the nominal device parameters indicated by the white circle. For $R_\text{Pt}$ = 16\,µm (already larger than the fabricated value of 15.2\,µm), the simulated absorption rate is of order $10^{-14}$\,MHz, about fourteen orders of magnitude below the measured value. Only when $R_\text{Pt}$ approaches 19\,µm, so that the disk edge overlaps the ring annulus, does $\kappa_{\rm abs}^{\rm FEM}$ reach the experimentally observed range. This dramatic discrepancy rules out direct evanescent absorption as the dominant mechanism, and the following interpretation is likely the most plausible one. In scattering-dominated resonators such as ours, where intrinsic linewidths substantially exceed simulated bending-loss contributions, a significant fraction of the intracavity power is continuously transferred into radiation and leaky modes by sidewall roughness. A portion of this scattered field reaches the platinum disk and is absorbed there, constituting an effective absorption channel invisible to overlap-based guided-mode simulations such as those in Fig.~\ref{fig:kappa_abs}(c). In this picture, $\kappa_{\rm abs}$ can be interpreted as the fraction of the total intrinsic loss rate $\kappa_i$ whose dissipated power ultimately heats the absorber, which reconciles the large measured value with the negligible direct evanescent overlap.

\subsection{Cavity-optomechanical sensing at large amplitudes}
\label{subsec:largeAmplitude}

At the large coherent oscillation amplitudes, $X_\mathrm{osc}$, used for frequency tracking, the mechanical motion modulates the cavity resonance by more than a half-linewidth, $GX_\mathrm{osc}\gtrsim\kappa/2$, and the transduction of displacement into transmitted intensity becomes strongly nonlinear (main text, Fig.~2). In this regime the operating point that maximizes the power carried by the demodulated first harmonic, $P_{N=1}$, no longer coincides with the maximum-slope detuning of the linear regime. Following Ref.~\cite{javid_cavity_2021}, we derive the harmonic content of the transmitted field as a function of detuning and modulation depth and identify the optimum reported in Fig.~2(e) of the main text.

We describe the driven cavity by the intracavity field amplitude $a(t)$, normalized such that $|a|^2$ is the intracavity photon number. In a frame rotating at the laser frequency, with the resonance modulated by the mechanical displacement $x(t)$ through the frequency-pull parameter $G$, the equation of motion is
\begin{equation}
\label{eq:eom_largeamp}
\dot{a}(t) = \left[\,i\big(\bar\Delta - G x(t)\big) - \frac{\kappa}{2}\,\right] a(t) + i\sqrt{\kappa_e}\,s_\mathrm{in},
\end{equation}
where $\bar\Delta = \omega_\mathrm{L} - \omega_0$ is the mean laser--cavity detuning, $\kappa$ ($\kappa_e$) the total (external) linewidth, and $s_\mathrm{in}$ the input field amplitude. For a coherent single-tone drive, $x(t) = X_\mathrm{osc}\cos(\Omega_\mathrm{m}t)$, Eq.~(\ref{eq:eom_largeamp}) admits an exact steady-state (Floquet) solution~\cite{javid_cavity_2021},
\begin{equation}
\label{eq:floquet}
a(t) = -i\sqrt{\kappa_e}\,s_\mathrm{in}\sum_{n,p} \frac{J_n(\beta)\,J_{n-p}(\beta)}{\,i(\bar\Delta - n\Omega_\mathrm{m}) - \kappa/2\,}\,e^{i p\Omega_\mathrm{m}t},
\qquad \beta \equiv \frac{G X_\mathrm{osc}}{\Omega_\mathrm{m}},
\end{equation}
where $J_n$ is the Bessel function of the first kind of order $n$ and $\beta$ is the modulation index. The transmitted intensity collected at the output port,
\begin{equation}
\label{eq:Pout_largeamp}
P(t) \propto \big|\,s_\mathrm{in} + i\sqrt{\kappa_e}\,a(t)\,\big|^2,
\end{equation}
inherits the periodicity of the drive and thus carries harmonics of $\Omega_\mathrm{m}$ whose relative weights depend on both $\bar\Delta$ and $X_\mathrm{osc}$.

Equation~(\ref{eq:floquet}) holds for any ratio $\Omega_\mathrm{m}/\kappa$. Our device operates deep in the unresolved-sideband (adiabatic) regime, $\Omega_\mathrm{m,0}/2\pi = 549.6$\,kHz $\ll \kappa/2\pi = 1.16$\,GHz, more than three orders of magnitude apart, so the intracavity field follows the instantaneous detuning quasi-statically. In this limit the resonant part of the detected power reduces to the steady-state cavity lineshape evaluated at the instantaneous detuning,
\begin{equation}
\label{eq:instLorentzian}
P(t) \propto \mathcal{L}\!\big(\bar\delta + \xi\cos\Omega_\mathrm{m}t\big),
\qquad
\mathcal{L}(u) = \frac{1}{1+u^2},
\end{equation}
where $\bar\delta = 2\bar\Delta/\kappa$ is the mean normalized detuning and
\begin{equation}
\label{eq:xi_def}
\xi \equiv \frac{2 G X_\mathrm{osc}}{\kappa}
\end{equation}
is the peak frequency excursion normalized to the cavity half-linewidth. Constant and slowly varying offsets of the transmission do not contribute to the harmonics and are omitted.

\begin{figure}[!htbp]
    \centering
    \includegraphics[width=\textwidth]{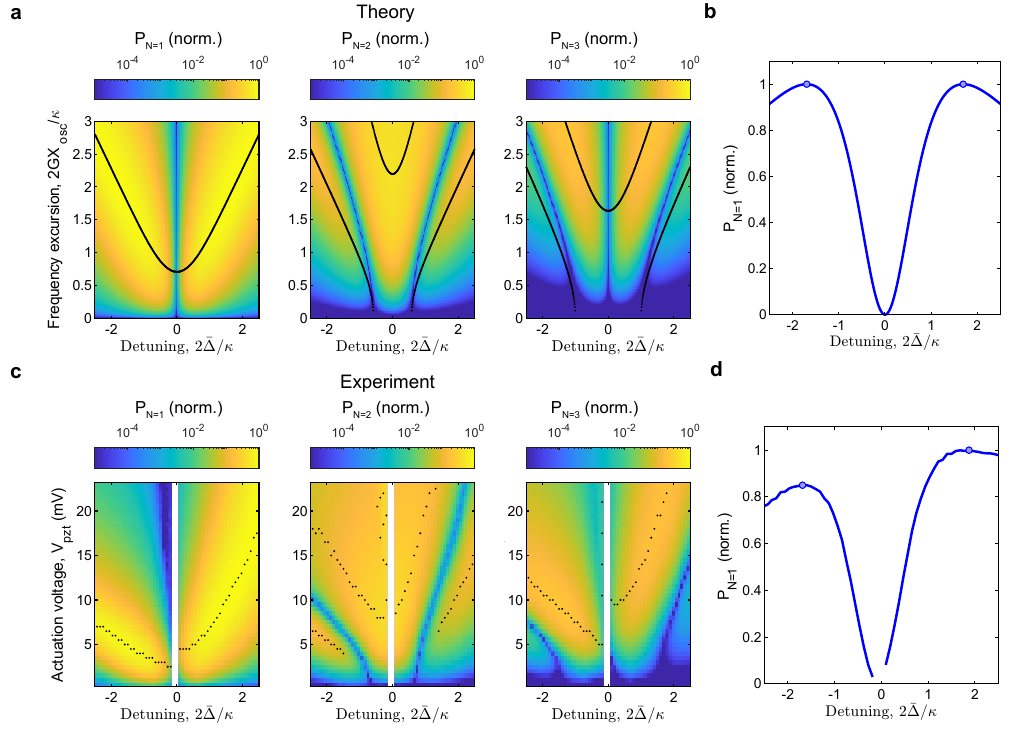}
    \caption{\textbf{Harmonic content of the nonlinear cavity transduction.} \textbf{a}, Theoretical normalized power in the first three harmonics of the transmitted intensity, $P_{N}$ ($N = 1,2,3$), computed from Eq.~(\ref{eq:BN_integral}) as a function of the normalized mean detuning $2\bar\Delta/\kappa$ and the normalized frequency excursion $2GX_\mathrm{osc}/\kappa$. Black markers trace, at each detuning, the excursion that maximizes each harmonic. \textbf{b}, Maximum over excursion of the first-harmonic power, $\max_\xi P_{N=1}$, versus detuning (the envelope of the $N=1$ map in \textbf{a}). The markers denote the optimal detuning $\pm\bar\delta_*$ and correspond to the blue curve in Fig.~2(e) of the main text. \textbf{c}, Measured counterpart of \textbf{a}, with the coherent drive set by the piezoelectric actuation voltage $V_\mathrm{pzt}$ (proportional to $X_\mathrm{osc}$) in place of the normalized excursion. \textbf{d}, Measured maximum over drive of $P_{N=1}$ versus detuning, with the optimal detuning marked. All maps are normalized to their maximum and shown on a logarithmic colour scale.}
    \label{fig:largeAmplitude}
\end{figure}

Since $P(t)$ is periodic and even in time, it can be expanded as $P(t)\propto\sum_{N\ge0}\mathcal{B}_N(\bar\delta,\xi)\cos(N\Omega_\mathrm{m}t)$, with Fourier coefficients
\begin{equation}
\label{eq:BN_integral}
\mathcal{B}_N(\bar\delta,\xi) = \frac{\epsilon_N}{2\pi}\int_0^{2\pi} \frac{\cos(N\theta)}{1+\big(\bar\delta+\xi\cos\theta\big)^2}\,d\theta,
\qquad \epsilon_0 = 1,\;\epsilon_{N\ge1} = 2 .
\end{equation}
This integral admits a closed form. Writing the Lorentzian as a pair of simple poles, $\mathcal{L}(u)=\tfrac{1}{2i}\big[(u-i)^{-1}-(u+i)^{-1}\big]$, and using the standard result $\frac{1}{2\pi}\int_0^{2\pi}\cos(N\theta)/(a+\xi\cos\theta)\,d\theta = (a^2-\xi^2)^{-1/2}\,r^{N}$, with $r = \big(\sqrt{a^2-\xi^2}-a\big)/\xi$ the root satisfying $|r|<1$, one obtains
\begin{equation}
\label{eq:BN_closed}
\mathcal{B}_N(\bar\delta,\xi) = \epsilon_N\,\mathrm{Im}\!\left[\frac{1}{\sqrt{a^2-\xi^2}}\left(\frac{\sqrt{a^2-\xi^2}-a}{\xi}\right)^{\!N}\right]_{a\,=\,\bar\delta-i},
\end{equation}
where the branch of the square root is fixed by $|r|<1$. The power demodulated at the $N$-th harmonic is $P_N\propto\mathcal{B}_N^2$; in particular the first-harmonic power tracked by the lock-in is $P_{N=1}\propto\mathcal{B}_1^2(\bar\delta,\xi)$, and maximizing it maximizes the transduction of mechanical motion into intensity modulation. With a detection apparatus limited by the photodetector noise floor, maximizing this transduction also maximizes the detection signal-to-noise ratio in the nonlinear regime.

Figure~\ref{fig:largeAmplitude}(a) shows the coefficients $\mathcal{B}_N$ for the first three harmonics as a function of $\bar\delta$ and $\xi$. At fixed detuning, $P_{N=1}$ first grows with the excursion, reaches a maximum, and then decreases as the transduced energy is redistributed into higher harmonics once the swing crosses the transmission minimum. The ridges of maxima for the three harmonics are traced by the black markers. The maximum attainable first-harmonic power at each detuning, $\max_\xi P_{N=1}(\bar\delta)$, is plotted in Fig.~\ref{fig:largeAmplitude}(b). It peaks at a finite detuning $\bar\delta_* = 2\bar\Delta_*/\kappa$, with $|\bar\Delta_*|\approx 0.85\,\kappa$, which sets the optimal operating point for nonlinear transduction and corresponds to the blue curve in Fig.~2(e) of the main text. The measured harmonic powers, acquired by step-scanning the detuning at a series of piezoelectric drive voltages, are shown in Fig.~\ref{fig:largeAmplitude}(c). The dataset is acquired while employing the PLL tracking scheme, which fails for very low signal-to-noise ratio (very low $P_{N=1}$ for our demodulation strategy), as evidenced by the lack of data around $\delta$ = 0 in Fig.~\ref{fig:largeAmplitude}(c). For the rest, the experimental data qualitatively reproduce the structure of Fig.~\ref{fig:largeAmplitude}(a), including the suppression of the first/third harmonic on resonance and the off-resonance maxima of the first-harmonic power (Fig.~\ref{fig:largeAmplitude}(c,d)). The quantitative differences between theory and experiment result from the deviation of the cavity response from a Lorentzian: the observed asymmetry is due to a slight Fano-like behaviour of the response. The consequence of such asymmetry in the cavity response is an asymmetry in the maximum first harmonic power as a function of detuning, with the global maximum occurring on the blue-detuned side (Fig.~\ref{fig:largeAmplitude}(d)). The same is seen for the detection noise contribution to the frequency noise budget in Fig. 2(e) in the main text. %
There, the measured inverse detection noise vanishes on resonance and is largest for $|\bar\Delta|\approx|\bar\Delta_*|$, peaking on the blue-detuned branch and forming a broad plateau on the red-detuned branch, as expected from this asymmetry.

Importantly, since $\mathcal{B}_N$ depends only on the reduced variables $\bar\delta$ and $\xi$ in the sideband unresolved regime, the optimal operating point is universal: it always occurs at the same normalized detuning $\bar\delta_*$ ($|\bar\Delta_*|\approx 0.85\,\kappa$, so that $\bar\Delta_*\propto\kappa$), while the frequency-pull parameter $G$ only sets the coherent amplitude $X_\mathrm{osc}^*=\xi_*\kappa/(2G)$ required to reach it. However, the implications for the frequency noise are a bit more subtle: while a cavity with a larger $\kappa$ can in principle lead to the same $P_{N=1}$ and detection frequency noise, the drive amplitude $X_\mathrm{osc}^*$ required to reach that point becomes much larger, which typically implies that strong amplitude-to-frequency noise conversion due to the Duffing nonlinearity appears before achieving that amplitude, as discussed in the main text and in Section~\ref{subsec:drive_induced_frequency_noise} here. Experimentally, the asymmetry of the cavity lineshape shifts the first-harmonic optimum on the blue-detuned side to larger drive voltages (Fig.~\ref{fig:largeAmplitude}(c)), where drive-induced excess frequency noise appears (Section~\ref{subsec:drive_induced_frequency_noise}). For this reason, despite the larger optimal $P_{N=1}$ on the blue-detuned side (Fig.~\ref{fig:largeAmplitude}(d)), we operate the OMTD at $\bar\Delta\approx-|\bar\Delta_*|$ rather than $+|\bar\Delta_*|$.

\subsection{Duffing nonlinearity}
\label{subsec:duffing}

\begin{figure}[t]
    \centering
    \includegraphics[width=\textwidth]{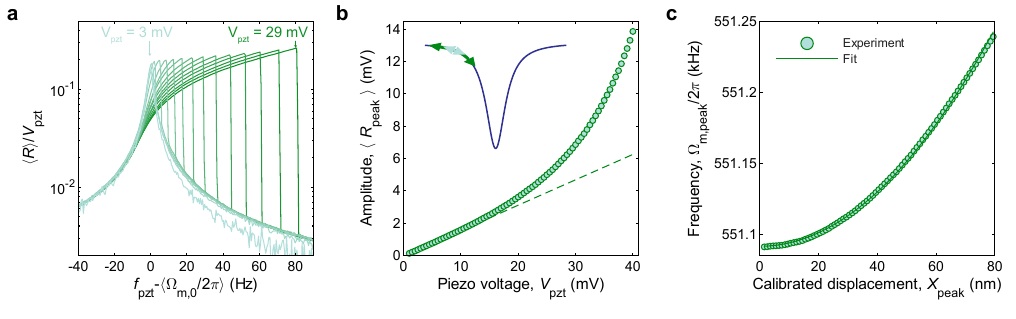}
    \caption{\textbf{Measurement of the Duffing response.} \textbf{a}, Coherent amplitude response normalized by the drive voltage, $\langle R \rangle/V_\mathrm{pzt}$, of the fundamental trampoline mode at increasing drive voltage $V_\mathrm{pzt}$. \textbf{b}, Demodulated peak amplitude $\langle R_\mathrm{peak} \rangle$ as a function of $V_\mathrm{pzt}$. The inset shows the cavity-transduction nonlinearity. The dashed line is a linear fit to the low-drive region, used to extract the amplitude-to-drive-voltage calibration factor. \textbf{c}, Backbone curve: peak frequency $\Omega_\mathrm{m,peak}/2\pi$ versus calibrated peak amplitude $X_\mathrm{peak}$, extracted from the family of resonance sweeps in panel~\textbf{a}. The solid line is a fit to the Duffing backbone relation, Eq.~(\ref{eq:backbone}).}
    \label{fig:Duffing_response}
\end{figure}

The high-$Q$ fundamental mode of the trampoline resonator exhibits, as shown in Fig.~\ref{fig:Duffing_response}, a hardening nonlinearity. We attribute this nonlinearity to a cubic Duffing-type geometric nonlinearity~\cite{lifshitz_nonlinear_2008}. The dynamics of the driven Duffing resonator are described by the equation of motion
\begin{equation}
    \ddot{x} + \Gamma_m\,\dot{x} + \Omega_{m,0}^2\,x + \alpha\,x^3 = \frac{F_\mathrm{pzt}}{m_\mathrm{eff}}\cos(2\pi f_\mathrm{pzt} t),
\end{equation}
where $\alpha$ is the Duffing coefficient (positive for a hardening spring), $m_\mathrm{eff}$ is the effective mass, and $F_\mathrm{pzt}$ is the drive force that results from the off-chip piezo-actuator. Figure~\ref{fig:Duffing_response}a shows the steady-state coherent amplitude response of the fundamental trampoline mode as the drive amplitude increases. The resonance peak shifts to higher frequencies and eventually exhibits the characteristic jump instability when the system becomes bistable, confirming hardening nonlinearity ($\alpha > 0$). To correct for a slow linear drift of $\Omega_{m,0}$ over the course of the measurement, the lowest-drive scan is repeated at the end of the acquisition sequence, and the drift estimated from the frequency offset between the first and last low-drive traces is subtracted linearly from all intermediate sweeps.

To calibrate the detected voltage amplitude into a physical displacement, we combine two independent measurements: (i) a Lorentzian fit to the thermomechanical noise PSD recorded with the same lock-in amplifier, which yields $V_\mathrm{rms}$ and, via the equipartition theorem, its correspondence to the thermal motion, and (ii) a linear fit to the amplitude peak, $\langle R_\mathrm{peak} \rangle$, which gives the amplitude-to-drive-voltage transduction factor without the strong cavity-transduction nonlinearity (Fig.~\ref{fig:Duffing_response}(b), dashed line). This factor remains valid at large drive because, in the absence of nonlinear damping, the peak amplitude of a Duffing resonator equals that of the linear resonator at the same force, up to the factor $\Omega_{m,0}/\Omega_\mathrm{m,peak}$, which differs from unity by less than $3\times10^{-4}$ here. At the peak, the drive is in quadrature with the displacement, as for a linear resonator at resonance, and the conservative cubic force does no net work over a cycle, so the amplitude is still set by the balance between the work done by the drive and the linear damping. Nonlinear damping, if present, would instead lower the peak amplitude below this linear extrapolation at large drive. With the normalization chosen for the definition of $m_\mathrm{eff}$, these two measurements determine $X_\mathrm{osc}$ as the oscillation amplitude at the centre of the trampoline. From the frequency sweeps in Fig.~\ref{fig:Duffing_response}(a), we also extract the frequency at which the amplitude attains its maximum ($\Omega_\mathrm{m,peak}$) and plot it against the corresponding calibrated amplitude ($X_\mathrm{peak}$), yielding the so-called \emph{backbone curve}~\cite{lifshitz_nonlinear_2008}, which is shown in Fig.~\ref{fig:Duffing_response}(c). We note that all sweeps are performed from low to high frequencies and that the jump frequency and the peak frequency coincide up to the acquisition resolution. This is a result of the small damping of the mechanical mode ($\Gamma_m \ll \Omega_{m,0}$). In the weakly nonlinear regime, the backbone follows the analytical expression~\cite{lifshitz_nonlinear_2008}
\begin{equation}
    \Omega_\mathrm{m,peak} = \Omega_{m,0} + \frac{3\alpha}{8\Omega_{m,0}}\,X_\mathrm{peak}^2,
    \label{eq:backbone}
\end{equation}
Fitting Eq.~\eqref{eq:backbone} to the data yields the Duffing coefficient $\alpha = (1.387\pm0.003)\times 10^{24}$~s$^{-2}$~m$^{-2}$, where the uncertainty is the standard error of the fit and does not include that of the displacement calibration. For comparison, treating the pad as a rigid mass suspended by four straight tethers of length $L_\text{t}\approx 55$\,\si{\micro\meter}, width $w_\text{t} = 8$\,\si{\micro\meter} and thickness $h = 100$\,nm, whose stretching under a pad displacement $x$ adds a restoring force $2Ew_\text{t}h\,x^3/L_\text{t}^3$~\cite{lifshitz_nonlinear_2008,schmid_fundamentals_2023}, gives $\alpha\approx 2Ew_\text{t}h/(m_\mathrm{eff}L_\text{t}^3)\approx 3.3\times10^{24}$~s$^{-2}$~m$^{-2}$, with $E = 250$\,GPa and the simulated $m_\mathrm{eff}$ of Section~\ref{subsec:broadband_mech}. The measured value lies a factor of about 2.4 below this estimate, which neglects the compliance of the clamping regions and relies on the simulated $m_\mathrm{eff}$, which also enters the displacement calibration.

\subsection{Mechanical characterization at varying chamber pressure}

\begin{figure}[!htbp]
    \centering
    \includegraphics[width=\textwidth]{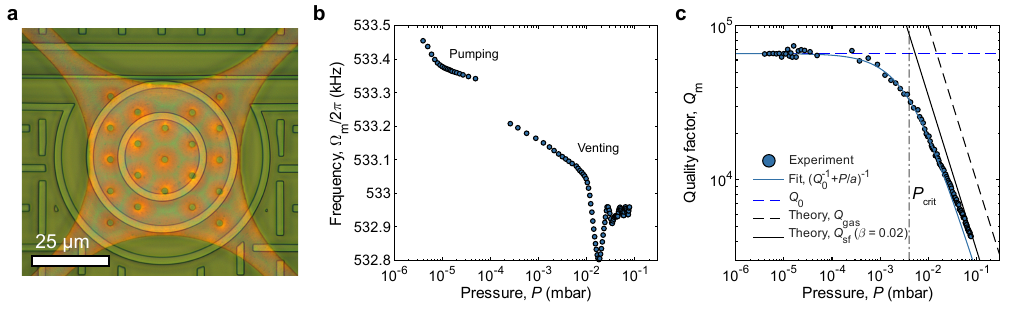}
    \caption{\textbf{Pressure dependence of the mechanical resonance.} \textbf{a}, Optical micrograph of the trampoline resonator pad for the resonator used for this specific study. \textbf{b}, Mechanical resonance frequency $\Omega_\mathrm{m}/2\pi$ and \textbf{c} quality factor $Q_\mathrm{m}$ versus chamber pressure $P$, recorded in two datasets, once for decreasing pressure (pumping down) and once for increasing pressure (venting). The dashed line in \textbf{c} marks the pressure-independent value $Q_0$, the black lines the free-molecular-flow prediction $Q_\mathrm{gas}$ (Eq.~(\ref{eq:Qgas})) and its squeeze-film correction $Q_\mathrm{sf}$ (Eq.~(\ref{eq:Qsz}), $\beta = 0.02$), and the blue line a fit to $Q_\mathrm{m}=(Q_0^{-1}+P/a)^{-1}$. The critical pressure $P_\mathrm{crit}$ is defined as the pressure at which the gas damping equals the intrinsic loss ($Q=Q_0/2$).}
    \label{fig:pressure}
\end{figure}

To establish the pressure range over which the detector operates without degradation, we record thermomechanical spectra of the fundamental trampoline mode while pumping down and venting the vacuum chamber, extracting the resonance frequency $\Omega_\mathrm{m}/2\pi$ and the quality factor $Q_\mathrm{m}$ as functions of pressure $P$. These measurements are performed on a representative resonator of nominally identical geometry but without the platinum disk. Since the gas damping is governed by the membrane geometry rather than the absorber, this does not affect the conclusions herein.

As shown in Fig.~\ref{fig:pressure}(b), the resonance frequency varies by about $0.1\%$ across the entire range, decreasing slightly with increasing pressure. This shift is consistent with mass loading by adsorbed gas, $\Delta\Omega_\mathrm{m}/\Omega_\mathrm{m,0} = -\Delta m/2m_\mathrm{eff}$, whose coverage grows with pressure. The quality factor (Fig.~\ref{fig:pressure}(c)) displays a far stronger dependence: below a critical pressure $P_\mathrm{crit}$ it saturates to a pressure-independent value $Q_0$, while above $P_\mathrm{crit}$ it decreases as $1/P$, as evidenced by the good agreement with the model $Q_\mathrm{m} = (Q_0^{-1} + P/a)^{-1}$ over the full range (Fig.~\ref{fig:pressure}(c), blue line). The low-pressure plateau reflects the intrinsic (clamping and material) losses, whereas the $1/P$ roll-off is the signature of gas damping in the free-molecular-flow regime. There, an isolated vibrating membrane exhibits a gas-limited quality factor~\cite{bao_energy_2002}
\begin{equation}
\label{eq:Qgas}
Q_\mathrm{gas} = \sqrt{\frac{\pi}{2}}\,\sqrt{\frac{RT}{M_\mathrm{m}}}\,\frac{\rho\, h\, \Omega_\mathrm{m}}{4P},
\end{equation}
where $R$ is the gas constant, $T$ the temperature, $M_\mathrm{m}$ the molar mass of the gas, and $\rho$ and $h$ the mass density and thickness of the membrane. When the resonator sits close to the substrate, molecules trapped in the gap undergo repeated collisions (squeeze-film damping), lowering the quality factor to~\cite{bao_energy_2002}
\begin{equation}
\label{eq:Qsz}
Q_\mathrm{sf} \simeq \frac{Q_\mathrm{gas}}{1 + \beta\, L/d},
\end{equation}
with $L$ a characteristic in-plane size, $d$ the gap to the substrate, and $\beta$ a geometry-dependent factor ($\beta = 0.02$ for a square membrane). For our resonator, we use the pad size $L = L_\text{p}\approx 42$~\si{\micro\meter}, $d = 465$~\si{\nano\meter}, $h = 100$~\si{\nano\meter} (Methods of the main text) and $T = 300$~\si{\kelvin}. Both the free-molecular-flow prediction and the squeeze-film correction are shown for reference (black lines), with the latter within 35\% of the fitted slope. We define the critical pressure $P_\mathrm{crit}$ as the point at which the gas damping equals the intrinsic loss, i.e.\ $Q(P_\mathrm{crit}) = Q_0/2$, which for this resonator ($Q_0\approx 6.6\times10^4$, $a\approx 260$~mbar) gives $P_\mathrm{crit}\approx 4\times10^{-3}$~mbar. Scaling to the device of the main text ($Q\approx 2\times10^5$) lowers $P_\mathrm{crit}$ to about $1\times10^{-3}$~mbar, still three orders of magnitude above the operating pressure. Since the detector operates at $P < 10^{-6}$~mbar, the mechanical quality factor---and hence the frequency stability---remains at its pressure-independent value. The vacuum requirement is therefore modest and compatible with wafer-level encapsulation with integrated getters, as discussed in the main text.

\section{Mechanical frequency noise characterization}
\label{sec:frequencyNoise}

\subsection{Theoretical model for the measured frequency noise power spectral density}

We derive Eq.~(1) of the main text in three steps: we describe how the lock-in amplifier extracts the quadratures of the detected signal, compute the noise in the measured phase of a driven oscillator read out linearly in the presence of thermomechanical and imprecision noise, and convert this phase noise into frequency noise.

The lock-in amplifier extracts the quadratures $X_1(t)$ and $X_2(t)$ of the detected signal $x(t)$, defined such that
\begin{equation}
    x(t) = X_1(t)\cos(\Omega_\mathrm{LO}t) + X_2(t)\sin(\Omega_\mathrm{LO}t)
\end{equation}
The quadratures are then given by
\begin{align}
    X_1(t) &= \mathrm{LP}_B\{2x(t)\cos(\Omega_\mathrm{LO}t)\}\\
    X_2(t) &= \mathrm{LP}_B\{2x(t)\sin(\Omega_\mathrm{LO}t)\}
\end{align}
or, in the frequency domain,
\begin{align}
    X_1(\omega) &= H_\mathrm{LP}(\omega)\left[ x(\omega - \Omega_\mathrm{LO}) + x(\omega + \Omega_\mathrm{LO})\right]\\
    X_2(\omega) &= iH_\mathrm{LP}(\omega)\left[ x(\omega - \Omega_\mathrm{LO}) - x(\omega + \Omega_\mathrm{LO})\right]
\end{align}
where $H_\mathrm{LP}(\omega)$ is the response function of the low-pass filter. The Zurich Instruments lock-in amplifier used here instead defines the quadratures through
\begin{equation}
    x(t) = \sqrt{2}(X_1(t)\cos(\Omega_\mathrm{LO}t) + X_2(t)\sin(\Omega_\mathrm{LO}t))
\end{equation}

We now apply this to an oscillator of frequency $\Omega$, driven to an amplitude $x_0$ and read out linearly, whose displacement reads
\begin{equation}\label{eq:osc_model}
    x(t) = x_0\cos(\Omega t + \theta(t)) + x_\mathrm{th}(t)+ x_\mathrm{imp}(t)
\end{equation}
where $x_\mathrm{th}$ is the thermomechanical term given by
\begin{equation}
    x_\mathrm{th}(\omega) = \chi(\omega)F_\mathrm{th}(\omega)
\end{equation}
with the mechanical susceptibility defined as
\begin{equation}
    \chi(\omega) = \frac{1/m}{\Omega^2-\omega^2-i\omega\Gamma}
\end{equation}
and the Langevin force defined with the correlation functions
\begin{align}
    &\avg{F_\mathrm{th}(t')F_\mathrm{th}(t)} = 2k_BTm\Gamma\delta(t'-t),\\
    &\avg{F_\mathrm{th}(\omega')^*F_\mathrm{th}(\omega)} = 2\pi\times 2k_BTm\Gamma\delta(\omega'-\omega).
\end{align}
The imprecision noise $x_\mathrm{imp}$, expressed in units of displacement, is a white-noise process (for example shot noise, but we keep it general here), with
\begin{align}
    &\avg{x_\mathrm{imp}(t')x_\mathrm{imp}(t)} = S_{xx}^\mathrm{imp}\delta(t'-t),\\
    &\avg{x_\mathrm{imp}(\omega')^*x_\mathrm{imp}(\omega)} = 2\pi\times S_{xx}^\mathrm{imp}\delta(\omega'-\omega).
\end{align}
Using the definition of the quadratures, we now derive their expressions for the model of Eq.~\eqref{eq:osc_model}, assuming $\Omega_\mathrm{LO} = \Omega$ and $|\theta(t)|\ll 1$. For the first term of Eq.~\eqref{eq:osc_model},
\begin{equation}
    x(t) = x_0\cos\left(\Omega t + \theta (t)\right) = \frac{x_0}{2}\left[e^{i\Omega t + i\theta(t)} + e^{-i\Omega t -i \theta(t)}\right] \approx \frac{x_0}{2}\left[e^{i\Omega t}(1+i\theta(t)) + e^{-i\Omega t}(1-i\theta(t))\right]
\end{equation}
In the Fourier domain,
\begin{equation}
    x(\omega) = \frac{x_0}{2}\left[ 2\pi\delta(\omega+\Omega) + i\theta(\omega+\Omega) + 2\pi\delta(\omega-\Omega)-i\theta(\omega-\Omega) \right]
\end{equation}
and the quadratures follow as
\begin{align}
    X_1(\omega) &= 2\pi x_0 H_\mathrm{LP}(\omega)\delta(\omega)\\
    X_2(\omega) &= -x_0H_\mathrm{LP}(\omega)\theta(\omega)
\end{align}
Including the thermomechanical and imprecision contributions,
\begin{align}
    X_1(t) &=  x_0 H_\mathrm{LP}(0) + X_1^\mathrm{th}(t) + X_1^\mathrm{imp}(t)\\
    X_2(t) &= -x_0\,\mathrm{LP}_B\{\theta(t)\} + X_2^\mathrm{th}(t) + X_2^\mathrm{imp}(t)
\end{align}
The phase is retrieved as
\begin{equation}
    \thmeas(t) = -\frac{X_2(t)}{x_0}
\end{equation}
which gives
\begin{equation}
    \thmeas(t) = \mathrm{LP}_B\{\theta(t)\} - \frac{X_2^\mathrm{th}(t)}{x_0}- \frac{X_2^\mathrm{imp}(t)}{x_0}
\end{equation}
Since the three noise contributions are uncorrelated,
\begin{equation}
    \Sthmeas(\omega) = |H_\mathrm{LP}(\omega)|^2S_{\theta\theta}(\omega) + \frac{1}{x_0^2}S_{X_2^\mathrm{th}X_2^\mathrm{th}}(\omega)+ \frac{1}{x_0^2}S_{X_2^\mathrm{imp}X_2^\mathrm{imp}}(\omega)
\end{equation}
We now evaluate the spectra of $X_2^\mathrm{th}$ and $X_2^\mathrm{imp}$. For the thermomechanical term,
\begin{align}
    X_2^\mathrm{th}(\omega) &= iH_\mathrm{LP}(\omega)\left[x_\mathrm{th}(\omega-\Omega) - x_\mathrm{th}(\omega+\Omega) \right]\\
    &= iH_\mathrm{LP}(\omega)\left[ \chi(\omega-\Omega)F_\mathrm{th}(\omega-\Omega) - \chi(\omega+\Omega)F_\mathrm{th}(\omega+\Omega) \right]
\end{align}
\begin{multline}
    \avg{X_2^\mathrm{th}(\omega')^*X_2^\mathrm{th}(\omega)} = H_\mathrm{LP}(\omega')^*H_\mathrm{LP}(\omega)[\chi(\omega'-\Omega)^*\chi(\omega-\Omega)\avg{F_\mathrm{th}(\omega'-\Omega)^*F_\mathrm{th}(\omega-\Omega)}\\+\chi(\omega'+\Omega)^*\chi(\omega+\Omega)\avg{F_\mathrm{th}(\omega'+\Omega)^*F_\mathrm{th}(\omega+\Omega)}\\
    -\chi(\omega'-\Omega)^*\chi(\omega+\Omega) \avg{F_\mathrm{th}(\omega'-\Omega)^*F_\mathrm{th}(\omega+\Omega)}\\
    -\chi(\omega'+\Omega)^*\chi(\omega-\Omega) \avg{F_\mathrm{th}(\omega'+\Omega)^*F_\mathrm{th}(\omega-\Omega)}]\\
    = 2\pi(2k_BTm\Gamma)H_\mathrm{LP}(\omega')^*H_\mathrm{LP}(\omega)[\chi(\omega'-\Omega)^*\chi(\omega-\Omega)\delta(\omega'-\omega)\\
    +\chi(\omega'+\Omega)^*\chi(\omega+\Omega)\delta(\omega'-\omega)\\
    -\chi(\omega'-\Omega)^*\chi(\omega+\Omega)\delta(\omega'-\omega-2\Omega)\\
    -\chi(\omega'+\Omega)^*\chi(\omega-\Omega)\delta(\omega'-\omega+2\Omega)]
\end{multline}
from which the spectrum follows as
\begin{multline}
    S_{X_2^\mathrm{th}X_2^\mathrm{th}}(\omega) = \frac{1}{2\pi}\int d\omega' \avg{X_2^\mathrm{th}(\omega')^*X_2^\mathrm{th}(\omega)}=\\2k_BTm\Gamma[|H_\mathrm{LP}(\omega)|^2|\chi(\omega-\Omega)|^2 + |H_\mathrm{LP}(\omega)|^2|\chi(\omega+\Omega)|^2\\
    -H_\mathrm{LP}(\omega+2\Omega)^*H_\mathrm{LP}(\omega)|\chi(\omega+\Omega)|^2
    -H_\mathrm{LP}(\omega-2\Omega)^*H_\mathrm{LP}(\omega)|\chi(\omega-\Omega)|^2]
\end{multline}
For a low-pass filter much narrower than $\Omega$, the last two terms are negligible, and
\begin{equation}
    S_{X_2^\mathrm{th}X_2^\mathrm{th}}(\omega) = |H_\mathrm{LP}(\omega)|^2\left[|\chi(\omega-\Omega)|^2+|\chi(\omega+\Omega)|^2\right]2k_BTm\Gamma
\end{equation}
Similarly, for the imprecision noise,
\begin{equation}
    S_{X_2^\mathrm{imp}X_2^\mathrm{imp}}(\omega) = |H_\mathrm{LP}(\omega)|^2 2S_{xx}^\mathrm{imp}
\end{equation}
In summary,
\begin{equation}
    \Sthmeas(\omega) = |H_\mathrm{LP}(\omega)|^2\left[S_{\theta\theta}(\omega) +  \frac{|\chi(\omega-\Omega)|^2+|\chi(\omega+\Omega)|^2}{2}\frac{2k_BTm\Gamma}{X_\mathrm{osc}^2} + \frac{S_{xx}^\mathrm{imp}}{X_\mathrm{osc}^2} \right]
\end{equation}
where we have used $X_\mathrm{osc}^2 = x_0^2/2$, the RMS value of the oscillator's amplitude. For a high-$Q$ oscillator and sideband frequencies $|\omega|\ll \Omega$,
\begin{equation}
    \Sthmeas(\omega) = |H_\mathrm{LP}(\omega)|^2\left[S_{\theta\theta}(\omega) + \frac{X_\mathrm{th}^2}{X_\mathrm{osc}^2} \frac{\Gamma/2}{\omega^2+\Gamma^2/4} + \frac{S_{xx}^\mathrm{imp}}{X_\mathrm{osc}^2}\right]
\end{equation}
with $X_\mathrm{th}^2 = k_BT/m\Omega^2$ and the low-pass response
\begin{equation}
    |H_\mathrm{LP}(\omega)|^2 = \frac{B^2/4}{\omega^2+B^2/4}
\end{equation}
Finally, we convert this phase noise into frequency noise. In the experiment, the PLL keeps the drive in phase with the resonator response, so that the oscillation frequency follows the resonance frequency within the loop bandwidth ($\Gamma_\text{PLL}/2\pi = 10$\,kHz). The frequency estimate reported in the main text is the time derivative of the phase measured by the out-of-loop demodulator, after removal of the constant 3\,kHz offset of its reference. Within the loop bandwidth, such closed-loop tracking has the same frequency noise as the open-loop conversion of the resonator phase through the inverse of its baseband response, $H_{BB}(\omega) = 1/(-i\omega + \Gamma/2)$, which accounts for the time $\sim\Gamma^{-1}$ that the phase of a driven resonator takes to follow a change of its resonance frequency~\cite{besic_schemes_2023}. We therefore write
\begin{equation}
    \Ogmeas(\omega) = H_{BB}(\omega)^{-1} \thmeas(\omega)
\end{equation}
With this, the double-sided frequency noise PSD is given by
\begin{equation}
    \label{eq:S_OmegaOmega_ds_theory}
    \SOgmeas(\omega) = |H_\mathrm{LP}(\omega)|^2\left[S_{\Omega\Omega}(\omega) +  \frac{X_\mathrm{th}^2}{X_\mathrm{osc}^2} \frac{\Gamma}{2} + (\omega^2 + \Gamma^2/4)\frac{S_{xx}^\mathrm{imp}}{X_\mathrm{osc}^2} \right]
\end{equation}
and the single-sided frequency noise PSD is given by
\begin{equation}
    S_{\Ogmeas}(\omega) = |H_\mathrm{LP}(\omega)|^2\left[S_{\Omega}(\omega) +  \frac{X_\mathrm{th}^2}{X_\mathrm{osc}^2} \Gamma + (\omega^2 + \Gamma^2/4)\frac{S_{x}^\mathrm{imp}}{X_\mathrm{osc}^2} \right]
\end{equation}
where $S_{x}^\mathrm{imp}$ is the single-sided imprecision noise PSD in the measurement of $x$, i.e., displacement.

\subsection{Characterization of excess low-frequency noise due to power fluctuations}
\label{subsec:power_fluctuations}

As discussed in Subsection~\ref{subsec:kappa_abs}, the Pt absorber we employ leads to unwanted optical absorption at the detection wavelength used in our experiments ($\lambda \approx 1550$ nm), which leads to a steady-state frequency shift proportional to the intra-cavity power. While this is not per se problematic, the same residual absorption also transduces any intra-cavity power fluctuations into mechanical frequency fluctuations. In particular, intensity noise of the readout light can manifest as excess frequency fluctuations that scale with optical power. This has been shown to be an important source of noise in previous works using optically-detected frequency shifts~\cite{kanellopulos_comparative_2025} and most recently an opening in the absorber has been employed to suppress such deleterious effects~\cite{martini_uncooled_2025}. In Fig.~2(f) and (g) of the main text, beyond a certain optical power, the frequency-noise PSD at sideband frequencies $\omega \le \omega_\text{th}$ is dominated by an excess contribution that scales as $P_\text{det}^2$ and rolls off around the thermal cutoff $\omega_\text{th}$. This is what one expects if the excess frequency fluctuations are driven by the absorbed readout power and filtered by the thermal response of the resonator, $H_\mathrm{th}(\omega) = [1+i\omega/\omega_\mathrm{th}]^{-1}$.

\begin{figure}[t]
    \centering
    \includegraphics[width=\textwidth]{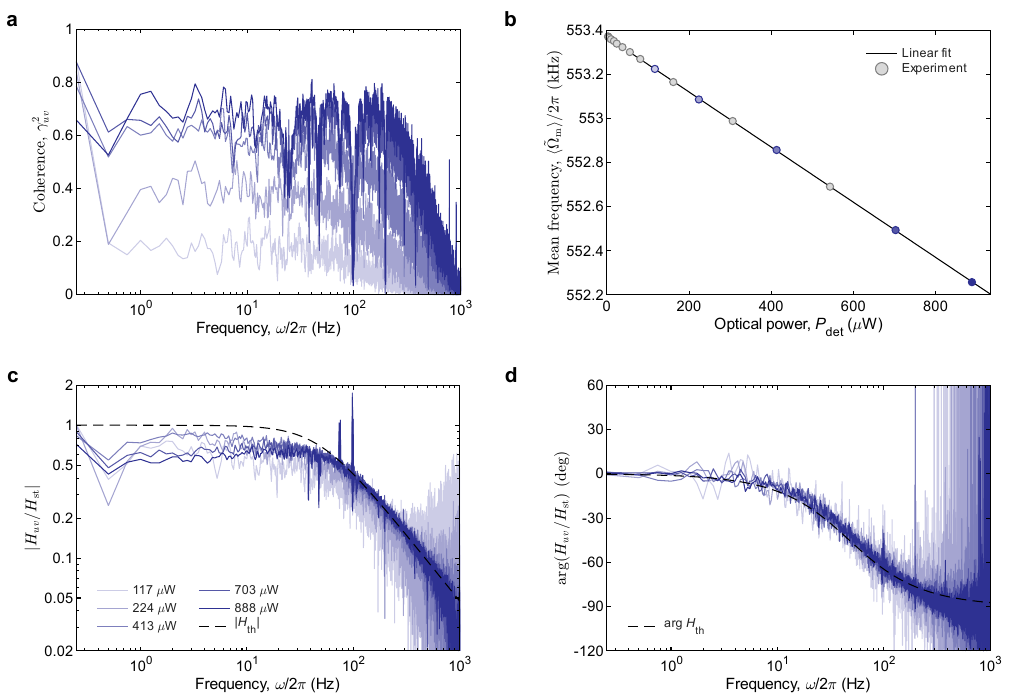}
    \caption{\textbf{Coherence between amplitude and frequency fluctuations.} Power sweep of main text Fig.~2(f,g) ($V_\mathrm{pzt} = 30$\,mV, $\Delta\approx-|\Delta_*|$), for $P_\text{det}$ between 117 and 888\,\si{\micro\watt} (light to dark), computed with 4\,s segments. \textbf{a}, Coherence between the relative fluctuations of the detected amplitude, $\delta R/R$, and the frequency fluctuations, $\delta\Omeas$. \textbf{b}, Mean frequency versus $P_\text{det}$ for all powers of the sweep (coloured: those shown in the other panels), with a linear fit for $P_\text{det} > 50$\,\si{\micro\watt}, whose slope defines the static photothermal shift $H_\mathrm{st}$. \textbf{c}, Magnitude of the transfer function $H_{uv}$ from $\delta R/R$ to $\delta\Omeas$, normalized to $H_\mathrm{st}$, and magnitude of the thermal response $H_\mathrm{th}$ for $\omega_\mathrm{th}/2\pi = 47.4$\,Hz (dashed), measured with mid-infrared illumination. \textbf{d}, Phase of $H_{uv}/H_\mathrm{st}$ and of $H_\mathrm{th}$ (dashed).}
    \label{fig:photothermalCoherence}
\end{figure}

We test this attribution directly. Fluctuations of the power reaching the cavity also modulate the magnitude $R$ of the demodulated signal, so that, for a large enough coherent signal, its relative fluctuations $\delta R/R$ track the fluctuations of the intracavity power that, through the residual absorption, would drive the frequency. For the power sweep of main text Fig.~2(f, g), recorded at $V_\mathrm{pzt} = 30$\,mV and $\Delta\approx-|\Delta_*|$, we compute the coherence, $\gamma_{uv}^2(\omega)$, and the transfer function, $H_{uv}(\omega)$, between $u = \delta R/R$ and the frequency fluctuations $v = \delta\Omeas$. In terms of their PSDs, $S_{uu}(\omega)$ and $S_{vv}(\omega)$, and their cross-spectral density, $S_{uv}(\omega) = \langle\tilde u^*(\omega)\,\tilde v(\omega)\rangle$, with $\tilde u$ and $\tilde v$ their Fourier transforms, these are defined as
\begin{equation}
    \gamma_{uv}^2(\omega) = \frac{|S_{uv}(\omega)|^2}{S_{uu}(\omega)\,S_{vv}(\omega)}, \qquad H_{uv}(\omega) = \frac{S_{uv}(\omega)}{S_{uu}(\omega)}.
    \label{eq:coherence_transfer}
\end{equation}
The coherence, bounded between 0 and 1, is the fraction of the frequency-noise PSD that is linearly related to $\delta R/R$, and $H_{uv}$ is the complex gain of that linear relation, $v = H_{uv}u$. The two are connected through $\gamma_{uv}^2 S_{vv} = |H_{uv}|^2 S_{uu}$, i.e., the part of the frequency noise coherent with $\delta R/R$ is the amplitude noise transferred through $H_{uv}$. The coherence is negligible at low power, 0.07 at the power of the NEP measurements ($P_\text{det} = 56$\,\si{\micro\watt}), and rises above $P_\text{max}$ to about 0.7 at $P_\text{det} = 888$\,\si{\micro\watt} (Fig.~\ref{fig:photothermalCoherence}(a)). We normalize $H_{uv}$ to the static photothermal shift, $H_\mathrm{st} = (\partial\langle\Omeas\rangle/\partial P_\text{det})\,P_\text{det}$, that is, the frequency change produced by a quasi-static relative change of the power. The slope $\partial\langle\Omeas\rangle/\partial P_\text{det} = -2\pi\times1.25$\,Hz\,\si{\micro\watt}$^{-1}$ is obtained from a linear fit of the measured mean frequency versus $P_\text{det}$ for $P_\text{det} > 50$\,\si{\micro\watt} (Fig.~\ref{fig:photothermalCoherence}(b)), and agrees, within its uncertainty, with the static shift expected at this detuning from the detuning scans of Section~\ref{subsec:kappa_abs}, $-2\pi\times(1.4\pm0.1)$\,Hz\,\si{\micro\watt}$^{-1}$. If the frequency follows the absorbed power through the thermal response and $\delta R/R$ equals the relative power fluctuations, $H_{uv}/H_\mathrm{st}$ reduces to $H_\mathrm{th}(\omega)$. The normalized transfer functions indeed follow $H_\mathrm{th}(\omega)$ with the cutoff $\omega_\mathrm{th}/2\pi = 47.4$\,Hz measured with mid-infrared illumination (main text, dashed lines), without any adjustable parameter, both in magnitude up to several hundred hertz (Fig.~\ref{fig:photothermalCoherence}(c)) and in phase (Fig.~\ref{fig:photothermalCoherence}(d)). Fitting the measured phase alone with a first-order low-pass response gives a cutoff of 52\,Hz, within about 10\% of the mid-infrared value. This difference is not explained by where the readout light is absorbed: finite-element simulations of heat transport (Section~\ref{subsec:thermal_bandwidth}) give the same thermal cutoff for a heat source concentrated at the edge of the Pt disk as for the mid-infrared spot, to within 0.5\%. Towards 1\,kHz, the coherence vanishes as detection noise dominates the frequency fluctuations. The low-frequency magnitude of $H_{uv}/H_\mathrm{st}$, 0.55--0.85, is a lower bound, because fluctuations of $R$ that do not change the absorbed power, such as detection noise and fluctuations of the drive, enter $S_{uu}$ but not $S_{uv}$ and therefore bias $|H_{uv}|$ low, by a factor between $\gamma_{uv}^2$ and 1. Conversely, the large coherence and $|H_{uv}/H_\mathrm{st}|$ close to 1 at intermediate powers show that the power fluctuations that drive the photothermal noise arise before the cavity, since fluctuations added after it, for instance at the output coupling or in the detection, would modulate $R$ without heating the resonator and reduce both. At high power, the excess low-frequency noise is therefore caused by fluctuations of the absorbed readout power, transduced through the same thermal response as the mid-infrared signal. Removing this noise would allow operation at higher optical power, with lower detection imprecision, while remaining limited by the intrinsic frequency noise at low frequencies. However, since the coherence remains limited, a correction of the measured frequency from $\delta R/R$, analogous to the amplitude-based correction demonstrated for the Duffing nonlinearity~\cite{brown_high_2025}, removes only part of it. Any linear correction can at best reduce the frequency-noise PSD to $(1-\gamma_{uv}^2)\,S_{vv}$, and we find reductions of the low-frequency noise by at most a factor of about 3, at $P_\text{det} = 888$\,\si{\micro\watt}.

To separate this photothermal backaction noise from the intrinsic resonator frequency noise, we model the real frequency fluctuations in Eq.~(\ref{eq:S_OmegaOmega_ds_theory}) as
\begin{equation}
\label{eq:SOmega_model}
S_{\Omega_\mathrm{m}\Omega_\mathrm{m}}(\omega) = S^\mathrm{int}_{\Omega_\mathrm{m}\Omega_\mathrm{m}}(\omega) + |H_\mathrm{th}(\omega)|^2\left(\Omega_\mathrm{m,0}\,\mathcal{R}_0^\mathrm{abs}\right)^2 S_{P_\mathrm{abs}P_\mathrm{abs}}(\omega),
\end{equation}
where $S^\mathrm{int}_{\Omega_\mathrm{m}\Omega_\mathrm{m}}(\omega)$ represents the power- and amplitude-independent frequency noise contribution we observe for low-enough optical power and piezo-drive amplitude, and the second term, which is governed by the fluctuations $S_{P_\mathrm{abs}P_\mathrm{abs}}(\omega)$, is the photothermal backaction noise. Here $\mathcal{R}_0^\mathrm{abs}$ is the fractional responsivity to absorbed power and $H_\mathrm{th}(\omega)$ is the thermal transfer function, both taken equal to those of the mid-infrared signal, as supported by Fig.~\ref{fig:photothermalCoherence}.

To identify the origin of the absorbed-power fluctuations, which can arise from intensity fluctuations of the light reaching the chip or from its frequency fluctuations converted into intensity fluctuations by the cavity, we write the input field as $s_\mathrm{in}(t) = s_0[1+\epsilon(t)]e^{i\phi(t)} \approx s_0[1+\epsilon(t)+i\phi(t)]$, with $\epsilon(t)$ and $\phi(t)$ classical amplitude and phase, and use coupled-mode theory for the resonator to obtain
\begin{equation}
\label{eq:SPP}
S_{P_\mathrm{abs}P_\mathrm{abs}}(\omega) = \left(\hbar\omega_c\kappa_\mathrm{abs}\right)^2\left[\,4\,n_c^2(\Delta)\,S_{\epsilon\epsilon}(\omega) + \left(\frac{\mathrm{d}n_c}{\mathrm{d}\Delta}\right)^2\!\omega^2 S_{\phi\phi}(\omega) + 4\,n_c(\Delta)\frac{\mathrm{d}n_c}{\mathrm{d}\Delta}\,\omega\,\mathrm{Im}\,S_{\epsilon\phi}(\omega)\,\right],
\end{equation}
which is valid in the adiabatic regime ($\omega\ll\kappa$). Here $S_{\epsilon\epsilon}(\omega)$ is the amplitude noise spectrum, $S_{\phi\phi}(\omega)$ the phase noise spectrum, and $S_{\epsilon\phi}(\omega)$ their cross spectrum. The first term is direct amplitude modulation (AM); the second is laser frequency noise ($\omega^2 S_{\phi\phi}(\omega)$) converted into intensity by the cavity slope $\mathrm{d}n_c/\mathrm{d}\Delta$ (phase-to-amplitude modulation, PAM); the third is their correlation. Crucially, these channels carry distinct detuning signatures, which we exploit in a detuning sweep at $P_\text{det}$ = 200\,\si{\micro\watt}. This sweep is recorded at a small drive, $V_\mathrm{pzt} = 2$\,mV, for which the detected amplitude is dominated by a detuning-independent relative noise of about $6\times10^{-6}$\,Hz$^{-1}$ at 0.5--5\,Hz, incoherent with the frequency fluctuations, so that the coherence analysis above cannot be applied. We therefore analyse the frequency-noise spectra directly.

\begin{figure}[!htbp]
    \centering
    \includegraphics[width=\textwidth]{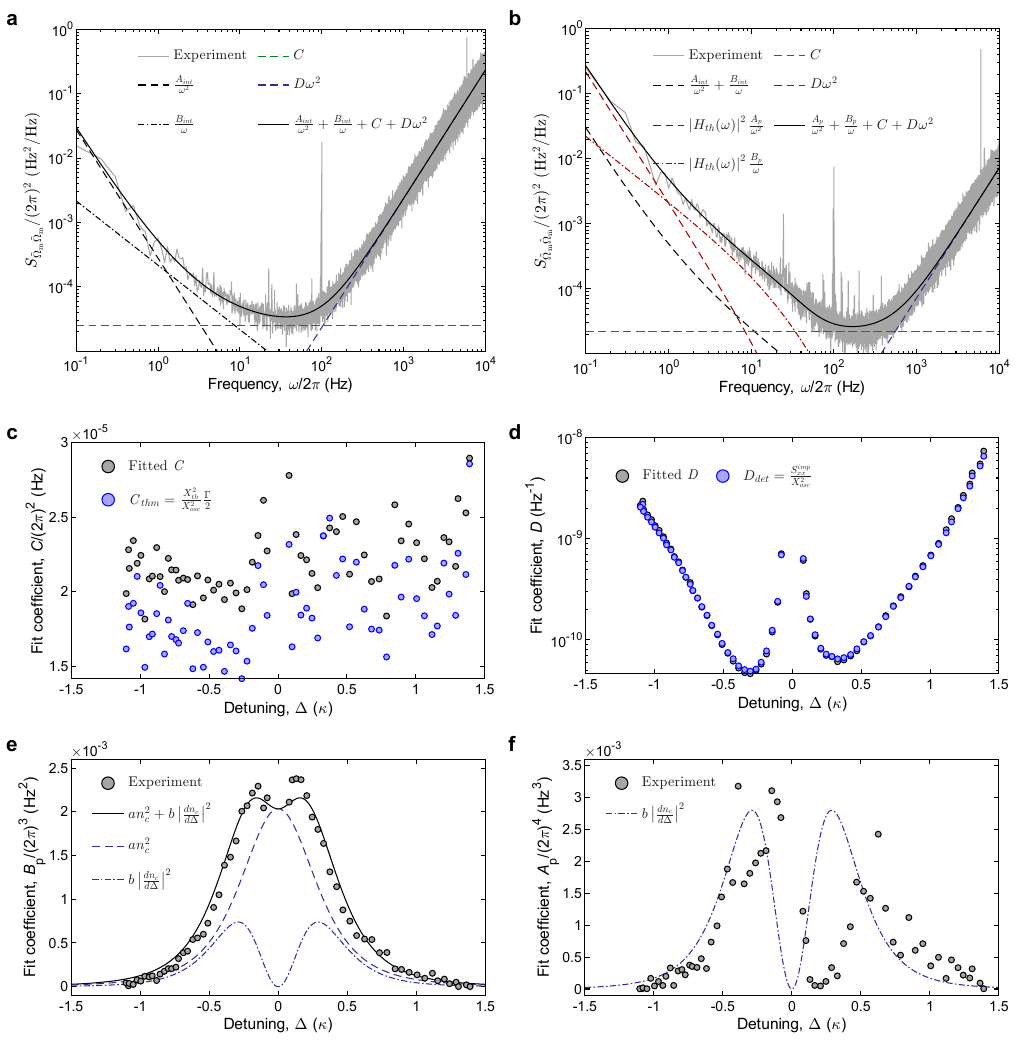}
    \caption{\textbf{Detuning dependence of the frequency noise.} \textbf{a}, Frequency-noise PSD (grey) at a detuning where the photothermal contribution is negligible, used to extract the intrinsic noise. The fit (black), $A_\mathrm{int}/\omega^2+B_\mathrm{int}/\omega+C+D\omega^2$, is decomposed into its terms (dashed). \textbf{b}, Representative PSD when the photothermal terms $A_p/\omega^2$, $B_p/\omega$ (red), filtered by the thermal response $|H_\mathrm{th}(\omega)|^2$, are also included in the fit ($\Delta = -0.2\kappa$). \textbf{c,d}, Fitted $C$ and $D$ versus detuning $\Delta$ (grey dots) and their expected values (blue dots), $C_\mathrm{thm}=(X_\mathrm{th}^2/X_\mathrm{osc}^2)\Gamma/2$ and $D_\mathrm{det}=S_{xx}^\mathrm{imp}/X_\mathrm{osc}^2$. \textbf{e}, Fitted $B_p$ versus $\Delta$, fitted to $a\,n_c^2+b\,|\mathrm{d}n_c/\mathrm{d}\Delta|^2$. \textbf{f}, Fitted $A_p$ versus $\Delta$, with $|\mathrm{d}n_c/\mathrm{d}\Delta|^2$ as a guide to the eye.}
    \label{fig:excessNoise}
\end{figure}

At this power, the photothermal term in Eq.~(\ref{eq:SOmega_model}) is appreciable, and we fit each frequency-noise PSD measured at varying laser--cavity detuning $\Delta$ to the phenomenological model
\begin{equation}
\label{eq:excessFit}
\SOmeas(\omega) = \frac{A_\mathrm{int}}{\omega^2} + \frac{B_\mathrm{int}}{\omega} + |H_\mathrm{th}(\omega)|^2\left(\frac{A_p}{\omega^2} + \frac{B_p}{\omega}\right) + C + D\omega^2 .
\end{equation}
The first two terms reproduce the intrinsic, power-independent frequency noise $S^\mathrm{int}_{\Omega_\mathrm{m}\Omega_\mathrm{m}}$ of Eq.~(\ref{eq:SOmega_model}) and are left unfiltered, while the photothermal terms $A_p/\omega^2 + B_p/\omega$ (the transduction of $S_{P_\mathrm{abs}P_\mathrm{abs}}$) are filtered by the thermal response $|H_\mathrm{th}(\omega)|^2 = [1+(\omega/\omega_\mathrm{th})^2]^{-1}$. The choice of these particular power-law terms is empirical, motivated by the shape of the measured spectra. The other two terms correspond respectively to the thermomechanical white noise ($C$) and detection noise ($D\omega^2$). We first extract the intrinsic coefficients $A_\mathrm{int}$, $B_\mathrm{int}$ from the spectrum of the dataset measured at the largest negative detuning, where the photothermal contribution is negligible. The fit and the different contributions are represented in Fig.~\ref{fig:excessNoise}(a). For the rest of the spectra, we hold $A_\mathrm{int}$, $B_\mathrm{int}$ and $\omega_\text{th}$ fixed and fit the remaining terms ($A_p$, $B_p$, $C$ and $D$). A representative fit, including the filtered photothermal contributions, is shown in Fig.~\ref{fig:excessNoise}(b). We see that for $\omega\le\omega_\text{th}$ the noise is dominated by the photothermal backaction noise, with the $B_p/\omega$ term dominating over the $A_p/\omega^2$ term at all frequencies above approximately 1 Hz. We note that all fits in the dataset quantitatively agree with the experimental data over 5 orders of magnitude in sideband frequency, with the exception of a series of narrow resonant peaks that are excluded from the fitted data to prevent biasing the fit to the broadband response.

Equation~(\ref{eq:S_OmegaOmega_ds_theory}) predicts $C_\mathrm{thm} = (X_\mathrm{th}^2/X_\mathrm{osc}^2)\Gamma/2$ and $D_\mathrm{det} = S_{xx}^\mathrm{imp}/X_\mathrm{osc}^2$, which we use as initial values for the fits. The fitted coefficients $C$ and $D$ agree with these predictions, to within about 20\% for $C$ (Fig.~\ref{fig:excessNoise}(c) and (d)). We measure $C$ to be detuning-independent, indicating no dynamical backaction effects on the mechanical linewidth at the employed power, while $D$ adequately tracks the inverse of the transmission lineshape derivative, with optima at $\Delta/\kappa\approx\pm\frac{1}{2\sqrt{3}}$. The photothermal coefficients carry the laser and fibre-path noise signatures. The detuning dependence of the coefficient $B_p$, which is shown in Fig.~\ref{fig:excessNoise}(e), is quantitatively captured by a model that includes AM and PAM, i.e., $a\,n_c^2(\Delta) + b\,(\mathrm{d}n_c/\mathrm{d}\Delta)^2$, with the first of the two terms dominating. The presence of these two terms in $B_p$ implies that the amplitude noise $S_{\epsilon\epsilon}(\omega)\propto\omega^{-1}$ and the phase noise $S_{\phi\phi}(\omega)\propto\omega^{-3}$. In contrast, the coefficient $A_p$, shown in Fig.~\ref{fig:excessNoise}(f), exhibits a detuning dependence that cannot be explained from the three terms in Eq.~(\ref{eq:SPP}). While the double peak structure at finite detunings resembles a pure PAM form (dash-dotted guide $\propto(\mathrm{d}n_c/\mathrm{d}\Delta)^2$), its minimum is displaced from $\Delta = 0$. This shift might be the fingerprint of the amplitude--phase cross term in Eq.~(\ref{eq:SPP}), which is odd in $\Delta$ ($\propto n_c(\Delta)\,\mathrm{d}n_c/\mathrm{d}\Delta$) and moves the minimum away from resonance, but we have been unable to explain the full detuning dependence from the three terms in Eq.~(\ref{eq:SPP}).

\begin{figure}[!htbp]
    \centering
    \includegraphics[width=\textwidth]{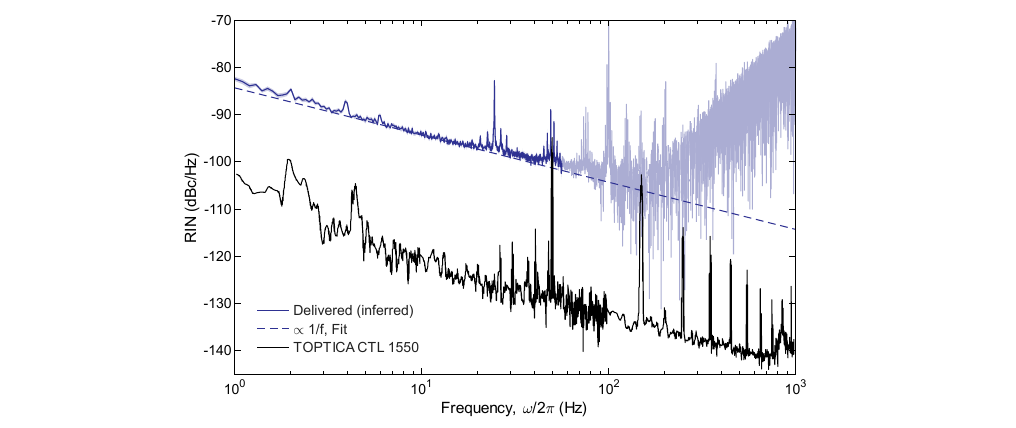}
    \caption{\textbf{Relative intensity noise of the delivered light.} RIN of the light delivered to the resonator, inferred from Eq.~(\ref{eq:RIN_inversion}) for the 13 spectra with $|\Delta| < 0.3\kappa$ (blue line: mean over the spectra at their 0.1\,Hz resolution, with the standard error of the mean as shading), the corresponding $1/\omega$ term of the fits (dashed), and the RIN measured for a Toptica CTL 1550 laser of the same model (black). Above $\approx 55$\,Hz (lighter colour, without error band), the photothermal contribution is less than half of the measured spectrum and the inferred RIN is no longer reliable.}
    \label{fig:deliveredRIN}
\end{figure}

Taken together, these measurements indicate that the dominant extrinsic low-frequency noise originates from amplitude fluctuations of the input field. Since the AM channel of Eq.~(\ref{eq:SPP}) reads $S_{P_\mathrm{abs}P_\mathrm{abs}}(\omega) = 4P_\mathrm{abs}^2\,S_{\epsilon\epsilon}(\omega)$, the measured spectra can be inverted, frequency by frequency, into the relative intensity noise (RIN) of the light that reaches the resonator,
\begin{equation}
\label{eq:RIN_inversion}
\mathrm{RIN}(\omega) = \frac{2\,\xi_\mathrm{AM}\left[\SOmeas(\omega) - A_\mathrm{int}/\omega^2 - B_\mathrm{int}/\omega - C - D\omega^2\right]}{|H_\mathrm{th}(\omega)|^2\left(\Omega_\mathrm{m,0}\,\mathcal{R}_0^\mathrm{abs}P_\mathrm{abs}\right)^2},
\end{equation}
where the factor 2 converts to the single-sided convention customary for laser RIN, and $\xi_\mathrm{AM} = 1 - b\,(\mathrm{d}n_c/\mathrm{d}\Delta)^2/B_p$ is the AM share of the photothermal noise, obtained from the fit of $B_p$ in Fig.~\ref{fig:excessNoise}(e). The denominator contains only the static photothermal shift $\Omega_\mathrm{m,0}\mathcal{R}_0^\mathrm{abs}P_\mathrm{abs}$ characterized in Section~\ref{subsec:kappa_abs}, with $P_\mathrm{abs} = \hbar\omega_c\kappa_\mathrm{abs}n_c(\Delta) \approx 295$\,nW on resonance (0.48\% of the on-chip power), so that the inversion does not depend separately on $\mathcal{R}_0^\mathrm{abs}$ or on the fibre-to-chip coupling efficiency.

We apply Eq.~(\ref{eq:RIN_inversion}) to the 13 spectra with $|\Delta|<0.3\kappa$, for which $\xi_\mathrm{AM}$ ranges from 0.59 to 0.93, assuming $\xi_\mathrm{AM}$ to be frequency-independent, and average the results. As shown in Fig.~\ref{fig:deliveredRIN}, the inferred RIN follows a $1/\omega$ dependence between 1 and $\approx 50$\,Hz, with $-94.1$\,dBc/Hz at $\omega/2\pi = 10$\,Hz (averaged over 9.5--10.5\,Hz, with a statistical standard error of 0.1\,dB), in agreement with the fitted $B_p$ coefficients. Below 1\,Hz (not shown), the $A_p/\omega^2$ term, whose detuning dependence we cannot explain, contributes, while above $\approx 50$\,Hz the photothermal contribution falls below the other terms of Eq.~(\ref{eq:excessFit}). We therefore restrict the comparison to the $1/\omega$ region. Above $\approx 5$\,Hz, where both spectra fall roughly as $1/\omega$, the inferred RIN lies 26--28\,dB above that measured for a Toptica CTL 1550 laser of the same model (black line in Fig.~\ref{fig:deliveredRIN}), which reaches $-121$\,dBc/Hz at 10\,Hz. Below 5\,Hz, the laser RIN rises more steeply and shows features near 2 and 4.5\,Hz, which reduce the gap to 14--25\,dB. Away from discrete lines, the intrinsic laser RIN would thus produce photothermal frequency noise at least a factor of 30, and above 5\,Hz several hundred times, below the measured one, indicating that the intensity fluctuations are added between the laser and the resonator.

Since our chips are photonically packaged, fluctuations of the fibre-to-chip alignment are unlikely, and the most likely origin is instead polarization fluctuations in the input fibre delivery, which are converted into power fluctuations by the polarization dependence of the fibre-to-chip coupling that we observe. This is consistent with the coherence analysis of Fig.~\ref{fig:photothermalCoherence}, which places the power fluctuations before the cavity. At the output, the polarization in the waveguide is fixed, and polarization fluctuations in the output fibre do not change the power on the photodetector. We observe this excess noise even though all fibres in the delivery path, namely the ultra-high-numerical-aperture (UHNA) fibre leading out of the vacuum chamber, the single-mode fibre (SMF) spliced to it and the additional SMF patch cords, are taped down to minimize their movement. The fibre-delivery contribution would be absent with an on-chip laser source, provided that its own low-frequency RIN is sufficiently low.

\subsection{Characterization of drive-induced excess low-frequency noise}
\label{subsec:drive_induced_frequency_noise}

\begin{figure}[!htbp]
    \centering
    \includegraphics[width=\textwidth]{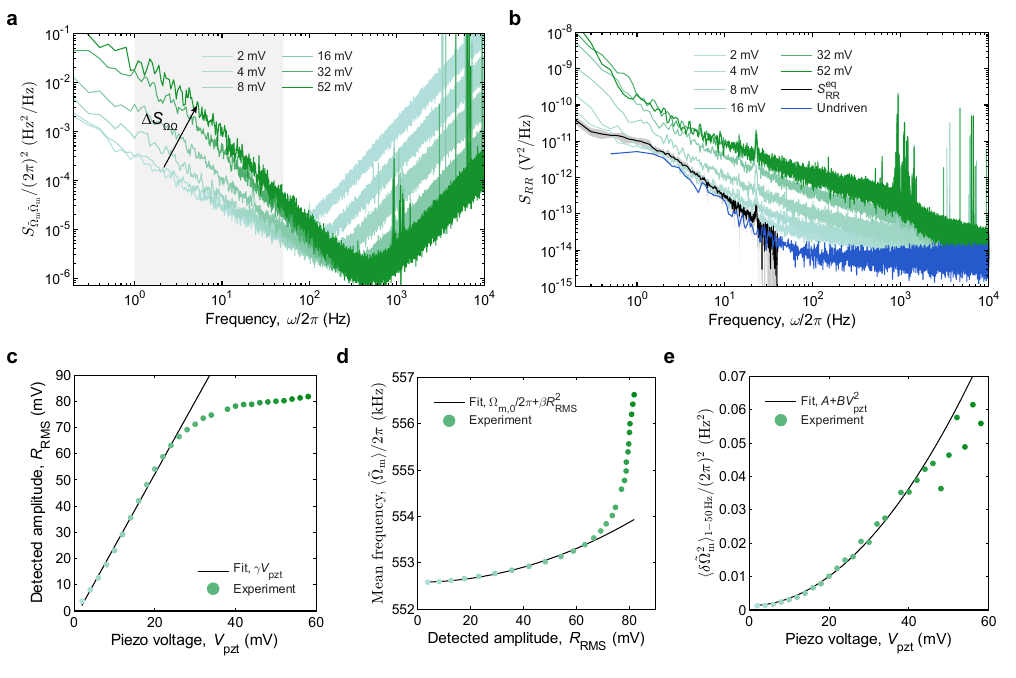}
    \caption{\textbf{Drive-induced excess frequency noise and nonlinear amplitude transduction.} \textbf{a}, PSD of the measured frequency fluctuations, $\SOmeas(\omega)/(2\pi)^2$, versus sideband frequency $\omega/2\pi$ for different piezo-drive voltages $V_{\mathrm{pzt}}$. The shaded band (1--50\,Hz) is integrated in \textbf{e}, and $\Delta\SOmeas$ denotes the excess relative to the lowest drive, Eq.~(\ref{eq:excess_frequency_noise}). \textbf{b}, PSD $S_{RR}(\omega)$ of the magnitude of the demodulated signal, and amplitude noise $S_{RR}^{\mathrm{eq}}$ that would produce $\Delta\SOmeas$ through Eq.~(\ref{eq:duffing_frequency_noise}), averaged over $V_{\mathrm{pzt}} = 8$--26\,mV (black line, grey band: $\pm1$ standard deviation) and shown up to 40\,Hz. Blue: amplitude fluctuations of the undriven resonator, obtained from the in-phase quadrature of its thermomechanical motion recorded without drive, demodulated at the mechanical resonance frequency. \textbf{c}, Root-mean-square magnitude of the demodulated signal, $R_{\mathrm{RMS}}=\sqrt{\langle X^2+Y^2\rangle}$, versus $V_{\mathrm{pzt}}$, with the linear fit of Eq.~(\ref{eq:linear_amplitude_response}) to the low-drive data. \textbf{d}, Mean resonance frequency versus $R_{\mathrm{RMS}}$, with a quadratic fit $\langle\Omeas\rangle/2\pi=\Omega_\mathrm{m,0}/2\pi+\beta R_{\mathrm{RMS}}^2$ to the same data. The fits in \textbf{c} and \textbf{d} are extrapolated beyond the fitted range. The frequency keeps rising after $R_{\mathrm{RMS}}$ saturates, so that $R_{\mathrm{RMS}}$ no longer measures the displacement at large drive. \textbf{e}, Band-limited variance of the measured frequency, $\langle\delta\Omeas^2\rangle_{1-50\,\mathrm{Hz}}$ (Eq.~(\ref{eq:band_limited_frequency_variance})), versus $V_{\mathrm{pzt}}$, with a fit $\langle\delta\Omeas^2\rangle_{1-50\,\mathrm{Hz}}/(2\pi)^2=A+BV_{\mathrm{pzt}}^2$.}
    \label{fig:excess_duffing}
\end{figure}

In the main text, we operate on the red-detuned branch, at $\Delta\approx-|\Delta_*|$, although the blue-detuned branch yields lower detection noise (main text Fig.~2(e)). On the blue-detuned branch, the first-harmonic signal is maximized only at larger drive amplitudes (Section~\ref{subsec:largeAmplitude}), at which the frequency stability of the resonator deteriorates. To understand this deterioration, we measure the frequency noise of the device of the main text on the blue-detuned branch, at $P_\text{det} = 200\,\si{\micro\watt}$ and for $V_{\mathrm{pzt}}$ between 2 and 58\,mV. As shown in Fig.~\ref{fig:excess_duffing}a, increasing the piezo-drive voltage produces a pronounced increase in the low-sideband-frequency component of the frequency-noise spectrum.

A natural candidate for this excess noise is the conversion of amplitude fluctuations into frequency fluctuations by the Duffing nonlinearity of the resonator (Section~\ref{subsec:duffing}). Since the resonance frequency depends on the oscillation amplitude $X$ through the backbone of Eq.~(\ref{eq:backbone}), a fluctuation $\delta X$ around the operating amplitude shifts the frequency by
\begin{equation}
    \delta\Omega_\mathrm{m} \simeq \frac{3\alpha X}{4\Omega_{m,0}}\,\delta X,
    \label{eq:am_fm_conversion}
\end{equation}
so that amplitude noise with PSD $S_{XX}(\omega)$ adds frequency noise $(3\alpha X/4\Omega_{m,0})^2 S_{XX}(\omega)$. The conversion coefficient grows linearly with the operating amplitude, so that amplitude fluctuations that are negligible at low drive, for instance of the drive force, can dominate the frequency noise at large drive. Because the amplitude is measured together with the frequency, this contribution can in principle be predicted from the measured amplitude noise and a calibrated Duffing coefficient, and removed. This approach was recently demonstrated on a high-stress silicon nitride trampoline resonator, enabling high-stability operation beyond the linear regime~\cite{brown_high_2025}. Here we rely on the same relation between amplitude and frequency fluctuations, but to identify the origin of the drive-induced excess rather than to remove it, using the magnitude $R$ of the demodulated signal as a proxy for the mechanical amplitude. Its fluctuations, $S_{RR}(\omega)$, grow with the drive (Fig.~\ref{fig:excess_duffing}b), but, as shown below, mostly because of the readout rather than the resonator.

To calibrate the conversion in detected units, we restrict the analysis to the low-drive measurements for which the detected lock-in magnitude remains proportional to piezo-drive voltage. Within this regime, the detected amplitude is well described by
\begin{equation}
    R_{\mathrm{RMS}}
    =
    \gamma V_{\mathrm{pzt}},
    \label{eq:linear_amplitude_response}
\end{equation}
as shown in Fig.~\ref{fig:excess_duffing}c. Over exactly the same set of measurements, the mean mechanical resonance frequency follows
\begin{equation}
    \frac{\langle\Omeas\rangle}{2\pi}
    =
    \frac{\Omega_\mathrm{m,0}}{2\pi}
    +
    \beta R_{\mathrm{RMS}}^2,
    \label{eq:duffing_frequency_shift}
\end{equation}
as shown in Fig.~\ref{fig:excess_duffing}d, which is the backbone of Eq.~(\ref{eq:backbone}) expressed in detected units. Fitting the measurements with $V_{\mathrm{pzt}}\leq 26$\,mV gives $\gamma = 2.77$ and $\beta = 0.20$\,Hz\,mV$^{-2}$, with $R_{\mathrm{RMS}}$ and $V_{\mathrm{pzt}}$ in mV.

Linearizing Eq.~\eqref{eq:duffing_frequency_shift} around an operating amplitude gives
\begin{equation}
    \frac{\delta\Omeas}{2\pi}
    \simeq
    2\beta R_{\mathrm{RMS}}\,\delta R,
    \label{eq:linearized_amplitude_frequency_conversion}
\end{equation}
and therefore
\begin{equation}
    \SOmeas^{\mathrm{Duff}}(\omega)
    =
    (2\pi)^2\left(2\beta R_{\mathrm{RMS}}\right)^2
    S_{RR}^{\mathrm{mech}}(\omega),
    \label{eq:duffing_frequency_noise}
\end{equation}
where $S_{RR}^{\mathrm{mech}}$ is the mechanical part of the detected amplitude noise.

To identify the origin of the excess, we invert this relation: from the excess frequency noise we infer the amplitude noise that the conversion would require, and compare it with the amplitude fluctuations of the resonator. For each drive of the low-drive subset, we compute the excess frequency noise relative to the lowest drive, $V_\mathrm{min} = 2$\,mV,
\begin{equation}
    \Delta\SOmeas(\omega;V_{\mathrm{pzt}})
    =
    \SOmeas(\omega;V_{\mathrm{pzt}})
    -
    \SOmeas(\omega;V_\mathrm{min}),
    \label{eq:excess_frequency_noise}
\end{equation}
and the detected-amplitude noise that would produce it through Eq.~\eqref{eq:duffing_frequency_noise},
\begin{equation}
    S_{RR}^{\mathrm{eq}}(\omega;V_{\mathrm{pzt}})
    =
    \frac{\Delta\SOmeas(\omega;V_{\mathrm{pzt}})}
    {(2\pi)^2\left(2\beta R_{\mathrm{RMS}}(V_{\mathrm{pzt}})\right)^2}.
    \label{eq:equivalent_amplitude_noise}
\end{equation}
For $V_{\mathrm{pzt}}$ between 8 and 26\,mV, $S_{RR}^{\mathrm{eq}}$ does not depend on the drive within its scatter. Its mean is shown in Fig.~\ref{fig:excess_duffing}b (black line, with the standard deviation as a grey band) up to 40\,Hz, above which the excess frequency noise is not resolved. It has the same Lorentzian spectrum, of half-width $\Gamma_\mathrm{m}/4\pi$, as the thermomechanical amplitude fluctuations of the undriven resonator (blue line), obtained from the thermomechanical spectrum recorded without drive, and exceeds them by up to a factor of about 2. The excess frequency noise is therefore consistent with the Duffing conversion of the thermomechanical amplitude fluctuations of the resonator.

The measured $S_{RR}$, in contrast, grows with the drive, because the detected magnitude also contains fluctuations of the readout that scale with the coherent signal but do not change the mechanical amplitude. Above $V_\mathrm{pzt}\approx20$\,mV, $S_{RR}/R^2$ is independent of the drive, and above a few hertz it agrees to within about 30\% with the relative fluctuations $\delta R/R$ measured in the power sweep of Section~\ref{subsec:power_fluctuations} at a similar optical power on the red-detuned branch. The coherence analysis of Fig.~\ref{fig:photothermalCoherence} attributes these mostly to fluctuations of the optical power reaching the cavity. The readout fluctuations in $S_{RR}$ therefore most likely originate in the delivery of the readout light, with possible smaller contributions from fluctuations of the laser--cavity detuning. At large drive, the nonlinear optical transduction further distorts the relation between $R$ and the mechanical amplitude, as the detected amplitude saturates above about 25\,mV (Fig.~\ref{fig:excess_duffing}c) while the resonance frequency keeps rising (Fig.~\ref{fig:excess_duffing}d). Unlike in that work~\cite{brown_high_2025}, the measured amplitude therefore cannot be used to correct the frequency for the Duffing conversion with this readout.

We quantify the drive-induced excess with the band-limited frequency variance
\begin{equation}
    \langle\delta\Omeas^2\rangle_{1-50\,\mathrm{Hz}}(V_{\mathrm{pzt}})
    =
    \int_{2\pi\times1\,\mathrm{Hz}}^{2\pi\times50\,\mathrm{Hz}}
    \SOmeas(\omega;V_{\mathrm{pzt}})\,\frac{\mathrm{d}\omega}{2\pi} .
    \label{eq:band_limited_frequency_variance}
\end{equation}
The resulting variance increases approximately quadratically with the piezo-drive voltage (Fig.~\ref{fig:excess_duffing}e), as expected from Eq.~\eqref{eq:duffing_frequency_noise} for a drive-independent amplitude noise, since $R_{\mathrm{RMS}}\propto V_{\mathrm{pzt}}$ at low drive. At the largest drive amplitudes, the measurements exhibit increased scatter and depart from the extrapolation of the lower-drive dependence. Since this excess appears to originate in the thermomechanical motion of the resonator itself, it grows with the drive irrespective of the readout, which is why measurements requiring maximum frequency stability are performed on the red-detuned branch, at the drive that maximizes the first-harmonic signal there (main text), rather than at the larger drives required on the blue-detuned branch.

\subsection{Balanced homodyne detection spectra}
\label{subsec:homodyne}

\begin{figure}[!htbp]
    \centering
    \includegraphics[width=\textwidth]{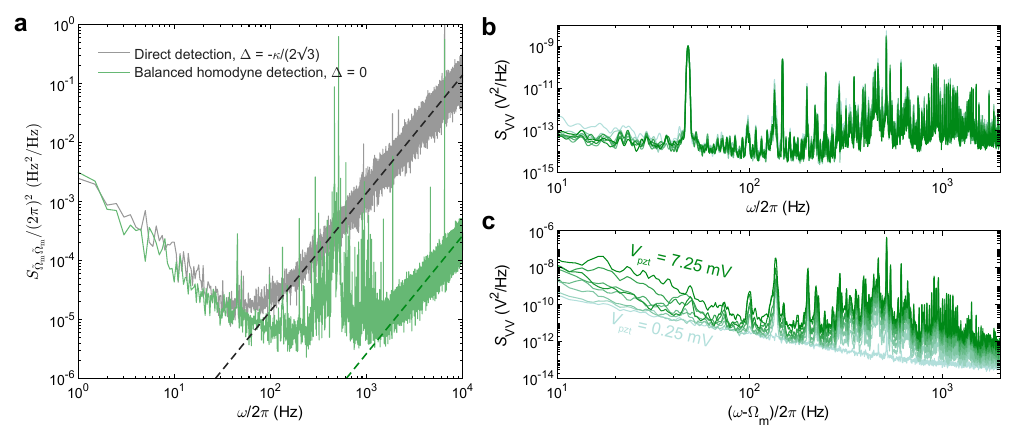}
    \caption{\textbf{Balanced homodyne detection.} \textbf{a}, Frequency noise PSD measured with direct detection at the maximum-slope detuning ($\Delta = -\kappa/(2\sqrt{3})$, grey) and with balanced homodyne detection on resonance ($\Delta = 0$, green), at the same drive ($V_\mathrm{pzt} = 4$\,mV) and optical power ($P_\mathrm{det} = 70$\,\si{\micro\watt}). Dashed lines: detection-noise contributions, $D\omega^2$, with $D$ estimated from the PSD between 2 and 6\,kHz. \textbf{b}, PSD of the balanced homodyne signal around DC, recorded with a real-time spectrum analyser on resonance ($P_\mathrm{det} = 100$\,\si{\micro\watt}) for drive voltages $V_\mathrm{pzt}$ from 0.25 to 7.25\,mV in steps of 1\,mV (light to dark green). \textbf{c}, Same as \textbf{b} around the mechanical carrier, as a function of the offset from the carrier frequency.}
    \label{fig:BHD}
\end{figure}

All other measurements use side-of-the-line (direct) detection of the intensity fluctuations caused by the motion of the mechanical resonator. As shown in Fig.~2(g) of the main text, the detection-noise coefficient $D$ scales as $P_\text{det}^{-2}$ across most of the power range we explore, which indicates that the imprecision is set by photodetector noise rather than by shot noise, with shot-noise-limited detection approached only at the highest powers, where the scaling crosses over to $P_\text{det}^{-1}$. To reduce the imprecision at the same optical power, we implement balanced homodyne detection (BHD) with a free-space interferometer and balanced detector, in which only the paths to and from the chip and a length-matching section of the local-oscillator arm are fibre-based, operating on resonance ($\Delta = 0$) with the local oscillator in phase quadrature ($\phi_\text{LO} = \pi/2$).

Figure~\ref{fig:BHD}(a) compares the frequency noise measured with both schemes at the same drive ($V_\mathrm{pzt} = 4$\,mV) and optical power (70\,\si{\micro\watt}), both in the linear transduction regime rather than at the optimal operating conditions discussed in the main text. BHD reduces the detection contribution, which scales as $D\omega^2$, about 500-fold, while the low-frequency noise remains unchanged, which confirms that these frequency fluctuations do not originate in the detection.

BHD also reveals a set of narrow peaks at frequencies $\{\Omega_i\}$ between about 100\,Hz and 2\,kHz in the frequency-noise spectrum. The same frequencies appear in the detected spectrum, $S_{VV}(\omega)$, recorded with a real-time spectrum analyser, both around DC (Fig.~\ref{fig:BHD}(b)) and as sidebands around the mechanical carrier at $\Omega_\mathrm{m}\pm\Omega_i$ (Fig.~\ref{fig:BHD}(c)). The baseband peaks do not depend on the drive, persist when the light bypasses the cavity and require interference between the two arms of the homodyne interferometer, and none of them is observed in direct detection. We therefore attribute them to phase fluctuations between the signal and local-oscillator arms of the interferometer, likely picked up in their non-common fibre sections and of acoustic origin. An additional series of baseband peaks at 48, 148, 248,~\ldots\,Hz is also present when the signal arm is blocked, and is therefore unrelated to the interference. Around the carrier, the sidebands grow with the drive, consistent with intermodulation products of the interferometric phase fluctuations with the coherently driven mechanical motion. This is analogous to the thermal intermodulation noise observed in cavity-optomechanical experiments~\cite{fedorov_thermal_2020,leijssen_nonlinear_2017}, but results here from a driven mechanical resonance and discrete interferometric noise tones rather than from the thermomechanical motion of multiple mechanical modes. An integrated homodyne receiver, with short and largely common optical paths for the signal and the local oscillator, would avoid this noise and allow BHD to deliver its full benefit for frequency tracking. For this reason, all measurements in the main text use direct detection.

\section{Mid-infrared response and noise-equivalent power}
\label{sec:MIR_details}

This section provides details on FEM simulations of the fractional frequency responsivity, on the properties and characterization of our platinum absorber and on the procedure to calibrate the spectrum that is incident onto our OMTD in the experimental conditions we employ.

\subsection{Finite-element simulations of the fractional frequency responsivity}
\label{subsec:Rf}

To determine the expected fractional frequency responsivity of the trampoline resonator to absorbed optical power, we perform numerical simulations using the finite-element method (FEM) that sequentially couple stationary heat diffusion, thermoelastic strain, stress redistribution and prestressed mechanical eigenmode analysis. The central quantity of interest is the absorbed-power fractional frequency responsivity of a mechanical resonance frequency $\Omega_\mathrm{m}$,
\begin{equation}
\mathcal{R}_0^{\mathrm{abs}} \equiv \frac{1}{\Omega_\mathrm{m,0}}\frac{\partial \Omega_\mathrm{m}}{\partial P_{\mathrm{abs}}},
\end{equation}
which quantifies the first-order relative change in mechanical frequency as a result of absorbed power. We consider only thermal mechanisms, neglecting for instance photo-induced electronic strain, and assume the material to be thermalized at all relevant timescales. The change of the mechanical resonance frequency can occur through several temperature-dependent mechanisms, such as changes in the constitutive elastic constants or changes in geometric dimensions and stress due to thermal expansion. In highly stressed trampoline resonators, the dominant contribution arises from temperature-induced relaxation of the tensile stress, while direct temperature dependence of Young’s modulus and geometric expansion play a secondary role. The modelling therefore separates naturally into three steps: (i) a stationary thermal problem determining the temperature field, (ii) a static mechanical equilibrium problem determining the thermally modified stress field, and (iii) a linearized eigenvalue problem determining the mechanical resonance frequencies about this equilibrium.\\

\noindent\emph{Thermal problem:}

\begin{figure}[!htbp]
    \centering
    \includegraphics[width=\textwidth]{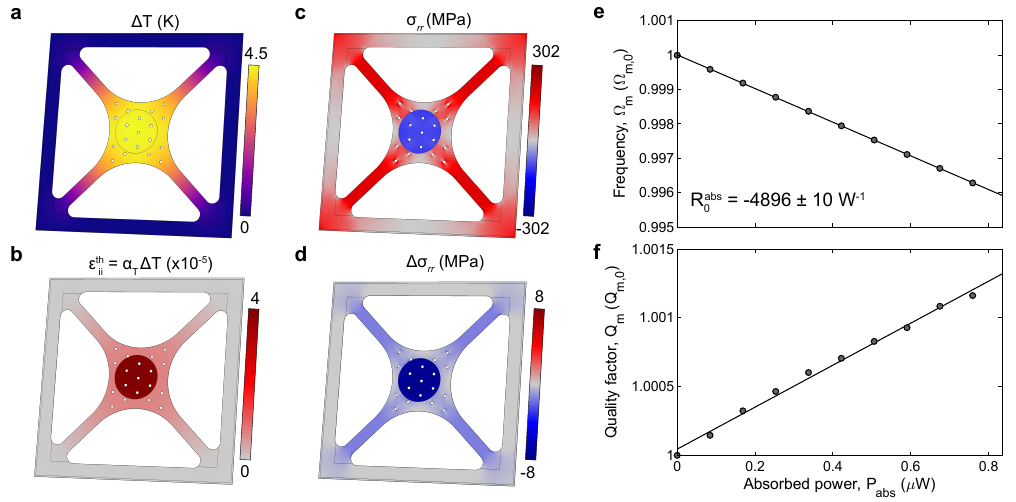}
        \caption{\textbf{Simulated photothermal response of the trampoline.} \textbf{a}, Temperature rise $\Delta T$ for the incident profile given above ($P_\text{inc} = 1.23$\,mW). \textbf{b}, Thermoelastic strain $\varepsilon^\text{th} = \alpha_\text{T}\Delta T$. \textbf{c}, Radial stress component $\sigma_{rr}$, including the built-in stress. \textbf{d}, Thermally induced change of $\sigma_{rr}$. \textbf{e}, Frequency of the fundamental mode, normalized to its unheated value, versus absorbed power, with a linear fit giving $\mathcal{R}_0^\text{abs}$. \textbf{f}, Same for the mechanical quality factor. All panels are computed for the central values of the material parameters (Table~\ref{tab:material_parameters}).}
    \label{fig:figRf}
\end{figure}

\noindent The steady-state temperature field $T(\mathbf{r})$ is obtained by solving the heat diffusion equation
\begin{equation}
\nabla \cdot \left[\kappa_\text{th}(\mathbf{r}) \nabla T(\mathbf{r})\right] = 0
\qquad \text{in } \Omega ,
\end{equation}
where $\kappa_\text{th}(\mathbf{r})$ is the thermal conductivity of the membrane material. The thermal boundary conditions are
\begin{align}
-\,\kappa_\text{th} \nabla T \cdot \mathbf{n} &= q_{\mathrm{abs}}(x,y)
&& \text{on } \partial\Omega_{\mathrm{top}}, \label{eq:heatDiffusionBD1}\\
T &= T_0
&& \text{on } \partial\Omega_{\mathrm{anch}}, \\
-\,\kappa_\text{th} \nabla T \cdot \mathbf{n} &= 0
&& \text{on } \partial\Omega_{\mathrm{free}},
\end{align}
where $\partial\Omega_{\mathrm{top}}$ is the top surface, exposed to the radiation, $\partial\Omega_{\mathrm{anch}}$ denotes the anchors held at bath temperature $T_0$, and $\partial\Omega_{\mathrm{free}}$ represents all remaining surfaces. Convective and radiative losses are neglected, and the radiative contribution is estimated below. The absorbed heat flux density in Eq.~(\ref{eq:heatDiffusionBD1}) is written as
\begin{equation}
q_{\mathrm{abs}}(x,y)=\eta_{\mathrm{abs}}(x,y)\,P(x,y),
\end{equation}
where $P(x,y)$ is the incident optical power density profile, normalized such that
\begin{equation}
\int_{\partial\Omega_{\mathrm{top}}} P(x,y)\,\mathrm{d}A = P_{\mathrm{inc}},
\end{equation}
and $\eta_{\mathrm{abs}}(x,y)$ is the local absorptance, taken as $\langle A\rangle_\text{S} = 0.42$ on the Pt disk (Section~\ref{subsec:thermal_source_calibration}) and zero on the bare Si$_x$N$_y$. Since $\mathcal{R}_0^\mathrm{abs}$ is defined per absorbed power, this value only sets $P_\mathrm{abs}$ for a given $P_\mathrm{inc}$. The total absorbed power is
\begin{equation}
P_{\mathrm{abs}}=
\int_{\partial\Omega_{\mathrm{top}}}
\eta_{\mathrm{abs}}(x,y)\,P(x,y)\,\mathrm{d}A .
\end{equation}
The resulting temperature field enters the mechanical problem through the thermoelastic eigenstrain
\begin{equation}
\boldsymbol{\varepsilon}^{\mathrm{th}}(\mathbf{r})
=
\alpha_\text{T}(\mathbf{r})\,[T(\mathbf{r})-T_0]\mathbf{I},
\end{equation}
where $\alpha_\text{T}$ is the thermal expansion coefficient. Figures~\ref{fig:figRf}(a) and (b), respectively, show the temperature field and the isotropic thermoelastic strain for the device shown in the main text subject to an incident field of the form
\begin{equation}
P(x,y)=P_{\mathrm{inc}}\,
\frac{S(r)}{\displaystyle \int_{\partial\Omega_{\mathrm{top}}} S(r)\,\mathrm{d}A},
\end{equation}
with
\begin{equation}
S(r)=\frac{1}{2}\left[1-\mathrm{erf}\!\left(\frac{r-R}{w}\right)\right]
\end{equation}
and $R = D_\text{MIR}/2 = 369.1$\,\si{\micro\meter}, $w$ = 151.1 \si{\micro\meter} and $P_\text{inc}$ = 1.23 \si{\milli\watt}, with $R$ and $w$ from the fit of the map of main text Fig.~3(e). The material parameters, with their central values and literature ranges, are listed in Table~\ref{tab:material_parameters}.\\

\noindent\emph{Static mechanical equilibrium:}

\noindent The static equilibrium of the suspended structure is governed by
\begin{equation}
\nabla \cdot \boldsymbol{\sigma}(\mathbf{r}) = \mathbf{0}
\qquad \text{in } \Omega ,
\end{equation}
with boundary conditions
\begin{align}
\mathbf{u}(\mathbf{r}) &= \mathbf{0}
& \text{for } \mathbf{r}\in\partial\Omega_{\mathrm{anch}}, \\
\boldsymbol{\sigma}(\mathbf{r})\cdot\mathbf{n} &= \mathbf{0}
& \text{for } \mathbf{r}\in\partial\Omega_{\mathrm{free}} ,
\end{align}
and constitutive relation written as
\begin{equation}
\boldsymbol{\sigma}(\mathbf{r})
= \boldsymbol{\sigma}_\text{dep}(\mathbf{r}) +
\mathsf{C}(\mathbf{r}):\bigl(
\boldsymbol{\varepsilon}(\mathbf{r})
-
\boldsymbol{\varepsilon}^{\mathrm{th}}(\mathbf{r})
\bigr),
\end{equation}
where $\mathsf{C}$ is the fourth-order elasticity tensor and $\boldsymbol{\sigma}_\text{dep}$ is the deposition stress of the film before release, which we take as in-plane, homogeneous and isotropic, with magnitude $\sigma_\text{init}$,
\begin{equation}
\boldsymbol{\sigma}_\text{dep}(\mathbf{r}) =
\begin{pmatrix}
\sigma_\text{init} & 0 & 0 \\
0 & \sigma_\text{init} & 0 \\
0 & 0 & 0
\end{pmatrix}
\end{equation}
in each of the materials. For the Si$_x$N$_y$ film we use $\sigma_\text{init} = 167$\,MPa (Table~\ref{tab:material_parameters}), the value that reproduces the measured frequency of the fundamental mode at the central density. It is lower than the as-deposited stress of 235\,MPa obtained from wafer-bow measurements (Methods of the main text), which suggests a further relaxation of the effective stress during processing. Solving this problem with $\boldsymbol{\varepsilon}^{\mathrm{th}}=\mathbf{0}$ yields the released, unheated equilibrium stress field. Including the thermoelastic term yields the thermally modified equilibrium stress field
\begin{equation}
\boldsymbol{\sigma}_0(\mathbf{r}) \equiv \boldsymbol{\sigma}(\mathbf{r};P_{\mathrm{abs}}),
\end{equation}
which is used as the reference state for the eigenmode analysis. Figures~\ref{fig:figRf}(c) and (d), respectively, show the radial stress component, $\sigma_{rr}$, and the contribution resulting from thermal expansion, $\Delta\sigma_{rr}$, calculated by subtracting the same stress component in the absence of the heat source. For the power used, the contribution from the built-in stress is much greater than the thermal one, so Fig.~\ref{fig:figRf}(c) can be largely understood from the former: the built-in stress in the central Si$_x$N$_y$ trampoline pad is partially released upon suspension, generating compression on the Pt and enhanced stress (up to around 300 MPa) on the trampoline tethers. As for the thermal contribution (Fig.~\ref{fig:figRf}(d)), $\Delta T$ leads to a release of the tensile stress in most of the trampoline, especially in the tethers, and additional compressive stress in the Pt ($\alpha_\text{T}^{\text{Pt}}>\alpha_\text{T}^{\SiNxy}$).\\

\begin{table}[h!]
\centering
\renewcommand{\arraystretch}{1.25}
\setlength{\tabcolsep}{5pt} %
\begin{tabular*}{\textwidth}{@{\extracolsep{\fill}}l c c c c}
\toprule
\textbf{Property} & \textbf{Symbol} & \textbf{Unit} & \textbf{Si$_x$N$_y$, central (range)} & \textbf{Pt}\\
\midrule
Thermal conductivity      & $\kappa_\text{th}$  & W\,m$^{-1}$\,K$^{-1}$   & 3.0 (2.5--3.5)~\cite{ftouni_thermal_2015,snell_heat_2022,piller_thermal_2020,martini_uncooled_2025}    & 22~\cite{martini_uncooled_2025}   \\
Heat capacity             & $c_\text{p}$   & J\,kg$^{-1}$\,K$^{-1}$  & 740 (700--780)~\cite{martini_design_2026,martini_uncooled_2025}   & 133~\cite{martini_uncooled_2025}  \\
Thermal expansion coefficient  & $\alpha_\text{T}$  & $10^{-6}$\,K$^{-1}$ & 1.7 (1.1--2.3)~\cite{kanellopulos_stress-dependent_2024,habermehl_coefficient_2018,sadeghi_thermal_2020,snell_heat_2022} & 9.1~\cite{martini_uncooled_2025} \\
Mass density               & $\rho$      & kg\,m$^{-3}$    & 3000 (2900--3100)~\cite{giesriegl_hf_2026,martini_uncooled_2025,khan_youngs_2004}  & 21450   \\
Young's modulus            & $E$      & GPa      & 250 (210--290)~\cite{martini_uncooled_2025,khan_youngs_2004}           & 145~\cite{martini_uncooled_2025}        \\
Poisson ratio              & $\nu$      & --       & 0.23~\cite{khan_youngs_2004}          & 0.39~\cite{martini_uncooled_2025}      \\
Intrinsic quality factor   & $Q_\text{int}$    & --      & $1.9\times10^3$ & $2\times10^3$\\
Initial stress (fitted)    & $\sigma_\text{init}$     & MPa                    & 167 ($\propto\rho$)           & 0         \\
\bottomrule
\end{tabular*}
\caption{Material parameters used in the FEM thermo-mechanical simulations. For Si$_x$N$_y$, the central value is the midpoint of the range of literature values. The range of $\alpha_\text{T}$ spans the values measured on silicon-rich LPCVD films, 1.1--1.8$\times10^{-6}$\,K$^{-1}$~\cite{kanellopulos_stress-dependent_2024,habermehl_coefficient_2018}, and those commonly used for Si$_x$N$_y$ resonators, 2.2$\times10^{-6}$\,K$^{-1}$~\cite{sadeghi_thermal_2020,snell_heat_2022}. The density of the film is taken from X-ray reflectivity on low-stress LPCVD films~\cite{giesriegl_hf_2026}.}
\label{tab:material_parameters}
\end{table}

\noindent\emph{Mechanical eigenmodes and linearization:}

\noindent Mechanical resonance frequencies are obtained by linearizing the equations of motion, $\rho\,\ddot{\mathbf{u}} = \nabla\cdot\boldsymbol{\sigma}$, for small vibrations $\mathbf{u}_1$ about the thermally modified equilibrium. Using the full Green--Lagrange strain to account for geometric nonlinearity and pre-stress, the equilibrium stress $\boldsymbol{\sigma}_0$ couples to the quadratic part of the strain and adds a geometric-stiffness term to first order in $\mathbf{u}_1$. Seeking time-harmonic solutions, $\mathbf{u}_1(\mathbf{r},t)=\mathbf{U}(\mathbf{r})e^{i\omega t}$, leads to the prestressed eigenvalue problem
\begin{equation}
-\,\omega^2\rho\,\mathbf{U}
=
\nabla\cdot\!\left(
\mathsf{C}:\boldsymbol{\varepsilon}[\mathbf{U}]
+
\boldsymbol{\sigma}_0 \cdot \nabla\mathbf{U}
\right),
\end{equation}
where $\boldsymbol{\varepsilon}[\mathbf{U}] = \tfrac12(\nabla\mathbf{U}+\nabla\mathbf{U}^{\mathrm T})$ is the linear strain,
with boundary conditions
\begin{align}
\mathbf{U} &= \mathbf{0}
&& \text{on } \partial\Omega_{\mathrm{anch}}, \\
\left(
\mathsf{C}:\boldsymbol{\varepsilon}[\mathbf{U}]
+
\boldsymbol{\sigma}_0 \cdot \nabla\mathbf{U}
\right)\mathbf{n} &= \mathbf{0}
&& \text{on } \partial\Omega_{\mathrm{free}} .
\end{align}

In this formulation, the equilibrium stress field, set by stress release and thermoelastic loading, enters the eigenfrequency problem as a geometric stiffness operator. In trampoline resonators, where membrane tension dominates over bending rigidity, this contribution provides the leading restoring force and therefore controls the mechanical resonance frequencies. The quality factor of each mode (Fig.~\ref{fig:figRf}(f)) is obtained from the intrinsic quality factor of the film, $Q_\text{int}$ (Table~\ref{tab:material_parameters}), multiplied by the dissipation dilution factor of the mode, i.e.\ the ratio of its total elastic energy to the part stored in bending~\cite{fedorov_generalized_2019,bereyhi_hierarchical_2022}. For the Pt film we set $Q_\text{int} = 2000$, but its value barely affects the simulated $Q$, since the central pad carrying the disk moves almost rigidly and stores little bending energy. This is consistent with the equal linewidths of Pt-coated and uncoated trampolines (Section~\ref{subsec:freespace_char}).

A linear fit of the simulated frequency of the fundamental mode versus absorbed power (Fig.~\ref{fig:figRf}(e)) gives $\mathcal{R}_0^\text{abs} = -4.9\times10^{3}$\,W$^{-1}$ at the central values of the material parameters. The thermal and mechanical parameters of thin Si$_x$N$_y$ films have not been calibrated independently on our devices, and the values reported in the literature spread considerably (Table~\ref{tab:material_parameters}). We propagate this spread by moving one parameter at a time to the ends of its range, with all others at their central values, and combine the resulting changes in quadrature, taking each half-range as one standard deviation. This gives $\mathcal{R}_0^\text{abs} = -(4.9\pm2.0)\times10^{3}$\,W$^{-1}$. The spread is dominated by the thermal expansion coefficient ($\pm1.7\times10^{3}$\,W$^{-1}$), followed by the thermal conductivity ($^{+1.0}_{-0.7}\times10^{3}$\,W$^{-1}$) and Young's modulus ($\pm0.8\times10^{3}$\,W$^{-1}$). The density contributes only $\pm0.15\times10^{3}$\,W$^{-1}$, because $\sigma_\text{init}$ is rescaled in proportion to $\rho$ to keep the measured frequency, which scales as $\sqrt{\sigma/\rho}$. The simulated responsivity is linear in $\alpha_\text{T}$ to within 0.3\% over its range and scales as $1/\kappa_\text{th}$, as expected when the stress relaxation in the tethers dominates and heat is conducted through them.

The measured $\mathcal{R}_0^\text{abs} = -(6.3\pm0.5)\times10^{3}$\,W$^{-1}$ (main text) lies within this range, in its upper part. A sharper comparison would require independent calibrations of $\alpha_\text{T}$, $\kappa_\text{th}$ and $E$ on the same films. Radiative losses, neglected in the simulations, add a thermal conductance in parallel with conduction through the tethers. Summing the emission of the Pt disk and of the bare Si$_x$N$_y$ of the pad, each with emissivity equal to its absorptance, gives $G_\text{rad}\approx 4\times10^{-9}$\,W\,K$^{-1}$ (Section~\ref{subsec:NEPth}), about 2\% of the simulated $G_\text{th}$ and well within the range above.

\subsection{Finite-element simulations of the frequency-dependent responsivity}
\label{subsec:thermal_bandwidth}

To define the frequency-dependent thermomechanical responsivity, we start from the linear relation between the temperature rise and the mechanical resonance frequency. For small perturbations, the frequency shift can be written as a causal linear functional of the temperature field,
\begin{equation}
\delta f(t)=\int dA \int_{-\infty}^{t} dt'\,K(\mathbf r,t-t')\,\Delta T(\mathbf r,t'),
\end{equation}
where the response kernel $K(\mathbf r,t-t')$ accounts for thermoelastic stress generation, its redistribution under global mechanical equilibrium, and the resulting perturbation of the eigenfrequency. In the present system, thermoelastic equilibration and mechanical stress redistribution occur on timescales much shorter than thermal diffusion, so that the kernel is effectively instantaneous on the timescales of interest,
\begin{equation}
K(\mathbf r,t-t')\approx K(\mathbf r)\,\delta(t-t').
\end{equation}
Under this condition, the frequency dynamics are governed entirely by thermal relaxation, while the mapping from temperature rise to frequency shift remains non-local in space. Because determining the full spatial kernel $K(\mathbf r)$ would require a coupled thermo-mechanical sensitivity analysis, we adopt a reduced description in which the frequency shift is parameterized by a single effective temperature rise,
\begin{equation}
\frac{\delta f(t)}{f_\mathrm{m,0}}\approx \mathcal{R}_T\Delta T_\mathrm{eff}(t),
\end{equation}
where $\mathcal{R}_T$ is the static fractional temperature coefficient of the mechanical frequency obtained from the DC thermo-mechanical analysis. In the following, $\Delta T_\mathrm{eff}$ is taken as the area-averaged temperature rise over the full suspended resonator (pad and tethers).

 \begin{figure}[!htbp]
    \centering
    \includegraphics[width=\textwidth]{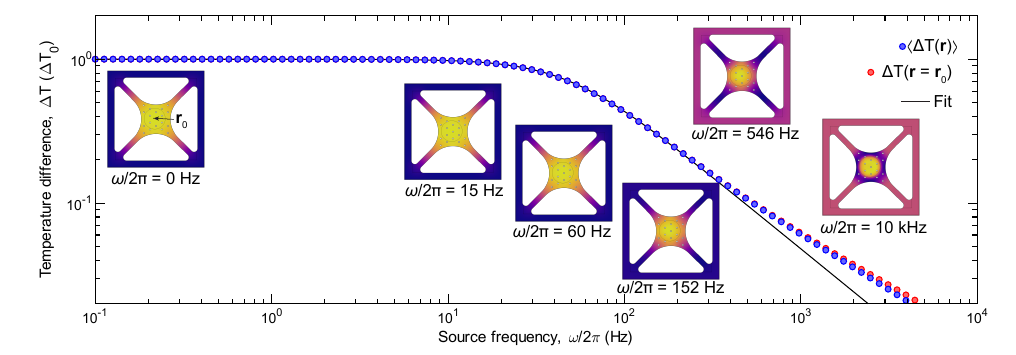}
        \caption{\textbf{Simulated temperature transfer function.} Amplitude of the harmonic temperature variation as a function of frequency, $\omega$, evaluated as an average over the full membrane (blue dots) and at a single spatial location, $\mathbf{r}_0$ (red dots). The black line is a fit to a single-pole first-order filter and insets show the temperature field on the top surface of the suspended device for selected $\omega$. Each inset has its own colour scale, from the minimum to the maximum of its temperature field. Computed for the central values of the material parameters (Table~\ref{tab:material_parameters}).}
    \label{fig:figOmegath}
\end{figure}

The thermal dynamics are described by the heat diffusion equation with temperature-independent material parameters,
\begin{equation}
\rho c_p\,\partial_t T(\mathbf r,t)-\nabla\cdot(\kappa_\text{th}\nabla T(\mathbf r,t))=q(\mathbf r,t),
\end{equation}
where $\rho$, $c_p$, and $\kappa_\text{th}$ are the mass density, specific heat capacity, and thermal conductivity, respectively. Treating these parameters as constants renders the problem linear and allows a frequency-domain formulation. For a harmonically modulated heat source, the resulting frequency-domain equation reads
\begin{equation}
-\nabla\cdot\big(\kappa_\text{th}\nabla T(\mathbf r,\omega)\big)-i\omega\rho c_p\,T(\mathbf r,\omega)=q(\mathbf r,\omega),
\end{equation}
and is solved numerically using the same boundary conditions as in the static thermal model of Section~\ref{subsec:Rf}. From the computed temperature field, the effective temperature rise $\Delta T_\mathrm{eff}(\omega)$ is extracted and normalized to its low-frequency value, defining the thermal transfer function
\begin{equation}
H_\mathrm{th}(\omega)=\frac{\Delta T_\mathrm{eff}(\omega)}{\Delta T_\mathrm{eff}(0)}.
\end{equation}
Figure~\ref{fig:figOmegath} shows the amplitude response extracted from the FEM both for our choice of $\Delta T_\text{eff}$ and for the temperature at a point $\mathbf{r}_0$ on the Pt. For the former, the numerical results are well described by a first-order low-pass response
\begin{equation}
H_\mathrm{th}(\omega)=\frac{1}{1-i\omega\tau_\mathrm{th}},
\end{equation}
characterized by a single thermal time constant $\tau_\mathrm{th}$ from DC up to about 300--400\,Hz. The origin of this single-pole behaviour can be understood within a lumped-element thermal model. In vacuum, heat loss is dominated by conduction through the suspension tethers, while the suspended pad and tethers together act as a thermal mass. The system can therefore be approximated as a single thermal node with heat capacity $C_\mathrm{th}$, coupled to the bath through an effective thermal conductance $G_\mathrm{th}$, yielding $\tau_\mathrm{th}\approx C_\mathrm{th}/G_\mathrm{th}$. For a simplified geometry of our trampoline \textemdash without filleting\textemdash these quantities can be estimated as
\begin{equation}
C_\mathrm{th} \approx \sum_i \rho_i c_{p,i} h_i A_\mathrm{susp,i},
\qquad
G_\mathrm{th} \approx 4\,\frac{\kappa_{\text{th},\SiNxy} w h_{\SiNxy}}{L_t},
\end{equation}
where $A_\mathrm{susp,i}$ is the total suspended area contributing to the thermal mass, $w$ and $L_t$ are the tether width and length, and $h_{\SiNxy}$ is the Si$_x$N$_y$ thickness. Using the device dimensions given in the main text and neglecting filleting, one obtains $G_\mathrm{th} \approx \SI{174}{\nano\watt/\kelvin}$. Depending on what fraction of the tether length is included in $A_{\mathrm{susp},\SiNxy}$, the thermal mass ranges between $C_\mathrm{th} \approx \SI{465}{\pico\joule/\kelvin}$ (pad and Pt absorber only) and $\approx \SI{857}{\pico\joule/\kelvin}$ (full tether length included), corresponding to a thermal pole bandwidth $f_\text{th}$ between 32\,Hz and 60\,Hz, all at the central values of the material parameters.

The first-order fit to the simulated $\Delta T_\mathrm{eff}$ gives $\tau_\mathrm{th} = 3.30$\,ms at the central values, i.e.\ $f_\text{th} = 1/(2\pi\tau_\mathrm{th}) = 48.2$\,Hz, which falls within this lumped estimate. As in the lumped model, the cutoff is set by $\kappa_\text{th}$ and by the product $\rho c_p$ only, and the simulations confirm that it scales as $\kappa_\text{th}/(\rho c_p)$, to within 0.5\% over the range of $\kappa_\text{th}$. Propagating the literature ranges of these parameters (Table~\ref{tab:material_parameters}) as for $\mathcal{R}_0^\text{abs}$ in Section~\ref{subsec:Rf} gives $f_\text{th} = 48\pm8$\,Hz, dominated by $\kappa_\text{th}$ ($\pm7.8$\,Hz), with $\rho$ and $c_p$ contributing about $\pm1.6$ and $\pm2.6$\,Hz. The measured cutoff, 47.4\,Hz (main text), lies close to the central value. Combined with the simulated heat capacity, it also gives the thermal conductance of the device, which we use in Section~\ref{subsec:NEPth}.

The lumped description holds up to about 300--400\,Hz. The temperature at a single point $\mathbf{r}_0$ on the Pt has a very similar 3\,dB cutoff, but both responses deviate from a single pole above these frequencies. There, heat no longer spreads uniformly over the pad within a modulation period, and thermal diffusion through the pad itself must be considered, as visible in the temperature maps in the insets of Fig.~\ref{fig:figOmegath}. More generally, because the thermoelastic mapping from temperature rise to resonance frequency is non-local, the effective weighting of the temperature field can depend on the spatial distribution of the heat source. Deviations from a single-pole response are then expected when heating is strongly localized, or when several thermal relaxation pathways with comparable timescales are present. This dependence on the size of the heat source is examined in Section~\ref{subsec:spot_size}.

 \begin{figure}[!htbp]
    \centering
    \includegraphics[width=\textwidth]{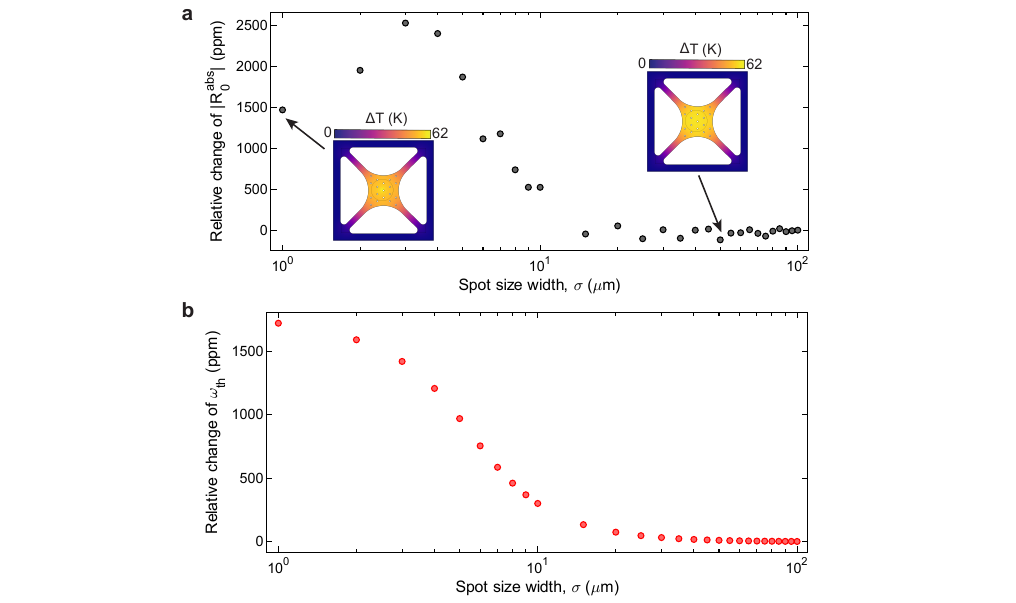}
        \caption{\textbf{Simulated dependence of the thermal and mechanical frequency response on the incident spot size.} \textbf{a}, Relative change of the magnitude of the steady-state responsivity to absorbed power, $|\mathcal{R}^\text{abs}_0|$, versus the width, $\sigma$, of a Gaussian heat source centred on the Pt disk, with respect to its value at $\sigma = 100$\,\si{\micro\meter}. Insets: steady-state temperature rise for $\sigma = 1$ and 50\,\si{\micro\meter} and 10\,\si{\micro\watt} absorbed, on a common colour scale. \textbf{b}, Same for the thermal bandwidth, $\omega_\text{th}$. Both are given in parts per million and computed for the central values of the material parameters (Table~\ref{tab:material_parameters}).}
    \label{fig:figRfSpotSize}
\end{figure}

\subsection{Effect of the incident spot size on the response}
\label{subsec:spot_size}

In the main text, the NEP referred to the absorber assumes that $\mathcal{R}^\mathrm{abs}(\omega)$ depends only weakly on the size of the incident spot. Figure~\ref{fig:figRfSpotSize} tests this with FEM simulations of the measured device at the central values of the material parameters (Section~\ref{subsec:Rf}), for a Gaussian boundary heat source of width $\sigma$ centred on the Pt disk. For each width, the incident power is set such that 10\,\si{\micro\watt} are absorbed, which raises the frequency shift well above the numerical precision of the eigenfrequency. At this power, the responsivity is about 2\% larger than its small-signal value, uniformly for all widths, so we show only its relative variation with $\sigma$, taking the largest width as reference. The thermal bandwidth is obtained from a first-order fit to the simulated response at 16 frequencies between 20 and 80\,Hz. Over two decades in $\sigma$ (1--100\,\si{\micro\meter}), $\mathcal{R}^\mathrm{abs}_0$ varies by less than 0.3\% and $\omega_\mathrm{th}$ by less than 0.2\%. Both are nearly constant once the spot is larger than the Pt disk, which is then heated uniformly, and change only for spots smaller than it, which heat its centre more strongly. This weak dependence arises because the steady-state temperature rise and thermal relaxation are dominated by heat conduction through the suspension tethers, whereas changing the spot size primarily affects the temperature distribution within the absorber without significantly modifying the global thermal resistance. These results justify treating $\mathcal{R}^\mathrm{abs}(\omega)$ as effectively independent of spot size for the purposes of NEP estimation.

\subsection{Characterization of the ultrathin Pt film as an impedance-matched mid-infrared absorber}
\label{subsec:FTIR_membranes}

 \begin{figure}[!htbp]
    \centering
    \includegraphics[width=0.90\textwidth]{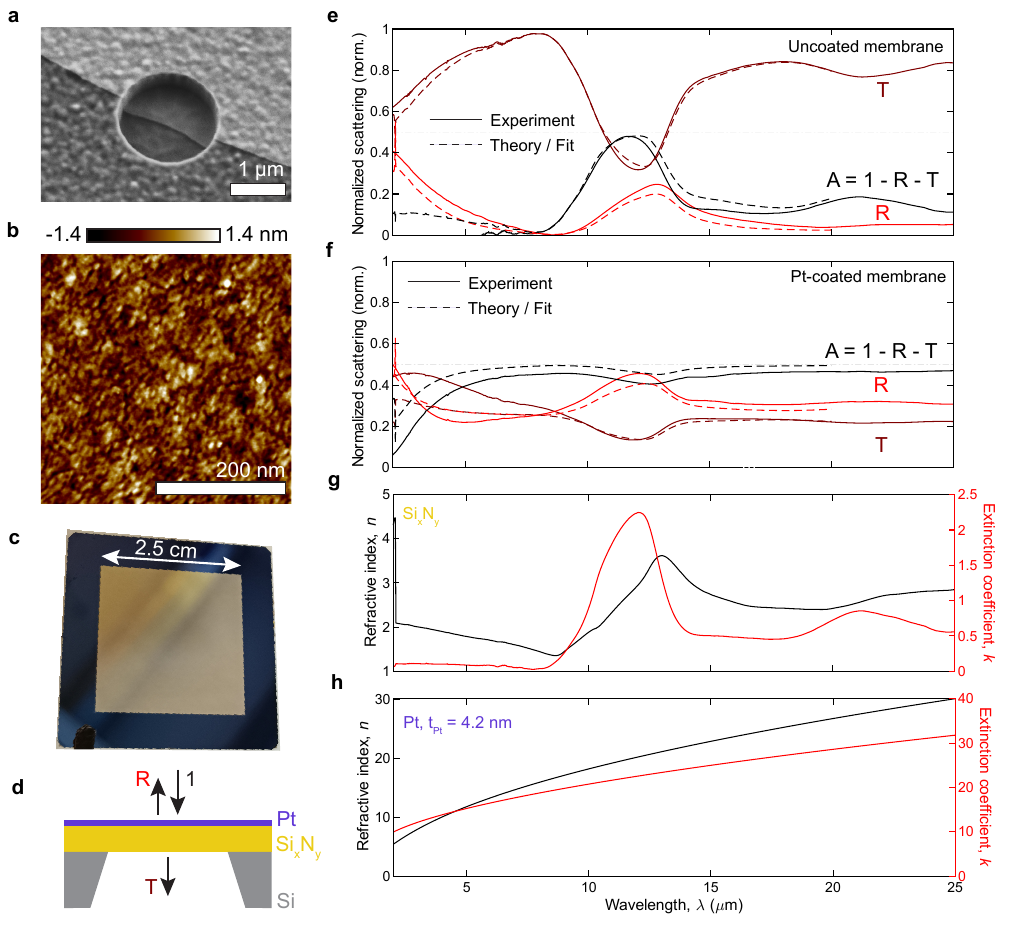}
        \caption{\textbf{Impedance-matched ultrathin Pt films for broadband mid-infrared absorption.} \textbf{a}, SEM image of the edge of a 4.2 nm thick Pt disk deposited over a patterned Si$_x$N$_y$ layer. \textbf{b}, AFM scan of a 4.5 nm thick Pt film, from the thickness series of the sheet-resistance study, deposited over a 205 nm thick Si$_x$N$_y$ film. \textbf{c}, Image of a Pt-coated Si$_x$N$_y$ $2.5 \times 2.5 \,\text{cm}^2$ square membrane. \textbf{d}, Schematic of the reflection-transmission FTIR measurement configuration. \textbf{e}, Reflectance ($R$), transmittance ($T$) and absorptance ($A$) spectra of the bare Si$_x$N$_y$ membrane. Solid lines are measurements and dashed lines are obtained from a transfer matrix model with fit parameters the refractive index ($n$) and extinction coefficient ($k$) of Si$_x$N$_y$. \textbf{f}, Same as \textbf{e} for the Pt-coated membrane with ($n$,$k$) for the Si$_x$N$_y$ layer as in \textbf{e}. \textbf{g}, \textbf{h}, Inferred optical constants for the Si$_x$N$_y$ and Pt layers, respectively. In \textbf{e} and \textbf{f}, $R$ and $T$ are normalized to the reference measurement without sample, and the dash-dotted line marks the 50\% maximum absorptance of a free-standing, impedance-matched deep-subwavelength film.}
    \label{fig:figRefMembranes}
\end{figure}

The Si$_x$N$_y$ trampoline resonators used as MTDs are coated with a 4.2 nm thick Pt disk. In all structurally characterized devices, the Pt film appears continuous, with no pinholes visible in SEM images or AFM scans, and its topography follows that of the surface it is deposited on, as shown by the SEM image in Fig.~\ref{fig:figRefMembranes}(a), in which deposition both over a Si$_x$N$_y$ layer and inside a trench are shown. The Si$_x$N$_y$ layer in that SEM is defective and contains an anomalous number of dome-shaped features that show up as effective roughness. An AFM scan of a 4.5\,nm thick film from the same deposition series, on an unprocessed Si$_x$N$_y$ surface (Fig.~\ref{fig:figRefMembranes}(b)), evidences an rms roughness of $\sigma_z$ = 0.384 nm, which is close to that of the Si$_x$N$_y$ surface prior to the Pt deposition ($\sigma_z$ = 0.308 nm). The Pt film exhibits a sheet resistance of $R_s = 185 \pm 10\,\Omega$ (Fig.~3(a) in the main text), with the uncertainty from wafer-scale inhomogeneities. This is close to the impedance-matching condition ($R_s = Z_0/2$) that leads to a theoretical maximum absorptance of 50\% for a free-standing deep-subwavelength metal film, or one on a dielectric layer much thinner than the wavelength~\cite{hadley_reflection_1947}. To verify this condition in the absence of the complex layer stack of the MTDs, we perform Fourier-transform infrared (FTIR) spectroscopy on a Pt-coated and an uncoated Si$_x$N$_y$ square membrane without backing substrate. The membranes are fabricated by LPCVD of a 205 nm thick low-stress Si$_x$N$_y$ layer on a silicon wafer, followed by through-wafer etching of the silicon substrate from the backside using a KOH release process~\cite{bereyhi_hierarchical_2022}. The Si$_x$N$_y$ deposition conditions are identical to those used for the MTDs. For the coated membrane, the Pt layer is deposited on the front side of the membrane after the release process. An image of the Pt-coated membrane after release is shown in Figure~\ref{fig:figRefMembranes}(c). The size of the membrane is $2.5 \times 2.5 \,\text{cm}^2$ to accommodate the full collimated beam of the FTIR spectrometer (Bruker Vertex 70) and therefore minimize beam truncation effects.

The fraction of power transmitted, $T$, and reflected, $R$, are first measured at normal incidence (see schematic in Fig.~\ref{fig:figRefMembranes}(d)), and the absorptance is computed as $A = 1-R-T$, with all three quantities shown for the uncoated and Pt-coated membranes respectively in Figs.~\ref{fig:figRefMembranes}(e) and (f). The uncoated membrane absorbs weakly except in a broad band peaking at \SI{11.6}{\micro\meter} with a maximum absorptance of 48\%, which is attributed to a TO-like Si–N asymmetric stretching phonon mode. In contrast, the Pt-coated membrane exhibits an absorptance of about 40\% or more from \SI{5}{\micro\meter} to \SI{25}{\micro\meter}, approaching the aforementioned 50\% theoretical maximum.  

We model both membranes with a transfer-matrix description of the multilayer stack~\cite{cheng_field_2013}, which requires the optical constants of the Si$_x$N$_y$ and of the Pt film. For the Si$_x$N$_y$, the refractive index $n$ and extinction coefficient $k$ are retrieved from the reflectance and transmittance of the bare membrane, independently at each wavelength and without imposing Kramers–Kronig consistency, as in prior work on ultrathin absorbers~\cite{luhmann_ultrathin_2020}. Normal-incidence data alone leave $k$ poorly constrained between \SI{10}{\micro\meter} and \SI{14}{\micro\meter}, so we add reflectance measurements at angles of incidence from 15$^\circ$ to 75$^\circ$ in steps of 15$^\circ$ to the fit. This makes the extracted $k$ more robust in that range and brings it closer to previously published values~\cite{kischkat_mid-infrared_2012}. The resulting $n$ and $k$ are shown in Fig.~\ref{fig:figRefMembranes}(g).

For the Pt film, we use a Drude permittivity, $\varepsilon(\omega) = \varepsilon_\infty - \omega_\text{p}^2/(\omega^2 + i\gamma\omega)$, with $\varepsilon_\infty = 1$ since interband transitions are neglected, and the plasma frequency of bulk Pt, $\hbar\omega_\text{p} = 9.59$\,eV ($\omega_\text{p}/2\pi \approx 2.3$\,PHz)~\cite{rakic_optical_1998}. The damping rate is set by the DC resistivity, $\gamma = \varepsilon_0\omega_\text{p}^2\rho_\mathrm{e}$, which for ultrathin films depends on the thickness $t$ as
\begin{equation}
    \rho_\mathrm{e}(t) = \rho_{\mathrm{e},0} + \frac{\beta}{t - t_0},
\end{equation}
as justified in Ref.~\cite{luhmann_ultrathin_2020}. A fit of $R_\mathrm{s} = \rho_\mathrm{e}/t$ to the thickness series of main text Fig.~3(a) gives $\rho_{\mathrm{e},0}=\SI{1.517e-7}{\ohm\meter}$, $\beta=\SI{1.396e-15}{\ohm \meter \squared}$ and $t_0=\SI{1.978e-9}{\meter}$. For $t = 4.2$\,nm, $\rho_\mathrm{e} = 7.8\times10^{-7}$\,\si{\ohm\meter}, about five times the value $\rho_{\mathrm{e},0}$ of thick films, so $\gamma$ is increased by the same factor. The resulting optical constants of the Pt film are shown in Fig.~\ref{fig:figRefMembranes}(h).

With these optical constants, the model reproduces the reflectance, transmittance and absorptance of both membranes (dashed lines in Fig.~\ref{fig:figRefMembranes}(e,f)). For the Pt-coated membrane no parameter is adjusted, since the Si$_x$N$_y$ constants are fixed by the bare membrane and those of Pt by the DC resistivity. The agreement holds even below about 3\,\si{\micro\meter}, where neglecting the interband transitions makes the Drude model least reliable. The same optical constants are used in Section~\ref{subsec:FTIR_microscopy} to compute the absorptance of each layer in the full device stack on the PIC.

\subsection{Fourier transform infrared microscopy}
\label{subsec:FTIR_microscopy}

\begin{figure}[!htbp]
    \centering
    \includegraphics[width=\textwidth]{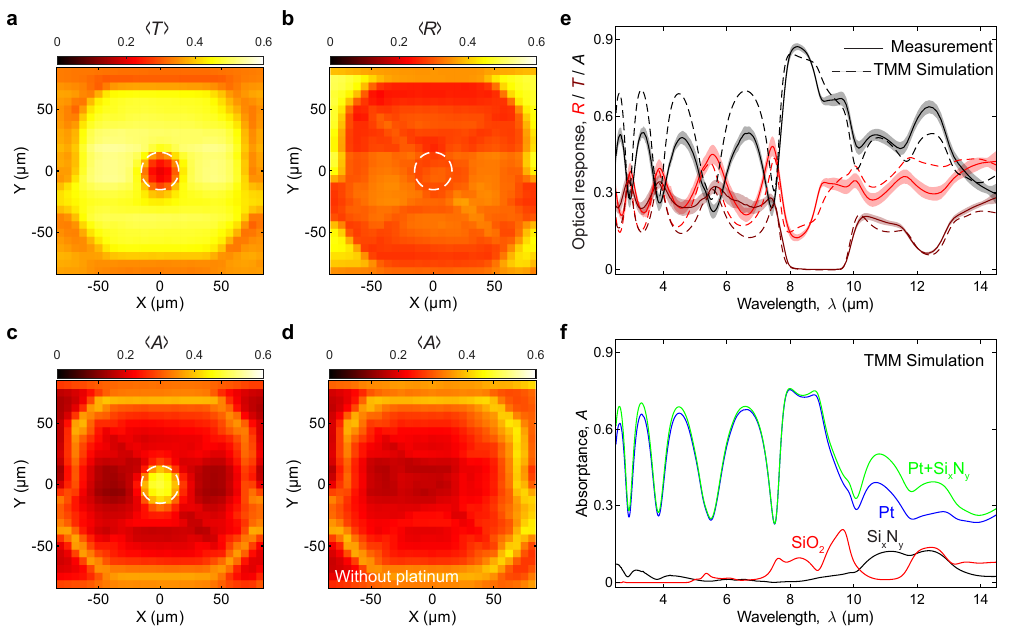}
    \caption{\textbf{FTIR microscopy of the OMTD.} \textbf{a}--\textbf{c}, Transmittance $\langle T\rangle$, reflectance $\langle R\rangle$ and absorptance $\langle A\rangle = 1-\langle R\rangle-\langle T\rangle$ maps of a device with Pt disk, averaged over $\lambda\in[2.5,14.5]$\,\si{\micro\meter}, on the native pixel grid of the measurement (6.25\,\si{\micro\meter}) and centred on the pixel of largest absorptance. The dashed circle marks the Pt disk (radius 15.2\,\si{\micro\meter}). \textbf{d}, Same as \textbf{c} for a device without Pt on the same chip. \textbf{e}, $R$ (red), $T$ (dark red) and $A$ (black) spectra at the centre pixel (solid lines, mean over eight devices, shading: $\pm2$ standard deviations) and normal-incidence transfer-matrix (TMM) simulation of the full stack (dashed lines). \textbf{f}, Simulated absorptance of each layer: Pt, Si$_x$N$_y$, their sum (resonator) and SiO$_2$ cladding.}
    \label{fig:FTIR_detailed}
\end{figure}

The absorptance of the OMTD shown in Fig.~3(b,c) of the main text is obtained from transmittance and reflectance hyperspectral maps, acquired separately by FTIR microscopy (Methods), as $A = 1-R-T$. Figure~\ref{fig:FTIR_detailed}(a--c) shows the three maps, averaged over $\lambda\in[2.5,14.5]$\,\si{\micro\meter}, on the native pixel grid and centred on the pixel of largest absorptance, which coincides with the Pt disk. Since $R$ and $T$ are separate scans, we check their registration from the centres of point symmetry of the two maps, which agree to within 1\,\si{\micro\meter}, about a sixth of a pixel. The Pt disk lowers the transmittance and raises the absorptance, which reaches a wavelength-averaged value $\langle A\rangle\approx 0.49$ at its centre, whereas the Pt-free pad and tethers show only weak contrast relative to their surroundings. A device without Pt on the same chip shows no central feature (Fig.~\ref{fig:FTIR_detailed}(d)). Both devices show an apparent increase of the absorptance along the edges of the released region (Fig.~\ref{fig:FTIR_detailed}(c,d)), which originates from a decrease of the measured reflectance there (Fig.~\ref{fig:FTIR_detailed}(b)). This is most likely because light reflects where the film is locally tilted, from stress release or slight buckling, and escapes the collection aperture of the objective. This missing reflected light is then counted as absorption in $A = 1-R-T$. Since it appears with and without Pt, it is unrelated to the absorber.

Figure~\ref{fig:FTIR_detailed}(e) shows the $R$, $T$ and $A$ spectra at the centre pixel, averaged over the 8 devices with Pt on this chip, including the one in Fig.~\ref{fig:FTIR_detailed}(a--c). This chip has the same nominal parameters as that of the detector characterized in the main text and was released at the same time. The spectra are compared with a normal-incidence transfer-matrix simulation of the full stack, air/Pt (4.2\,nm)/Si$_x$N$_y$ (140\,nm)/air gap (465\,nm)/SiO$_2$ (3.64\,\si{\micro\meter})/Si, with the dielectric layer and air-gap thicknesses extracted from UV--Vis reflectometry measurements and the nominal Pt thickness of the deposition, built from the optical constants of Pt and Si$_x$N$_y$ of Section~\ref{subsec:FTIR_membranes} and literature data for SiO$_2$~\cite{franta_optical_2016} and Si~\cite{franta_temperature-dependent_2017}. The SiO$_2$ thickness is that of the full cladding measured on the finished chip, after CMP, rather than the 4\,\si{\micro\meter} of thermal oxide grown at the start of the process (Methods of the main text). The simulation reproduces the slowly varying response above $\lambda\approx 8$\,\si{\micro\meter} and the fringes below it. These are Fabry--P\'erot interferences in the multilayer stack, whose spacing is set mainly by the thickness of the SiO$_2$ cladding. The measured fringes have a lower contrast than the simulated ones, which we attribute to scattering by the Si$_3$N$_4$ filler pattern in the cladding~\cite{pfeiffer_photonic_2016} and to the range of incidence angles of the reflective objective, neither of which is included in the model. Between about 8 and 10\,\si{\micro\meter}, the strong dispersion of SiO$_2$ near its Si--O--Si phonon band~\cite{kitamura_optical_2007} makes the cladding opaque ($T\approx0$) and reflective. Interference with the wave reflected by the cladding then raises the absorption in the Pt film above the 50\% bound of a free-standing film. The layer-resolved simulation (Fig.~\ref{fig:FTIR_detailed}(f)) separates the contributions. The Pt film dominates throughout, the Si$_x$N$_y$ adds a weak band around 11--12\,\si{\micro\meter}, and the SiO$_2$ absorbs at most about 0.2, near 9.6\,\si{\micro\meter}, a small contribution that should have only a minor thermo-optic effect on the microring. Subtracting the simulated SiO$_2$ absorptance from the measured $A$ gives the absorptance of the resonator (Pt+Si$_x$N$_y$), whose average over the same wavelength range is $0.441\pm0.010$ (standard deviation over the 8 devices, main text).

\begin{figure}[!htbp]
    \centering
    \includegraphics[width=\textwidth]{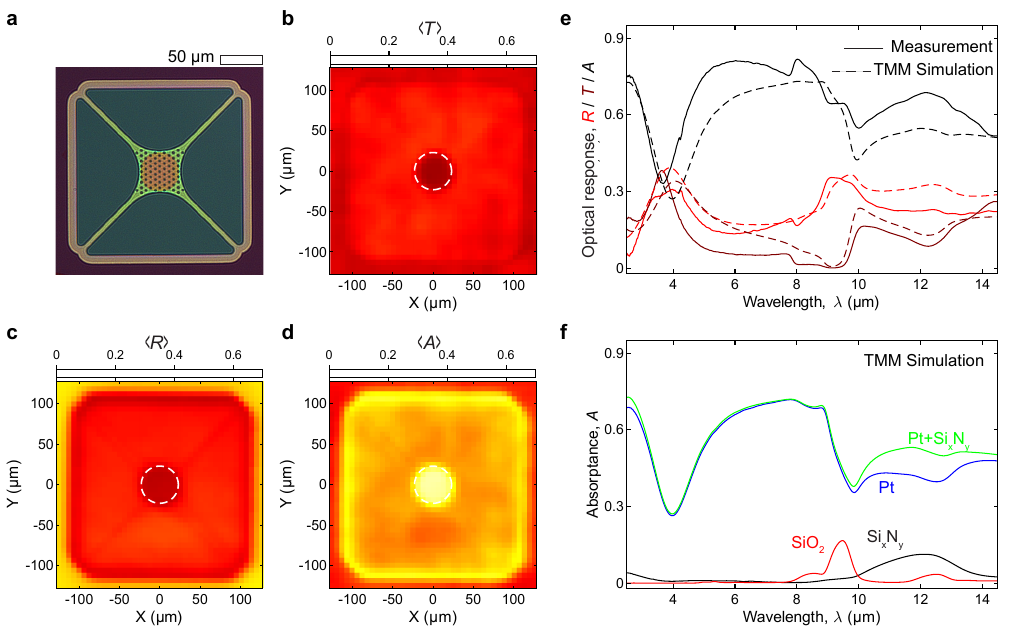}
    \caption{\textbf{FTIR microscopy of a device on a thin SiO$_2$ layer without PIC.} \textbf{a}, Optical microscope image of the device. \textbf{b}--\textbf{d}, Transmittance $\langle T\rangle$, reflectance $\langle R\rangle$ and absorptance $\langle A\rangle = 1-\langle R\rangle-\langle T\rangle$ maps averaged over $\lambda\in[2.5,14.5]$\,\si{\micro\meter}, on the native grid of the transmittance scan (6.25\,\si{\micro\meter}), with the reflectance scan interpolated onto it after registration, and centred on the pixel nearest to the centre of the transmission dip. The dashed circle marks the Pt disk (diameter 46\,\si{\micro\meter}). \textbf{e}, $R$, $T$ and $A$ spectra at the centre pixel (solid lines) and normal-incidence TMM simulation of the stack (dashed lines). \textbf{f}, Simulated absorptance of each layer: Pt, Si$_x$N$_y$, their sum (resonator) and SiO$_2$.}
    \label{fig:FTIR_thinOxide}
\end{figure}

The role of the SiO$_2$ cladding is seen by comparison with a device fabricated directly on a thin SiO$_2$ layer on silicon, without PIC (Fig.~\ref{fig:FTIR_thinOxide}(a)). This device has a larger trampoline window ($L = 200$\,\si{\micro\meter}) with a perforated pad, like the devices on the PIC, a Pt disk of 46\,\si{\micro\meter} diameter and the same layer sequence above the SiO$_2$. Here the reflectance and transmittance scans are acquired on grids offset by a fraction of a pixel, so the reflectance map is interpolated onto the transmittance grid after registering the two maps by their centres of point symmetry. As for the device on the PIC, the Pt disk produces a clear absorptance enhancement, the Pt-free pad and tethers show only weak contrast relative to their surroundings, and an apparent absorptance appears along the edges of the released region (Fig.~\ref{fig:FTIR_thinOxide}(b--d)).

Figure~\ref{fig:FTIR_thinOxide}(e) compares the spectra at the centre pixel with a TMM simulation of the stack air/Pt (4.2\,nm)/Si$_x$N$_y$ (82\,nm)/air gap (450\,nm)/SiO$_2$ (0.98\,\si{\micro\meter})/Si, with the dielectric layer and air-gap thicknesses again extracted from UV--Vis reflectometry. With an SiO$_2$ layer almost four times thinner, the dense Fabry--P\'erot fringes of the device on the PIC are replaced by a single broad minimum of the absorptance near 4\,\si{\micro\meter}, which confirms that the fringe spacing is set by the cladding thickness. The contrast of this minimum agrees better with the simulation, which is consistent with the absence of a filler pattern in this layer. Above 5\,\si{\micro\meter}, the measured absorptance exceeds the simulated one by about 0.1, mostly because the measured transmittance is lower. We attribute this mainly to a lower collection efficiency on this sample. The open window around the trampoline, where only the SiO$_2$ layer on Si remains, also shows an apparent absorptance of about 0.45--0.5 (Fig.~\ref{fig:FTIR_thinOxide}(d)), far above what this thin layer can absorb, which indicates that part of the light is not collected by the objective and is counted as absorption in $A = 1-R-T$.
In the layer-resolved simulation (Fig.~\ref{fig:FTIR_thinOxide}(f)), the Pt film again dominates, with an absorption of up to about 0.7 that exceeds the 50\% bound of a free-standing film because of interference with the reflection from the layers below. The Si$_x$N$_y$ contributes near 11--12\,\si{\micro\meter}, and the SiO$_2$ absorbs at most about 0.17, near 9.5\,\si{\micro\meter}. Subtracting the simulated SiO$_2$ absorptance from the measured $A$ gives a resonator absorptance averaging 0.64 over the same range, an upper bound given the reduced collection, against 0.55 in the simulation. Both exceed the value of 0.44 obtained for the device on the PIC, so the broadband absorptance of the absorber is sensitive to the layers below the resonator, which offers a handle to increase it in future devices.

\subsection{Calibration of the incident radiation on the device from the emission of a thermal source}
\label{subsec:thermal_source_calibration}

\begin{figure}[!htbp]
    \centering
    \includegraphics[width=\textwidth]{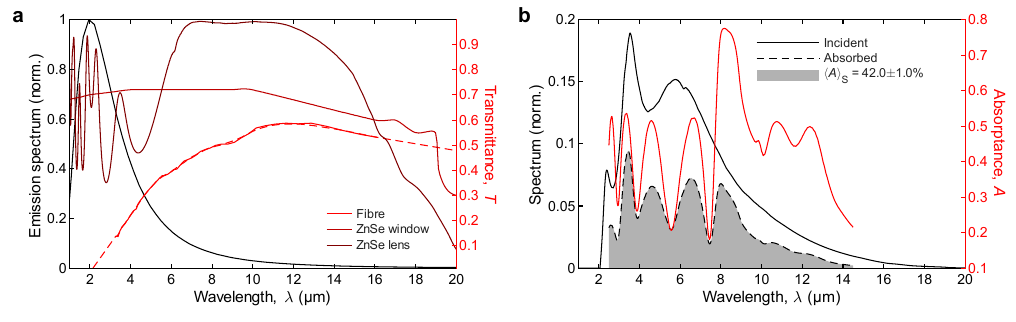}
    \caption{\textbf{Spectrum of the fibre-coupled thermal source incident on and absorbed by the device.} \textbf{a}, Emission spectrum of the SiC globar ($T = 1550$\,K), normalized to its maximum (black, left axis), and transmittance of the optical elements in the beam path (red, right axis): polycrystalline fibre (FIB-PIR-900-100, dashed where extrapolated), ZnSe vacuum window ($t = 2$\,mm) and ZnSe focusing lens (AL72550-E3, $t = 5$\,mm). \textbf{b}, Spectrum incident on the device, normalized to unit integral (solid black, left axis), spectrum absorbed by the resonator (dashed black, shaded), and resonator absorptance over its measured range (red, right axis).}
    \label{fig:FigS_IncidentSpectrumThermalSource}
\end{figure}

To determine the power incident on the MTD and the fraction it absorbs, we characterize the full optical chain from the thermal source to the device plane (Methods of the main text). The SiC globar ($T = 1550$\,K) is coupled into a polycrystalline infrared fibre (FIB-PIR-900-100, 900\,\si{\micro\meter} core), whose output is imaged onto the chip by a ZnSe aspheric lens (AL72550-E3, $f = 50$\,mm, clear aperture 23\,mm, $t = 5$\,mm) in a $2f$--$2f$ ($1\times$) relay, through a ZnSe vacuum viewport (VPZL-450UZ, $t = 2$\,mm). Figure~\ref{fig:FigS_IncidentSpectrumThermalSource}(a) shows the normalized emission spectrum of the globar together with the vendor-provided transmittance spectra of these three elements, the fibre transmittance being extrapolated where dashed.

The globar emission peaks near $\lambda \approx 2$\,\si{\micro\meter} and falls steeply at longer wavelengths, whereas the fibre transmits only about 25\% at 4\,\si{\micro\meter} and reaches its maximum of about 58\% only around 10--12\,\si{\micro\meter}, where the source emits a few percent of its peak spectral power. The chain is therefore spectrally mismatched to the source, and the power delivered to the device plane is well below what a better-matched fibre and source would give. The lens is anti-reflection coated for 7--12\,\si{\micro\meter}, where it transmits about 0.97, but below about 5\,\si{\micro\meter} its transmittance oscillates between about 0.35 and 0.95 because of interference in the coating. Since this is where the globar emits most, these oscillations are imprinted on the incident spectrum (Fig.~\ref{fig:FigS_IncidentSpectrumThermalSource}(b)), producing its peaks near 2.5 and 3.3\,\si{\micro\meter} and its dip around 4\,\si{\micro\meter}. The uncoated viewport transmits about 0.7 almost uniformly, as expected from the reflection at its two ZnSe surfaces.

Figure~\ref{fig:FigS_IncidentSpectrumThermalSource}(b) shows the resulting spectrum incident on the MTD (solid black line) and the spectrum absorbed by the resonator (dashed black line, shaded area), the product of the incident spectrum and the resonator absorptance (red, right axis). These are the two spectra shown in the inset of Fig.~3(d) of the main text. About 90\% of the incident power lies between 3 and 12\,\si{\micro\meter}. The absorptance is obtained from the FTIR reflectance and transmittance of the resonator (Fig.~3(c) of the main text and Section~\ref{subsec:FTIR_microscopy}), from which the simulated absorption of the SiO$_2$ cladding is subtracted. Its Fabry--P\'erot fringes, set by the cladding thickness (Section~\ref{subsec:FTIR_microscopy}), are narrow compared with the incident spectrum, which spans several of them. Integrating over the incident spectrum therefore gives an effective absorptance, $\langle A \rangle_\text{S} = 0.420 \pm 0.010$, close to the wavelength-averaged value of 0.44, where the uncertainty is the standard deviation of the values obtained with the absorptance of each of the 8 devices. The 3.9\% of the incident power that falls outside the measured range of the absorptance, $\lambda\in[2.5,14.5]$\,\si{\micro\meter}, is counted as incident but not absorbed.

Finally, the total power at the device plane, $P_\text{MIR}$, is measured with a thermal power sensor (S401C). The focused fibre output is much larger than the absorber and is well described by a smoothed top-hat profile of diameter $D_\text{MIR} = 738.1$\,\si{\micro\meter} (Fig.~3(e) of the main text), so that the Pt disk (radius $R_\text{Pt} = 15.2$\,\si{\micro\meter}) receives the fraction of $P_\text{MIR}$ obtained by integrating the fitted profile over its area, $1.63\times10^{-3}$, slightly below the sharp top-hat value $(2R_\text{Pt}/D_\text{MIR})^2\approx 1.7\times10^{-3}$ because the smoothed edge spreads part of the power beyond $D_\text{MIR}$. Dividing the responsivity to the total power, $\mathcal{R}_0^\text{inc,tot}$ (Fig.~1(i) of the main text), by this fraction gives the responsivity to the power incident on the disk, $\mathcal{R}_0^\text{inc}$, and dividing it further by $\langle A \rangle_\text{S}$ gives the responsivity to absorbed power, $\mathcal{R}_0^\text{abs}$, reported in the main text.

\subsection{Imaging of a resistively heated source}
\label{subsec:resistor}

\begin{figure}[!htbp]
    \centering
    \includegraphics[width=\textwidth]{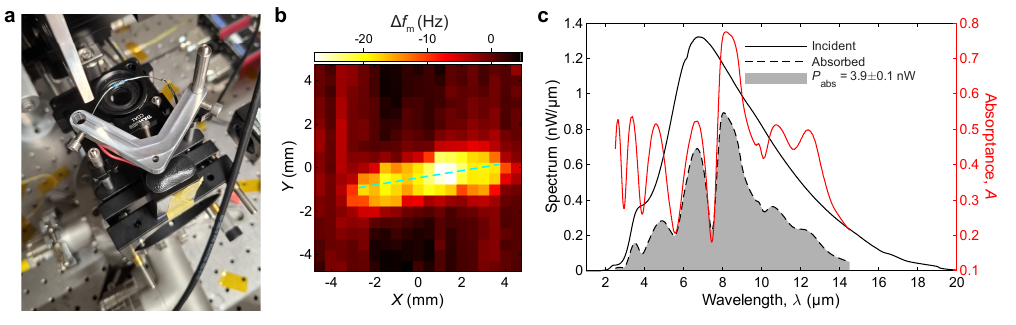}
    \caption{\textbf{Imaging of a resistively heated source.} \textbf{a}, Image of the $56~\Omega$ resistor above the vacuum chamber. \textbf{b}, Map of the steady-state frequency shift $\Delta f_\mathrm{m}$ obtained by raster-scanning the resistor across the object plane of the $2f$--$2f$ relay, after subtraction of a plane fitted to the background. The dashed line is the mid-line of the resistor, along which the shift is averaged. \textbf{c}, Estimated incident (solid black) and absorbed (dashed black, shaded) spectral power on the Pt disk for unit emissivity, and resonator absorptance over its measured range (red, right axis).}
    \label{fig:FigS_Resistor}
\end{figure}

We image a resistively heated element onto the detector with the same $2f$--$2f$ ($1\times$) relay, lens and viewport as in Section~\ref{subsec:thermal_source_calibration}, without the fibre. The source is a $56~\Omega$ metal-film resistor driven by a $9~\mathrm{V}$ battery (Fig.~\ref{fig:FigS_Resistor}(a)). The voltage across it during operation, $7.49~\mathrm{V}$, corresponds to $P_\mathrm{el}=V^2/R \approx 1.00~\mathrm{W}$, its maximum power rating, and its surface temperature, measured with a handheld thermal camera, is \SI{170}{\celsius} ($T\approx 443~\mathrm{K}$). Raster-scanning the resistor across the object plane gives the map of the steady-state frequency shift in Fig.~\ref{fig:FigS_Resistor}(b), from which a plane fitted to the background, excluding the resistor, is subtracted to remove a slow drift of the mechanical frequency during the scan. The map reproduces the elongated body of the resistor to scale. To evaluate the shift without selecting extreme values of the noisy map (background noise 2.3\,Hz per pixel), we fit a straight line through the positions of the largest shift in each column crossed by the resistor and average the map along this mid-line (dashed line in Fig.~\ref{fig:FigS_Resistor}(b)). This gives $\Delta f_\mathrm{m} = -20.0\pm1.0$\,Hz with respect to the background, the shift varying from about $-16$\,Hz near the ends of the resistor to $-27$\,Hz at its centre, which is presumably hotter.

To estimate the expected shift, we model the resistor as a Lambertian graybody with emissivity $\varepsilon$, whose net spectral radiance with respect to the room-temperature background is $\varepsilon\,[B_\lambda(T)-B_\lambda(T_0)]$, with $T_0 = 298$\,K. In the $2f$--$2f$ relay, source and image are at $2f = 100$\,mm from the lens, so the collection solid angle is $\Omega\simeq\pi\mathrm{NA}^2$ with $\mathrm{NA}\approx D/4f \approx 0.114$. The power incident on the Pt disk follows by multiplying with $\pi R_\text{Pt}^2\,\Omega$ and with the transmittances of the lens and viewport (Fig.~\ref{fig:FigS_IncidentSpectrumThermalSource}(a)), and the absorbed power by weighting further with the resonator absorptance (Fig.~\ref{fig:FigS_Resistor}(c)). Over the measured range of the absorptance, which contains 95\% of the incident power (the remaining 5\%, mostly at longer wavelengths, is counted as incident but not absorbed), this gives $P_\mathrm{abs} = (3.9\pm0.1)\,\varepsilon$\,nW, the uncertainty reflecting the spread of the absorptance between devices. Emissivities near unity are typical for paint- and lacquer-like coatings in the long-wave infrared, with $\varepsilon\simeq 0.92$--$0.96$ for oil paints~\cite{bramson_infrared_1968}, so $P_\mathrm{abs}\approx 3.6$--$3.7$\,nW. With $|\mathcal{R}_0^\text{abs}| = (6.3\pm0.5)\times10^3$\,W$^{-1}$ (main text) and $f_\mathrm{m}\approx 552$\,kHz, the expected shift is $|\Delta f_\mathrm{m}| = f_\mathrm{m}\,|\mathcal{R}_0^\text{abs}|\,P_\mathrm{abs}\approx 12.5$--$13.0$\,Hz.

The measured shift is about 1.6 times this estimate, which we consider reasonable for such a simple model. Neither the emissivity, which cannot exceed unity and whose realistic range changes the estimate by only 4\%, nor the 8\% uncertainty of the responsivity accounts for the difference. The estimate depends more strongly on the surface temperature, which is uncertain for a millimetre-sized object read with a handheld thermal camera, and it neglects the absorption outside the Pt disk and the heating of the surrounding chip, which the image of the resistor, much larger than the calibration spot, illuminates over several square millimetres.

\subsection{Wavelength-dependent noise-equivalent power}
\label{subsec:NEPmono}

\begin{figure}[!htbp]
    \centering
    \includegraphics[width=\textwidth]{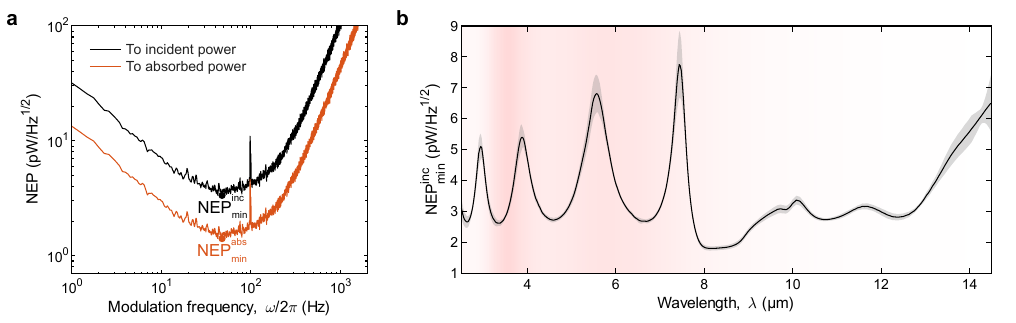}
    \caption{\textbf{Wavelength-dependent noise-equivalent power.} \textbf{a}, NEP to incident (black) and absorbed (orange) power as a function of the modulation frequency, reproduced from Fig.~3(j) of the main text. The dots mark the minima, $\text{NEP}^\text{inc}_\text{min}$ and $\text{NEP}^\text{abs}_\text{min}$. \textbf{b}, Minimum over $\omega$ of the NEP to incident power for monochromatic radiation (Eq.~(\ref{eq:NEPmono})), $\text{NEP}^\text{inc}_\text{min}(\lambda_\text{MIR})$, computed with the absorptance averaged over the 8 devices (black line). The grey band represents $\pm2$ standard deviations over these devices. It does not include the 6.8\% uncertainty of the responsivity calibration, which scales the whole curve. The red tint follows the spectrum incident on the device during the NEP measurement (Fig.~3(d) of the main text).}
    \label{fig:NEPmono}
\end{figure}

The incident-referred NEP reported in the main text is specific to the spectrum of the thermal source reaching the device (Fig.~3(d) of the main text). To characterize the detector independently of the source, we infer the NEP to incident power it would achieve for monochromatic radiation at wavelength $\lambda_\text{MIR}$ as
\begin{equation}
\text{NEP}^\text{inc}(\omega,\lambda_\text{MIR}) = \frac{\text{NEP}^\text{abs}(\omega)}{A(\lambda_\text{MIR})} = \text{NEP}^\text{inc}(\omega)\,\frac{\langle A\rangle_\text{S}}{A(\lambda_\text{MIR})},
\label{eq:NEPmono}
\end{equation}
where $\text{NEP}^\text{inc}(\omega)$ and $\text{NEP}^\text{abs}(\omega)$ are the NEP to incident and absorbed power for the thermal source (Fig.~3(j) of the main text and Fig.~\ref{fig:NEPmono}(a)), $A(\lambda_\text{MIR})$ is the absorptance of the resonator (Pt+Si$_x$N$_y$) of Section~\ref{subsec:FTIR_microscopy}, and $\langle A\rangle_\text{S}$ is its average over the incident spectrum (Section~\ref{subsec:thermal_source_calibration}). The absorptance enters only through the ratio $\langle A\rangle_\text{S}/A(\lambda_\text{MIR})$, so a common scaling of it cancels and only its spectral shape matters. The dependence on $\omega$, set by the frequency noise and the thermal transfer function, does not depend on the wavelength. The only wavelength dependence of the response itself could come from the size of the focused spot, which with conventional optics grows with $\lambda_\text{MIR}$, for instance through the diffraction limit. The finite-element simulations of Section~\ref{subsec:spot_size} show that both the responsivity and the thermal bandwidth depend only weakly on the spot size, so we neglect this effect. The wavelength dependence is therefore fully captured by the minimum over $\omega$, $\text{NEP}^\text{inc}_\text{min}(\lambda_\text{MIR}) = \text{NEP}^\text{abs}_\text{min}/A(\lambda_\text{MIR})$, shown in Fig.~\ref{fig:NEPmono}(b). We evaluate Eq.~(\ref{eq:NEPmono}) separately for each of the 8 devices of Section~\ref{subsec:FTIR_microscopy}, each with its own $\langle A\rangle_\text{S}$, and take their spread as the uncertainty arising from the absorptance (Fig.~\ref{fig:NEPmono}(b)). The resulting monochromatic NEP reaches $1.8\pm0.1$\,\si{\pico\watt/\sqrt{\hertz}} at $\lambda_\text{MIR}$ = 8.17\,\si{\micro\meter}, where the uncertainty (one standard deviation) combines the device-to-device spread and the 6.8\% uncertainty of the responsivity calibration, and its device-averaged value remains below 8\,\si{\pico\watt/\sqrt{\hertz}} across $\lambda_\text{MIR} \in [2.5,14.5]$\,\si{\micro\meter}.

\subsection{Fundamental thermal-fluctuation limit}
\label{subsec:NEPth}

For a given thermal conductance, the NEP of a thermal detector is ultimately limited by thermal-fluctuation noise. Random phonon and photon exchanges between the resonator and its environment drive temperature fluctuations, which appear as irreducible frequency noise through the thermal responsivity~\cite{zhang_enhanced_2025}. Referred to absorbed power, this limit is $\text{NEP}_\text{th} = \sqrt{4k_BT_0^2G_\text{th}}$~\cite{richards_bolometers_1994}, with $G_\text{th}$ the thermal conductance of the resonator to its surroundings and $T_0 = 298$\,K the ambient temperature. It is shown as the dash-dotted line in Fig.~3(j) of the main text.

We determine $G_\text{th}$ from the measured thermal cutoff rather than from the simulated conductance, since the cutoff constrains it more tightly than the literature values of $\kappa_\text{th}$ do (Section~\ref{subsec:thermal_bandwidth}). The heat capacity, which depends only on the geometry and on $\rho c_p$, is taken from the simulation at the central values of the material parameters, $C_\mathrm{th} = G_\mathrm{th}^\text{FEM}/(2\pi f_\text{th}^\text{FEM}) = 0.57$\,nJ\,K$^{-1}$. With the measured cutoff, $f_\text{th} = 47.4$\,Hz, this gives $G_\mathrm{th} = 2\pi f_\text{th} C_\mathrm{th} = 1.69\times10^{-7}$\,\si{\watt/\kelvin}, with an uncertainty of about 6\% set mainly by $\rho c_p$, and $\text{NEP}_\text{th} = 0.91\pm0.03$\,\si{\pico\watt/\sqrt{\hertz}}. Since it rests on the measured cutoff, this $G_\mathrm{th}$ also includes the radiative conductance that the simulations neglect.

This value is a lower bound on the thermal-fluctuation noise of the device. Fluctuations originate everywhere on the resonator, and the conductance that governs them is an average over all source positions~\cite{zhang_enhanced_2025}. Positions closer to the tethers dissipate faster, so this effective conductance, $G_\text{eff}$, is larger than the conductance $G_\text{th}$ that sets the response to the absorbed signal. At the minimum of the measured NEP to absorbed power, 1.4\,\si{\pico\watt/\sqrt{\hertz}}, thermal-fluctuation noise therefore accounts for about 40\% or more of the frequency-noise PSD, the rest coming mainly from the intrinsic frequency noise of the resonator (Section~\ref{subsec:NEP_pathways}).

The limit is not intrinsic to the detector, however, since $G_\text{th}$ is set mainly by conduction through the tethers. For vanishing tether width, the resonator exchanges heat with its surroundings only by radiation, and $G_\text{th}$ reduces to $G_\text{rad}=4\sigma_\text{SB} T_0^3\sum_i \varepsilon_i A_i$, with $\sigma_\text{SB}$ the Stefan--Boltzmann constant. The sum runs over the top face of the impedance-matched Pt disk and over the bare Si$_x$N$_y$ on both faces of the pad, each with emissivity equal to its absorptance by Kirchhoff's law (about 0.47 and 0.12, respectively). This gives $G_\text{rad}\approx 4\times 10^{-9}$\,\si{\watt/\kelvin}, about 2\% of the present $G_\text{th}$, and a radiation-limited floor $\text{NEP}_\text{rad}\approx 0.14$\,\si{\pico\watt/\sqrt{\hertz}}. Section~\ref{subsec:NEP_pathways} discusses how much reducing $G_\text{th}$ actually lowers the NEP in the presence of the intrinsic frequency noise and photothermal noise that we observe.

\subsection{Pathways to a lower noise-equivalent power}
\label{subsec:NEP_pathways}

The minimum NEP of the present device is set mainly by the intrinsic frequency noise of the resonator near the thermal cutoff, with thermal-fluctuation noise contributing about 40\% of the frequency-noise PSD (Section~\ref{subsec:NEPth}). Here we use the measured noise to estimate how far the NEP can be lowered by changing the design. Reducing the thermal conductance of the tethers is the most effective lever. It lowers the minimum NEP to absorbed power by a factor of about 2 before photothermal noise from the delivered readout light takes over, and by about 4 with the intensity noise of the laser alone. A thinner or less stressed film probably helps little, because the intrinsic $1/f$ noise is expected to grow with the responsivity. A better absorber stack would lower the NEP to incident power by up to a factor of about 1.5. Table~\ref{tab:NEP_pathways} summarizes these estimates.\\

\noindent\emph{Model:}

\noindent The responsivity to absorbed power is $\mathcal{R}^\text{abs}(\omega) = \mathcal{R}_T/(G_\text{th} - i\omega C_\text{th})$, with $\mathcal{R}_T = \Omega_\mathrm{m,0}^{-1}\,\partial\Omega_\mathrm{m}/\partial T$ the fractional temperature coefficient of the mechanical frequency, $G_\text{th}$ the thermal conductance and $C_\text{th}$ the heat capacity, which are $G_{\text{th},0} = 1.69\times10^{-7}$\,\si{\watt/\kelvin} and $C_\text{th} = 0.57$\,nJ/K for the present device (Section~\ref{subsec:NEPth}). We split the measured NEP to absorbed power of main text Fig.~3(j) into three contributions that respond differently to a change of design:
\begin{itemize}
\item[(i)] Frequency noise that does not depend on the tethers (intrinsic, thermomechanical and detection noise). Its NEP scales as $1/|\mathcal{R}^\text{abs}(\omega)|$. We obtain it as the measured NEP minus (ii) and (iii), subtracted in quadrature.
\item[(ii)] Photothermal noise from fluctuations of the absorbed readout power. These act on the resonator like the mid-infrared signal, so their NEP, $P_\text{abs}\sqrt{\mathrm{RIN}(\omega)}$, does not depend on $G_\text{th}$. We use the RIN of the delivered light (Section~\ref{subsec:power_fluctuations}) and $P_\text{abs}\approx 21$\,nW at the operating point of the NEP measurement (Section~\ref{subsec:kappa_abs}).
\item[(iii)] Thermal-fluctuation noise, $\text{NEP}_\text{th} = \sqrt{4k_BT_0^2G_\text{th}}$, which scales as $\sqrt{G_\text{th}}$.
\end{itemize}
The decomposition uses the measured spectrum averaged in logarithmic frequency bins, whose minimum is 1.5\,\si{\pico\watt/\sqrt{\hertz}} rather than the 1.4\,\si{\pico\watt/\sqrt{\hertz}} of the smoothed curve of main text Fig.~3(j). All changes below are quoted relative to 1.5\,\si{\pico\watt/\sqrt{\hertz}}.\\

\begin{figure}[!htbp]
    \centering
    \includegraphics[width=\textwidth]{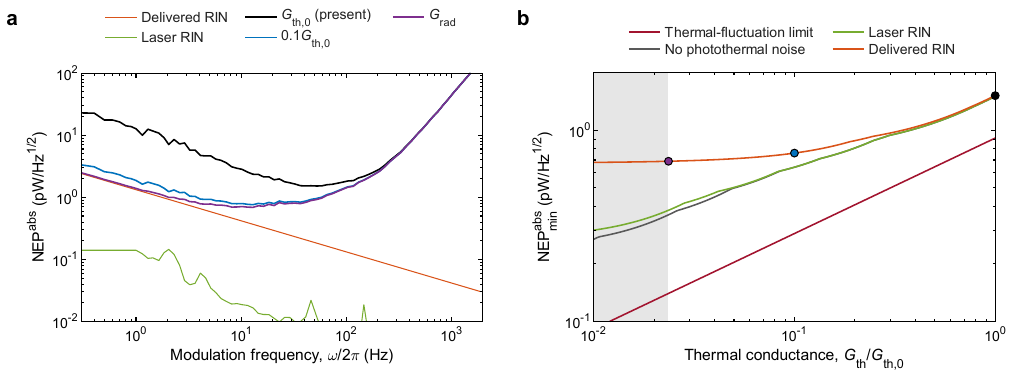}
    \caption{\textbf{Pathways to a lower NEP.} \textbf{a}, NEP to absorbed power for the present thermal conductance $G_{\text{th},0}$ (black, measured) and, from the model, for $G_\text{th} = 0.1\,G_{\text{th},0}$ (blue) and the radiative limit $G_\text{rad}$ (purple), with the photothermal (PT) noise of the delivered light. The orange and green lines show the photothermal noise for the RIN of the delivered light and of the laser, respectively. \textbf{b}, Minimum NEP to absorbed power over $\omega/2\pi\in[0.3,2000]$\,Hz versus $G_\text{th}/G_{\text{th},0}$ (grey: below the radiative limit, not reachable), with the photothermal noise of the delivered light (orange), with that of the laser (green) and without photothermal noise (grey), together with the thermal-fluctuation limit (dark red). The dots mark the three conductances of \textbf{a}.}
    \label{fig:NEP_pathways}
\end{figure}

\begin{table}[!htbp]
\centering
\renewcommand{\arraystretch}{1.25}
\begin{tabular*}{\textwidth}{@{\extracolsep{\fill}}l c l}
\toprule
\textbf{Change} & \textbf{NEP$^\text{abs}_\text{min}$ (pW/$\sqrt{\text{Hz}}$)} & \textbf{Comment}\\
\midrule
Present device & 1.5 & Measured\\
$G_\text{th} = 0.1\,G_{\text{th},0}$ & 0.76 (at 13\,Hz) & Photothermal floor\\
$G_\text{th} = G_\text{rad}$ & 0.69 & 0.55--0.85 for $P_\text{abs}/2$ to $2P_\text{abs}$\\
$G_\text{th} = G_\text{rad}$, laser RIN & 0.38 (at 1\,Hz) & Laser RIN only\\
$t/t_0 = 0.5$, 0.25, 0.1 & 1.10, 0.81, 0.58 & Volume fluctuators\\
 & $\approx 1.5$ & Surface fluctuators\\
$\sigma = 10$\,MPa & $\approx 1.5$ & Strain-coupled fluctuators\\
 & 0.93--0.97 & Other fluctuators\\
\bottomrule
\end{tabular*}
\caption{\textbf{Estimated minimum NEP to absorbed power for the levers of Section~\ref{subsec:NEP_pathways}.} All values follow from the decomposition into contributions (i)--(iii). Unless noted, the photothermal noise is that of the delivered light. $G_{\text{th},0}$: present thermal conductance, $G_\text{rad}$: radiative conductance, $t_0$: present film thickness, $\sigma$: tensile stress.}
\label{tab:NEP_pathways}
\end{table}

\noindent\emph{Thermal conductance:}

\noindent Narrower tethers lower $G_\text{th}$ while leaving $C_\text{th}$ essentially unchanged, since the tethers hold a small fraction of the heat capacity. The magnitude of the responsivity then increases by up to $G_{\text{th},0}/G_\text{th}$ well below the new cutoff and does not change well above the present one. Figure~\ref{fig:NEP_pathways}(a) shows the result for $G_\text{th} = 0.1\,G_{\text{th},0}$ and for the radiative limit, $G_\text{rad}\approx 4\times10^{-9}$\,\si{\watt/\kelvin} $\approx 0.024\,G_{\text{th},0}$ (Section~\ref{subsec:NEPth}). Above the present cutoff, all curves merge, since there the NEP is set by the heat capacity rather than by $G_\text{th}$. Below it, the NEP decreases, already by a factor of about 1.6 near the present optimum of $\approx 50$\,Hz, until the photothermal noise of the delivered light takes over at lower frequencies. The minimum NEP thus halves for $G_\text{th} = 0.1\,G_{\text{th},0}$ but barely improves further at the radiative limit (Fig.~\ref{fig:NEP_pathways}(b)). This floor scales with $P_\text{abs}$, which we estimate only to within a factor of about 2, since the large mechanical amplitude modulates the detuning. The thermal-fluctuation noise stays below the tether-independent frequency noise at all frequencies, and remains so as $G_\text{th}$ decreases, so it never sets the minimum. This pathway comes at a cost and rests on assumptions, so the gains above are upper estimates:
\begin{itemize}
\item The detector becomes slower, with $\omega_\text{th}/2\pi$ moving from 47\,Hz to 4.7\,Hz at $0.1\,G_{\text{th},0}$ and to 1.1\,Hz at the radiative limit.
\item The static self-heating by the absorbed readout power, $P_\text{abs}/G_\text{th}$, grows from 0.13\,K to $\approx 5$\,K at the radiative limit, which may require a lower readout power.
\item We assume that the fractional frequency noise, $S_{\Omega\Omega}/\Omega_\mathrm{m,0}^2$, does not change. A change of $\Omega_\mathrm{m,0}$ alone does not affect the NEP, since the signal is also a fractional frequency shift. Narrower tethers, however, reduce the strain-weighted volume of the mode, whose strain lies mostly in the tethers and near the clamps, and may thereby raise the intrinsic noise~\cite{roukes_1f_2026}.
\item Narrower tethers also redistribute the stress, which changes $\mathcal{R}_T$ and the mechanical properties.\\
\end{itemize}

\noindent\emph{Readout photothermal noise:}

\noindent The photothermal floor can itself be lowered. With the RIN of the laser, which lies 26--28\,dB below that of the delivered light above 5\,Hz (Section~\ref{subsec:power_fluctuations}), the photothermal NEP drops by about an order of magnitude or more, and the minimum NEP keeps decreasing with $G_\text{th}$. At the radiative limit it reaches 0.38\,\si{\pico\watt/\sqrt{\hertz}} at about 1\,Hz, a factor of about 4 below the present device and still above the radiation-limited floor, $\text{NEP}_\text{rad}\approx 0.14$\,\si{\pico\watt/\sqrt{\hertz}}. This requires removing the fibre delivery, for instance with a laser integrated on the same chip. The absorbed readout power can also be reduced without increasing the detection noise, for instance with an opening in the absorber~\cite{martini_uncooled_2025} or a smaller overlap of the Pt with the evanescent field of the ring.\\

\noindent\emph{Film thickness:}

\noindent A thinner film lowers $G_\text{th}$ and $C_\text{th}$ together, since the film holds about 98\% of the heat capacity. The cutoff then stays essentially unchanged and $|\mathcal{R}^\text{abs}(\omega)|$ increases as $1/t$ at all frequencies. Whether the NEP follows depends on how the intrinsic $1/f$ noise scales with $t$. Its coefficient scales as the inverse of the strain-weighted volume of the mode over many decades of device size~\cite{roukes_1f_2026}, and in Si$_x$N$_y$ fluctuators at the surfaces dominate below a film thickness of about 100\,nm~\cite{kanellopulos_1f_2026}. If the fractional frequency-noise PSD grew only as $1/t$, as for fluctuators in the volume, the minimum NEP would decrease as $\sqrt{t}$, together with $\text{NEP}_\text{th}$. For fluctuators at the surfaces, it grows as $1/t^2$ and cancels the gain. Since our 100-nm film lies near the crossover, the first estimate is optimistic.\\

\noindent\emph{Stress:}

\noindent A lower tensile stress $\sigma$ raises the responsivity while leaving $G_\text{th}$, $C_\text{th}$ and the cutoff unchanged. Heating relieves the tensile stress of the tethers by about $E\alpha_\text{T}$ per kelvin, so that $|\mathcal{R}_T|\approx E\alpha_\text{T}/(2\sigma)$. With the parameters of Table~\ref{tab:material_parameters}, this gives $1.3\times10^{-3}$\,K$^{-1}$, close to the measured $|\mathcal{R}_T| = |\mathcal{R}_0^\text{abs}|\,G_{\text{th},0}\approx 1.1\times10^{-3}$\,K$^{-1}$. Bending stiffness becomes comparable to the tension only around 1\,MPa, so that $|\mathcal{R}_T|$ could be raised about 17-fold at 10\,MPa. The intrinsic noise of Si$_x$N$_y$, however, is dominated by strain-coupled fluctuators, which are diluted by the prestress as $1/\sigma^2$~\cite{kanellopulos_1f_2026}. They raise the rms fractional frequency noise by the same factor as the responsivity, so that no net gain is expected. Even if other fluctuators dominated, the NEP could not fall below the thermal-fluctuation limit of the present tethers. A lower stress would also lower $Q$, strengthen the Duffing nonlinearity and make the pad more prone to curling by the stress of the Pt film.\\

\noindent\emph{Absorptance:}

\noindent For the NEP to incident power, the absorptance offers a further factor. The absorbed fraction of the present device, $\langle A\rangle_\text{S} = 0.42$, is shaped by the SiO$_2$ cladding and the air gap below the resonator, and the device fabricated directly on a thin SiO$_2$ layer reaches a wavelength-averaged resonator absorptance of up to 0.64 (Section~\ref{subsec:FTIR_microscopy}). A layer stack designed as a resonant absorber, with a reflector at a suitable distance below the resonator, could raise it further over part of the mid-infrared, at the expense of bandwidth.

\section{System-level demonstration in a commercial FTIR}
\label{sec:FTIR_demo}

This section characterizes the settings of the commercial spectrometer used in the power sweep of main text Fig.~4(e), calibrates the power incident on both detectors, and probes the frequency response of the OMTD through the interferometer scan rate.

\begin{figure}[!htbp]
    \centering
    \includegraphics[width=\textwidth]{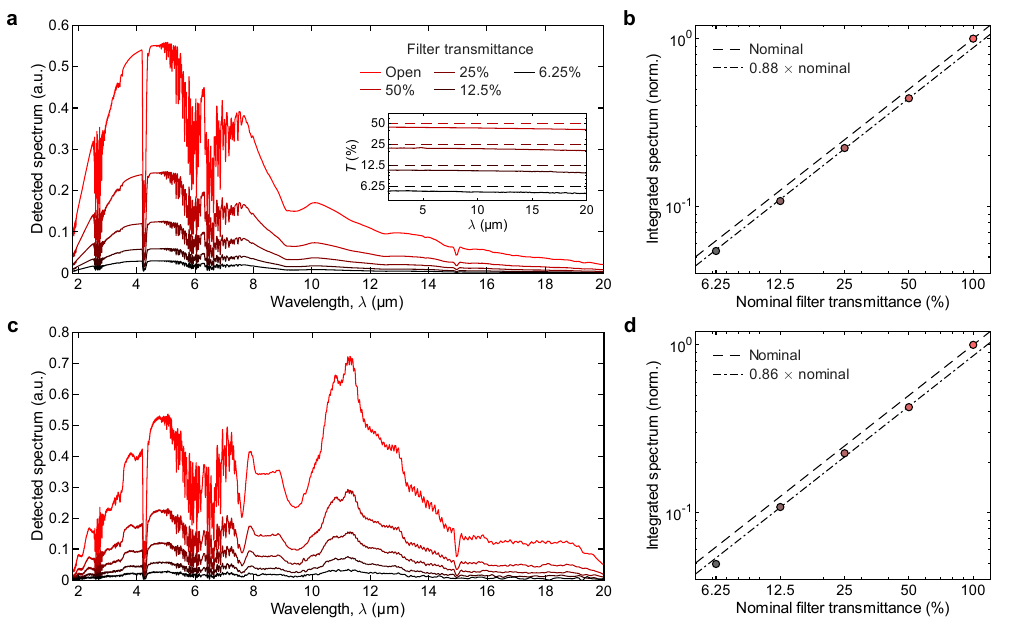}
    \caption{\textbf{Effect of the attenuation filters.} \textbf{a}, Spectra of the spectrometer's thermal source detected with the DLaTGS detector without sample for the filter settings indicated (aperture $D = 3$\,mm, scan rate 400\,Hz, resolution 4\,cm$^{-1}$), each averaged over 10 to 55 single-scan spectra. Inset: ratio of the filtered to the open spectra (solid) and nominal transmittance (dashed). \textbf{b}, Integrated spectrum over $\lambda\in[1.8,20]$\,\si{\micro\meter}, normalized to the open setting, as a function of the nominal filter transmittance. The dashed line is the nominal transmittance and the dash-dotted line a line through the origin fitted to the filtered settings ($0.88\times$ nominal). \textbf{c},\textbf{d}, Same as \textbf{a},\textbf{b} for the OMTD ($D = 4$\,mm, scan rate 1.6\,kHz, one spectrum per setting averaged over 4 scans), with a fitted slope of 0.86.}
    \label{fig:FTIR_filters}
\end{figure}

\subsection{Effect of the neutral-density filters}
\label{subsec:FTIR_filters}

The power sweep of main text Fig.~4(e) attenuates the beam with neutral-density filters of nominal transmittance 50\%, 25\%, 12.5\% and 6.25\%. Converting each setting into an incident power requires their actual transmittance and a linear response of the detectors, which we verify in Fig.~\ref{fig:FTIR_filters}. Figure~\ref{fig:FTIR_filters}(a) shows reference spectra detected with the DLaTGS detector (without any sample in the optical path), where a progressive drop in intensity is observed for decreasing filter transmittance. Figure~\ref{fig:FTIR_filters}(b) shows the integral of the spectra, normalized to the open setting, as a function of the nominal transmittance. The filtered settings lie on a line through the origin with a slope of 0.88. Successive filters attenuate by factors between 0.48 and 0.51, in agreement with their nominal ratio of 1/2, while all of them share an insertion loss of about 12\% with respect to the open position. The ratio of the filtered to the open spectra (inset of Fig.~\ref{fig:FTIR_filters}(a)) is flat to within about 1\% over $\lambda\in[4,14]$\,\si{\micro\meter} and decreases only slightly at longer wavelengths, so the attenuation is neutral. The same measurement with the OMTD (Fig.~\ref{fig:FTIR_filters}(c,d), $D = 4$\,mm) gives a slope of 0.86. Since two different detectors give the same factor, we attribute it to the filters, for instance to reflection at their substrates or to clipping of the beam by their mounts, rather than to a nonlinear detector response. Once this common insertion loss is accounted for, both detectors respond linearly over the 16-fold range of power covered by the open setting and the filters, and in the calibration of Section~\ref{subsec:FTIR_power} the transmittance of each filter is taken as $0.88\pm0.02$ times its nominal value, the uncertainty reflecting the spread between filters.

\begin{figure}[!htbp]
    \centering
    \includegraphics[width=\textwidth]{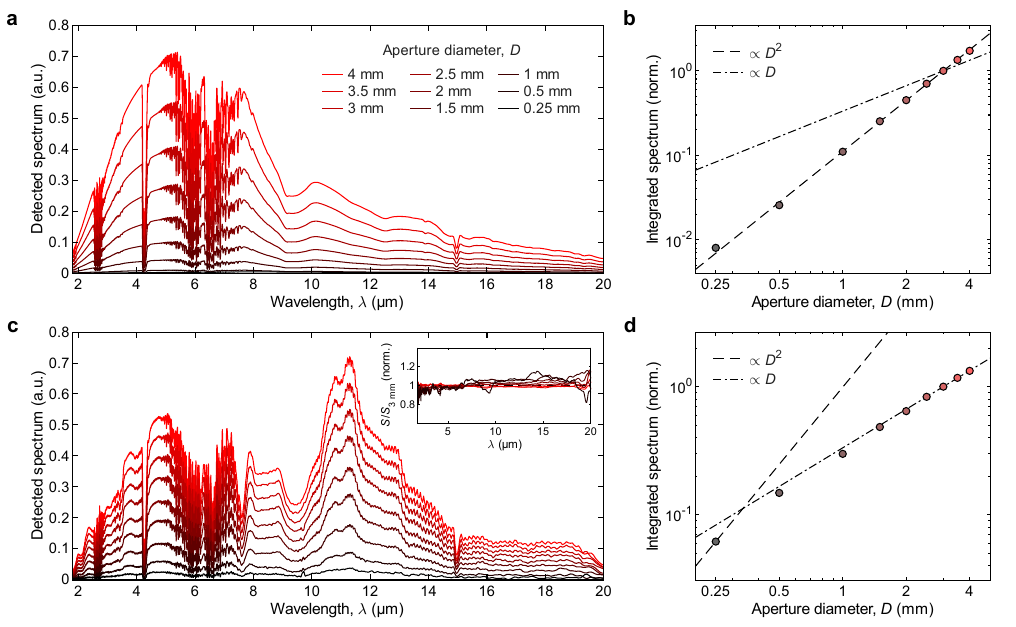}
    \caption{\textbf{Effect of the aperture diameter.} \textbf{a}, Spectra of the spectrometer's thermal source detected with the DLaTGS detector without sample for aperture diameters $D$ from 0.25 to 4\,mm (scan rate 1.6\,kHz, resolution 2\,cm$^{-1}$, one spectrum per aperture averaged over 4 scans). \textbf{b}, Integrated spectrum over $\lambda\in[1.8,20]$\,\si{\micro\meter}, normalized to its value at $D = 3$\,mm, as a function of $D$. The dashed and dash-dotted lines are proportional to $D^2$ and $D$, respectively, and pass through the point at $D = 3$\,mm. \textbf{c},\textbf{d}, Same as \textbf{a},\textbf{b} for the OMTD (resolution 4\,cm$^{-1}$). Inset of \textbf{c}: OMTD spectra for $D \geq 0.5$\,mm divided by the spectrum at $D = 3$\,mm and normalized to their mean over $\lambda\in[4,14]$\,\si{\micro\meter}. In \textbf{d}, the line proportional to $D^2$ passes through the point at $D = 0.25$\,mm.}
    \label{fig:FTIR_aperture}
\end{figure}

\subsection{Effect of the aperture diameter}
\label{subsec:FTIR_aperture}
The spectrometer aperture sets the size of the source image and thus the power in the beam. In the power sweep of main text Fig.~4(e) it is reduced from $D = 3$ to 0.5\,mm to extend the range of power on the DLaTGS detector beyond that of the filters. Figure~\ref{fig:FTIR_aperture}(a) shows the DLaTGS spectra for $D$ between 0.25 and 4\,mm, and Fig.~\ref{fig:FTIR_aperture}(b) their integral, normalized to $D = 3$\,mm. The integrated power closely follows the expected $D^2$ dependence (fitted exponent 1.97), confirming that the DLaTGS collects the full beam for all apertures. The ratio between $D = 0.5$ and 3\,mm is 0.0256, 8\% below $(0.5/3)^2 = 0.0278$, both in this sweep and in the reference scans of the power sweep. In Section~\ref{subsec:FTIR_power} we use the $D^2$ law and take this 8\% deviation as the uncertainty of the power at $D = 0.5$\,mm. For $D \geq 1.5$\,mm the spectral shape does not depend on the aperture to within 3\% over $\lambda\in[4,14]$\,\si{\micro\meter}. At smaller apertures the spectrum shifts towards longer wavelengths: at $D = 0.5$\,mm, the spectral density relative to that at 3\,mm is about 20\% lower near 3\,\si{\micro\meter} and 25\% higher near 10\,\si{\micro\meter}. The power-sweep points at $D = 0.5$\,mm therefore probe a somewhat different spectral distribution than those at 3\,mm.

For the OMTD (Fig.~\ref{fig:FTIR_aperture}(c,d)), the integrated signal instead scales approximately as $D$ (fitted exponent 1.10). The focused spot, of diameter comparable to $D$ (Methods), is much larger than the Pt disk, so the OMTD signal follows the irradiance at the disk rather than the total power in the beam. If the spot were a sharp image of the aperture, its area would scale as $D^2$, like the transmitted power, and this irradiance would not depend on $D$. The observed scaling therefore indicates that the spot size also contains a contribution that does not depend on $D$, for instance from aberrations of the focusing optics or a slight defocus. At the smallest aperture this contribution dominates, and the data bend towards the $D^2$ dependence expected for a spot of fixed size. In contrast with the DLaTGS, the spectral shape measured with the OMTD changes by less than 10\% down to $D = 0.5$\,mm (inset of Fig.~\ref{fig:FTIR_aperture}(c)). We have not established the origin of the spectral shift seen by the DLaTGS at small apertures. It does enter the comparison of main text Fig.~4(e), since the crossover lies between two DLaTGS settings at $D = 0.5$\,mm, whereas the OMTD was measured only at $D = 3$\,mm.

\subsection{Calibration of the power incident on the DLaTGS detector and on the OMTD}
\label{subsec:FTIR_power}
The horizontal axis of main text Fig.~4(e) is the ratio between the power incident on the DLaTGS detector, $P_\text{DLaTGS}$, and the power incident on the Pt disk of the OMTD during the measurement of Fig.~4(d), $P_\text{OMTD}$. For the DLaTGS, we measure the power of the beam at the sample position with a thermal power sensor (Thorlabs S401C) for the open setting at $D = 3$\,mm, obtaining 350\,\si{\micro\watt}, and assume that this power reaches the detector without further losses. The power at the other settings follows by scaling this value with the filter and $D^2$ laws of Sections~\ref{subsec:FTIR_filters} and~\ref{subsec:FTIR_aperture}. The $D^2$ dependence of Section~\ref{subsec:FTIR_aperture} shows that the $2\,\si{\milli\meter}\times 2\,\si{\milli\meter}$ active area of the DLaTGS is not overfilled up to $D = 4$\,mm, which implies that the detector optics image the aperture onto it with a magnification of at most about 0.5. The mirrors between the sample position and the detector, the detector window and the atmospheric absorption along this part of the unpurged beam path, however, reduce the power that reaches it, so the no-loss assumption overestimates $P_\text{DLaTGS}$ and therefore favours the OMTD in the comparison of Fig.~4(e).

For the OMTD, we infer $P_\text{OMTD}$ from the steady-state frequency shift away from zero path difference, $\Delta f_\text{m}\approx 560$\,Hz at $f_\text{m} = 551.1$\,kHz (Fig.~4(b) of the main text), as $P_\text{OMTD} = \Delta f_\text{m}/(f_\text{m}\,|\mathcal{R}_0^\text{inc}|)$, with $|\mathcal{R}_0^\text{inc}| = (2.65\pm0.18)\times10^3$\,W$^{-1}$ (main text). This assumes that the spectrum of the spectrometer reaching the device has the same absorbed fraction, $\langle A\rangle_\text{S} = 0.42$, as that of the fibre-coupled thermal source used for the calibration (Section~\ref{subsec:thermal_source_calibration}), and that the responsivity measured with the 0.74\,mm calibration spot also holds for the millimetre-scale spot of the spectrometer. Both spots are much larger than the 120\,\si{\micro\meter} resonator, which therefore sees a nearly uniform irradiance in both cases, and the responsivity depends only weakly on the size of the heat source (Section~\ref{subsec:spot_size}). We obtain $P_\text{OMTD}\approx 0.38$\,\si{\micro\watt}, about 900 times less than the power incident on the DLaTGS at the same aperture.

Figure~4(e) of the main text then compares the wavelength-averaged standard error $\langle\sigma_T\rangle_\lambda$ obtained with the DLaTGS at the different settings, as a function of $P_\text{DLaTGS}/P_\text{OMTD}$, with the value obtained with the OMTD. The horizontal error bars combine the uncertainty of $\mathcal{R}_0^\text{inc}$ (6.8\%), common to all points, with the uncertainties of the filter law (2\%) and of the $D^2$ law (8\%) for the settings where they apply. They do not include the uncertainty of the power-meter reading or the one-sided effect of the losses discussed above. The crossover is obtained by interpolating, on logarithmic scales, between the two DLaTGS points that bracket the OMTD value. Its uncertainty (shaded band) adds to the horizontal uncertainty of these points the 2\% uncertainty of the OMTD value, estimated by bootstrapping over its scans, converted into power through the local slope. The DLaTGS reaches the same $\langle\sigma_T\rangle_\lambda$ at $P_\text{DLaTGS}/P_\text{OMTD} = 12.0\pm1.3$, that is, with roughly an order of magnitude more power incident on its absorber.

\subsection{Effect of the interferometer scan rate}
\label{subsec:FTIR_scanrate}
\begin{figure}[!htbp]
    \centering
    \includegraphics[width=\textwidth]{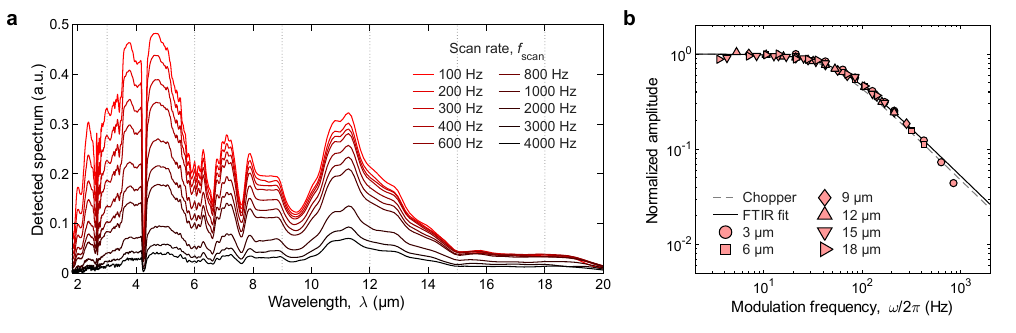}
    \caption{\textbf{Effect of the interferometer scan rate.} \textbf{a}, Spectra of the spectrometer's thermal source detected with the OMTD without sample for the scan rates $f_\text{scan}$ indicated (aperture $D = 3$\,mm, resolution 16\,cm$^{-1}$, one single-scan spectrum per scan rate). Dotted lines mark the wavelengths evaluated in \textbf{b}. \textbf{b}, Amplitude at these wavelengths, averaged over $\pm0.1$\,\si{\micro\meter}, as a function of the modulation frequency $\omega/2\pi = (\lambda_\text{HeNe}/\lambda)\,f_\text{scan}$ and divided by the gain of a joint fit to a first-order roll-off with a common cutoff (solid line, $\omega_\text{th}/2\pi = 52.5$\,Hz). The dashed line is the response measured with the chopper ($\omega_\text{th}/2\pi = 47.4$\,Hz, main text Fig.~3(h)).}
    \label{fig:FTIR_scanrate}
\end{figure}

In an FTIR spectrometer, scanning the optical path difference of the interferometer converts each wavelength into a modulation of the detected power at the frequency $\omega/2\pi = (\lambda_\text{HeNe}/\lambda)\,f_\text{scan}$, where $f_\text{scan}$ is the fringe frequency of the reference laser ($\lambda_\text{HeNe} = 632.8$\,nm). Varying $f_\text{scan}$ therefore probes the frequency response of the detector at all wavelengths at once, independently of the chopper-based measurement of main text Fig.~3(h). Figure~\ref{fig:FTIR_scanrate}(a) shows the spectra detected with the OMTD for $f_\text{scan}$ between 100\,Hz and 4\,kHz. The detected spectrum decreases at higher scan rates, first at short wavelengths, where the modulation frequency is highest. Figure~\ref{fig:FTIR_scanrate}(b) shows the amplitude at six wavelengths between 3 and 18\,\si{\micro\meter} as a function of $\omega/2\pi$. A joint fit to a first-order roll-off, $A_i/\sqrt{1+(\omega/\omega_\text{th})^2}$, with a common cutoff and wavelength-specific gains $A_i$, collapses all wavelengths onto a single curve and gives $\omega_\text{th}/2\pi = 52.5\pm1.4$\,Hz. This is about 11\% above the value obtained with the chopper, $47.4\pm0.9$\,Hz (main text), a difference whose origin we have not identified. The collapse confirms that the thermal response of the OMTD does not depend on the wavelength of the absorbed radiation. At the scan rate of main text Fig.~4 ($f_\text{scan} = 400$\,Hz), the interferogram occupies modulation frequencies between about 13 and 130\,Hz for $\lambda\in[2,20]$\,\si{\micro\meter}, around the minimum of the NEP at 40--50\,Hz (main text Fig.~3(j)).

\clearpage
\section{Comparison with the state of the art}
\label{sec:SoA}

\begin{figure}[t]
    \centering
    \includegraphics[width=\textwidth]{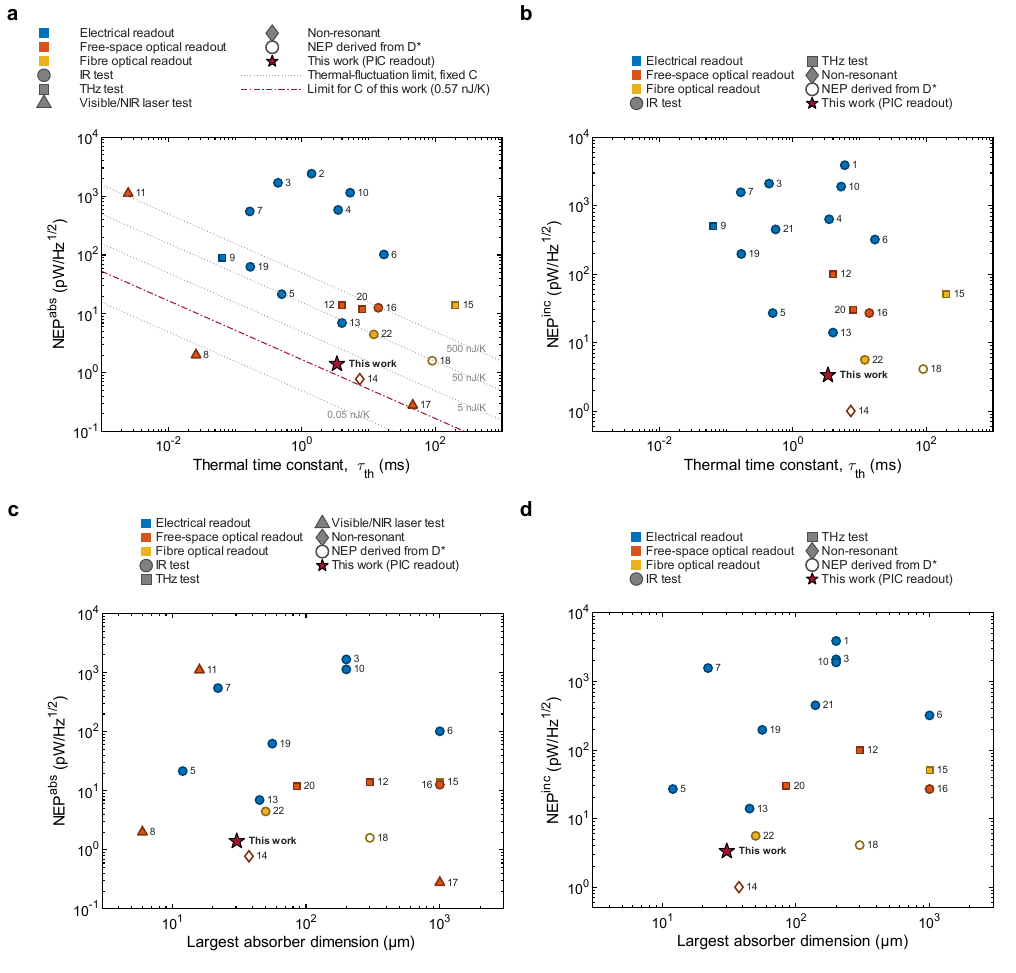}
    \caption{\textbf{Comparison with room-temperature mechanical thermal detectors.} NEP referred to absorbed (\textbf{a}, \textbf{c}) and incident (\textbf{b}, \textbf{d}) power, versus thermal time constant (\textbf{a}, \textbf{b}) and largest absorber dimension (\textbf{c}, \textbf{d}). Colours give the readout, and shapes the radiation used for the characterization, with a diamond for the non-resonant detector. Open symbols mark NEPs derived from the specific detectivity. Entries characterized with visible or near-infrared lasers appear in \textbf{a} and \textbf{c} only. In \textbf{a}, the dotted lines give the thermal-fluctuation limit for the indicated heat capacities, and the dash-dotted line the limit for the heat capacity of our resonator. Numbers refer to Table~\ref{tab:SoA}, and the star marks this work.}
    \label{fig:SoA}
\end{figure}

Figure~\ref{fig:SoA} and Table~\ref{tab:SoA} compare our detector with room-temperature mechanical thermal detectors. All of them transduce the absorbed power into a shift of a mechanical resonance frequency, read out electrically~\cite{pisani_application_2011,hui_high_2013,hui_plasmonic_2016,qian_narrowband_2016,laurent_12-m-pitch_2018,piller_nanoelectromechanical_2019,qian_high_2019,zhang_fast_2019,hui_uncooled_2021,piller_thermal_2023,xi_room-temperature_2025,gulseren_high_2026} or optically~\cite{blaikie_fast_2019,papas_optomechanical_2021,vicarelli_micromechanical_2022,zhang_high_2024,martini_uncooled_2025,kanellopulos_comparative_2025,zhang_enhanced_2025,alborghetti_enhanced_2026,he_uncooled_2026}, except one, a non-resonant detector whose deflection is read out interferometrically~\cite{das_thermodynamically_2023}. We do not include the thin-film lithium niobate optomechanical resonator of Ref.~\cite{yu_uncooled_2026}, whose NEP of 6.2\,\si{\nano\watt/\sqrt{\hertz}} at 10\,kHz in air is inferred from its temperature coefficient, measured by tuning the chip temperature with thermoelectric coolers, and from a simulated thermal conductance, rather than from absorbed radiation. For each reference, we take the device highlighted by the authors and its lowest reported NEP, as printed whenever possible. When the NEP is given referred to only one of the incident or absorbed powers, we obtain the other with the absorptance stated in the reference, and when only the specific detectivity $D^*$ is given, we use $\text{NEP}=\sqrt{A_\text{d}}/D^*$, with $A_\text{d}$ the detector area. The references define the NEP in different ways, from Allan deviations at a given integration time to noise spectral densities at a given modulation frequency, and do not always refer the incident power to the absorber area, so that the comparison is indicative at about the factor-of-two level.

For a thermal detector of heat capacity $C$ and thermal time constant $\tau_\text{th}=C/G_\text{th}$, temperature fluctuations set a lower bound on the NEP referred to absorbed power, $\text{NEP}_\text{th}=\sqrt{4k_BT_0^2G_\text{th}}=\sqrt{4k_BT_0^2C/\tau_\text{th}}$ (Section~\ref{subsec:NEPth}). The dotted lines in Fig.~\ref{fig:SoA}(a) show this bound for fixed heat capacities between 0.05 and 500\,\si{\nano\joule/\kelvin}. Along each line a faster detector has a higher NEP, and a smaller heat capacity shifts the line down. Since most references do not report the heat capacity, these lines serve as guides rather than as limits for individual devices. For our resonator ($C\approx 0.57$\,\si{\nano\joule/\kelvin}, dash-dotted line), the bound is 0.91\,\si{\pico\watt/\sqrt{\hertz}} at $\tau_\text{th}=3.36$\,ms, a factor of 1.5 below our measured value. All electrically read detectors lie above ours, the best of which reaches 7\,\si{\pico\watt/\sqrt{\hertz}}~\cite{piller_thermal_2023}. Lower or comparable values have been reported only with free-space or fibre-coupled interferometric readout: by SiN resonators with time constants 14 to 27 times longer than ours~\cite{kanellopulos_comparative_2025,zhang_enhanced_2025}, by a graphene drum 130 times faster but with a slightly higher NEP~\cite{blaikie_fast_2019}, and by the non-resonant detector~\cite{das_thermodynamically_2023}. Among resonant detectors, only the SiN resonators of Refs.~\cite{zhang_demonstration_2023,zhang_enhanced_2025} have been shown to operate at their thermal-fluctuation limit, with time constants of about 90\,ms.

Three entries were characterized with visible or near-infrared lasers~\cite{blaikie_fast_2019,papas_optomechanical_2021,kanellopulos_comparative_2025}. Two of them, a graphene drum and a SiN string, absorb a few percent or less at the test wavelength, as assumed or modelled by the authors~\cite{blaikie_fast_2019,kanellopulos_comparative_2025}. Turning such resonators into efficient infrared absorbers requires an absorber, which adds heat capacity and modifies their stress and damping, so that a low NEP to absorbed power does not carry over to the incident power. Figures~\ref{fig:SoA}(b) and (d) therefore show the NEP referred to incident power only for detectors characterized with infrared or THz radiation. Our value of 3.4\,\si{\pico\watt/\sqrt{\hertz}} includes the heat capacity of the Pt absorber and the photothermal noise from the residual absorption of the readout light. Referred to the area of the Pt disk, $A_\text{d} = \pi R_\text{Pt}^2$, it corresponds to a specific detectivity $D^* = \sqrt{A_\text{d}}/\text{NEP} \approx 8\times10^{8}$\,cm\,Hz$^{1/2}$\,W$^{-1}$, and about $1.5\times10^{9}$\,cm\,Hz$^{1/2}$\,W$^{-1}$ for monochromatic radiation at 8.17\,\si{\micro\meter} (Section~\ref{subsec:NEPmono}).

Referred to incident power (Fig.~\ref{fig:SoA}(b)), only the non-resonant detector reaches a lower NEP at comparable speed, about 1\,\si{\pico\watt/\sqrt{\hertz}} at $\tau_\text{th}=7.4$\,ms with a 37.5\,\si{\micro\meter} absorber whose deflection is imaged with a free-space interferometer~\cite{das_thermodynamically_2023}. Among resonant detectors, the closest are two SiN trampolines read out with fibre-coupled interferometers, at about 4\,\si{\pico\watt/\sqrt{\hertz}} with a 27 times longer time constant and a 300\,\si{\micro\meter} absorber~\cite{zhang_enhanced_2025}, and at 5.6\,\si{\pico\watt/\sqrt{\hertz}} with a 3.6 times longer time constant and a 50\,\si{\micro\meter} absorber~\cite{he_uncooled_2026}, while all electrically read detectors lie at or above 14\,\si{\pico\watt/\sqrt{\hertz}}. Figures~\ref{fig:SoA}(c) and (d) show the same NEPs versus the largest absorber dimension, which sets a lower bound on the pitch of an array. Among resonant detectors with dimensions between 10 and 100\,\si{\micro\meter}, ours has the lowest NEP, by factors of about 3 and 1.7 referred to absorbed and incident power, respectively~\cite{he_uncooled_2026}.

Overall, our detector reaches noise-equivalent powers that, among mechanical thermal detectors, have so far been obtained only with free-space or fibre-coupled interferometric readout, here with a photonic readout integrated on the same chip.

\begin{sidewaystable}
\centering
\caption{\textbf{Mechanical thermal detectors compared in Fig.~\ref{fig:SoA}.} One device per reference, the one highlighted by the authors. Readout class: E, electrical, O, free-space optical, F, fibre optical, I, photonic integrated circuit (PIC). $f_0$, resonance frequency. $A$, absorptance for the test radiation. Size: absorber dimensions ($D$, diameter). NEP in \si{\pico\watt/\sqrt{\hertz}}, referred to incident (inc.) and absorbed (abs.) power, with values in parentheses converted by us with the absorptance stated in the reference. Last column: Allan-deviation integration time, measurement bandwidth (BW) or modulation frequency $f_\text{mod}$ at which the NEP is evaluated. $^a$Absorptance assumed, calculated or simulated by the authors. $^b$Time constant simulated, calculated or measured on a similar device. $^c$Derived from the specific detectivity. $^d$Incident power referred to an area larger than the absorber. $^e$Read from a figure. $^f$Non-resonant, deflection read out interferometrically. $^g$Electrical (heater) calibration.}
\label{tab:SoA}
\scriptsize
\setlength{\tabcolsep}{4pt}
\renewcommand{\arraystretch}{1.2}
\begin{tabular}{rclllllllccl}
\toprule
\# & Ref. & Resonator, material (absorber) & Readout & $f_0$ (MHz) & Radiation & $A$ (\%) & Size (\si{\micro\meter}) & $\tau_\text{th}$ (ms) & NEP inc. & NEP abs. & NEP evaluated at \\
\midrule
1 & \cite{pisani_application_2011} & BAW pixel, quartz & Impedance, E & 241 & Broadband IR & -- & $D=200$ & 6$^b$ & 3900 & -- & BW 1\,Hz \\
2 & \cite{hui_high_2013} & Nano-plate, AlN (Si$_3$N$_4$) & Piezoelectric, E & 197 & 0.9--4.5\,\si{\micro\meter} & -- & -- & 1.4$^b$ & -- & 2400 & BW 1\,Hz \\
3 & \cite{hui_plasmonic_2016} & Plate, AlN (plasmonic) & Piezoelectric, E & 161 & 2--16\,\si{\micro\meter} & 80 & $200\times75$ & 0.44 & 2100 & (1680) & BW 100\,Hz \\
4 & \cite{qian_narrowband_2016} & Plate, AlN (plasmonic) & Piezoelectric, E & 179 & $\approx$5\,\si{\micro\meter} & 92 & -- & 3.5$^b$ & 633 & (582) & Allan, 1\,ms \\
5 & \cite{laurent_12-m-pitch_2018} & Paddle, Si/SiN (TiN) & Capacitive, E & $\approx$1 & 8--12\,\si{\micro\meter} & 80$^a$ & 12 (pitch) & 0.5 & 27 & (22) & Allan, 20\,ms \\
6 & \cite{piller_nanoelectromechanical_2019} & Drum, SiN (thin film) & Magnetomotive, E & 0.073 & 9.5\,\si{\micro\meter} & 32 & $1000\times1000$ & 17 & 320 & (102) & Allan, 40\,ms \\
7 & \cite{qian_high_2019} & Nano-plate, AlN (Pt) & Piezoelectric, E & $\approx$800 & Blackbody, \SI{900}{\celsius} & 30--40 & $20\times22$ & 0.17 & (1570) & 549 & Allan, 0.25\,ms \\
8 & \cite{blaikie_fast_2019} & Drum, graphene & Interferometric, O & 16.6 & 532\,nm & 2.3$^a$ & $D=6$ & 0.026 & (87) & 2 & Allan, 1\,ms \\
9 & \cite{zhang_fast_2019} & Beam, GaAs (NiCr) & Piezoelectric, E & 0.50 & 0.25--3\,THz & 20$^a$ & -- & 0.064 & 500$^d$ & 90$^g$ & $f_\text{mod}$ = 1\,kHz \\
10 & \cite{hui_uncooled_2021} & Plate, AlN (metasurface) & Piezoelectric, E & 172 & 3--5\,\si{\micro\meter} & 60 & $200\times75$ & 5.3$^b$ & 1900 & (1140) & BW 200\,Hz \\
11 & \cite{papas_optomechanical_2021} & Nanowires, SiN/Au & Transmission, O & 2.0 & 1.31\,\si{\micro\meter} & 37 & $16\times0.8$ & 0.0025 & 3000 & (1110) & Noise density \\
12 & \cite{vicarelli_micromechanical_2022} & Trampoline, SiN (Cr/Au) & Self-mixing, O & 0.080 & 0.14\,THz & 14$^a$ & $300\times300$ & 4 & 100$^d$ & (14) & $f_\text{mod}$ = 7--14\,Hz \\
13 & \cite{piller_thermal_2023} & Trampoline, SiN (Pt) & Magnetomotive, E & $\approx$0.025 & Broadband IR & 50 & $45\times45$ & 4 & (14) & 7 & Allan, 40\,ms \\
14 & \cite{das_thermodynamically_2023} & Membrane$^f$, Ti/SiN$_x$/Ni & Interferometric, O & -- & 8--12\,\si{\micro\meter} & 78 & $37.5\times37.5$ & 7.4 & 1.0$^c$ & (0.78)$^c$ & BW 10\,Hz \\
15 & \cite{zhang_high_2024} & Drum, SiN (Ti metasurface) & Interferometric, F & 0.124 & 0.5--4\,THz & 27 & $D=1000$ & 200$^e$ & 51 & (14) & Spectral minimum \\
16 & \cite{martini_uncooled_2025} & Membrane, SiN (Pt) & Laser Doppler, O & -- & Broadband IR & 47 & $1000\times1000$ & 14$^b$ & 27 & (12.7) & $f_\text{mod}=1/2\pi\tau_\text{th}$ \\
17 & \cite{kanellopulos_comparative_2025} & String, SiN & Laser Doppler, O & 0.18$^e$ & 633\,nm & 0.5$^a$ & $1000\times5$ & 46$^e$ & (56) & 0.28 & $f_\text{mod}=1/2\pi\tau_\text{th}$ \\
18 & \cite{zhang_enhanced_2025} & Trampoline, SiN & Interferometric, F & 0.060 & Broadband IR & 40$^a$ & $300\times300$ & 90 & 4.1$^c$ & (1.6)$^c$ & From $D^*$ \\
19 & \cite{xi_room-temperature_2025} & Phononic cavity, LN (Au) & Piezoelectric, E & 1000 & 6.3\,\si{\micro\meter} & 32$^a$ & $48\times56$ & 0.17 & 197 & (63) & $f_\text{mod}$ = 1.8\,kHz \\
20 & \cite{alborghetti_enhanced_2026} & Trampoline, SiN (PyC) & Interferometric, O & 0.52 & 0.14\,THz & 40 & $60\times85$ & 8 & 30 & (12) & Nonlinear drive \\
21 & \cite{gulseren_high_2026} & Contour mode, AlN (metasurface) & RF reflectometry, E & 202 & 5.94\,\si{\micro\meter} & -- & $140\times60$ & 0.55 & 450 & -- & $f_\text{mod}$ = 200--400\,Hz \\
22 & \cite{he_uncooled_2026} & Trampoline, SiN (Au mirror) & Fabry--P\'erot, F & 0.451 & 10.6\,\si{\micro\meter} & 79.8 & $50\times50$ & 12 & (5.6) & 4.48 & $f_\text{mod}\approx 9$\,Hz \\
\midrule
\multicolumn{2}{l}{This work} & Trampoline, Si$_x$N$_y$ (Pt) & Integrated microring, I & 0.55 & 1550\,K thermal source & 42 & $D=30.4$ & 3.36 & 3.4 & 1.4 & $f_\text{mod}$ = 40--50\,Hz \\
\bottomrule
\end{tabular}
\end{sidewaystable}

\end{document}